%% file: main.tex
\pdfoutput=1
\documentclass[aps,prd,reprint,superscriptaddress,nofootinbib,floatfix,amsmath,amssymb]{revtex4-2}
\usepackage{graphicx}
\usepackage{bm}
\usepackage{dsfont}
\usepackage{hyperref}
\hypersetup{colorlinks=true,linkcolor=blue,citecolor=blue,urlcolor=blue,
  pdftitle={A Casimir turning point as the beginning of a small flat universe},
  pdfauthor={Alan Zhang}}
\usepackage{orcidlink}
\graphicspath{{figures/}}

\input{dnec_macros}
\input{numbers}
\newcounter{paassump}%

\newcommand{\sNeff}{N_C^{\mathrm{eff}}}%
\newcommand{\sgam}{\gamma_C}%
\newcommand{\sVar}{\operatorname{Var}}
\newcommand{\sCov}{\operatorname{Cov}}
\newcommand{\sTm}{\mathcal{T}}%
\newcommand{\sGain}{\mathcal{G}}%
\newcommand{\sHeff}{H_{\mathrm{eff}}}%
\newtheorem{stheorem}{Theorem}

\begin{document}

\title{A Casimir turning point as the beginning of a small flat universe}

\author{Alan Zhang\,\orcidlink{0009-0000-1187-0755}}
\email[Contact author: ]{zhana142@McMaster.ca}
\affiliation{Department of Physics and Astronomy, McMaster University, Hamilton, Ontario L8S 4M1, Canada}
\affiliation{Perimeter Institute for Theoretical Physics, Waterloo, Ontario N2L 2Y5, Canada}

\begin{abstract}
We study a flat three-torus universe whose contraction is turned around by the negative Casimir energy of Standard Model fields, and ask what nonsingularity requires. The torus turns at an edge of $722$ reduced Planck lengths, where it can hold only about five quanta, so the turning point is a nearly empty, low-entropy beginning of both branches rather than a bounce. The negative Casimir energy cannot hold the shape of the torus. To stay regular on both branches, the universe must begin with nearly equal edges, and the uncertainty principle then leaves a fossil, a nearly rectangular box whose sides typically differ by $16$--$18\%$, which the microwave background can read if the torus is close to the size of the last-scattering sphere. Although the Higgs quartic coupling is negative at the turn, the electroweak vacuum survives with probability above one half if and only if the top-quark pole mass is below $172.96$~GeV, a criterion within reach of FCC-ee. %
\end{abstract}

\maketitle

\input{sections/v13_s01_intro}
\input{sections/v13_s02_turning_point}

\input{sections/v13_s03_beginning}

\input{sections/v13_s04_shape}

\input{sections/v13_s05_higgs}

\input{sections/v13_s06_observables}
\input{sections/v13_s07_discussion}

\begin{acknowledgments}
The author acknowledges graduate scholarship and research support through McMaster University. Research at Perimeter Institute is supported in part by the Government of Canada through the Department of Innovation, Science and Economic Development and by the Province of Ontario through the Ministry of Colleges, Universities, Research Excellence and Security. AI assistants, Anthropic Claude Code (model Claude Opus~5.5) and OpenAI Codex (model gpt-6-astra), were used for literature screening, for parts of the derivations, the development of the numerical code and the computations, and for manuscript drafting and English-language editing. Key results were reproduced from a written specification of the model by separately instructed AI agents without access to the author's code or derivations (the shape check was given the method sections of two derivations, but not their results); these are the blind reimplementations quoted in Secs.~\ref{sec:higgs} and~\ref{sec:shape}. OpenAI Codex also carried out, under the author's direction and with its own code and conventions, the separate line of calculation summarized in Ref.~\cite{v12Report}, Sec.~21. Symbolic steps were checked in SymPy or Wolfram Language, and every number quoted here is reproduced by the scripts of Ref.~\cite{DNECv12code} or contained in its archived outputs, including those of the blind reimplementations. The author checked all derivations, numbers and conclusions and takes full responsibility for this work.
\end{acknowledgments}

\section*{Data Availability}
The code and numerical data that support the findings of this article are openly available~\cite{DNECv12code}; a development version is maintained at \url{https://github.com/Ailun-Zhang/casimir-cosmology}. Step-by-step derivations, alternative methods, convergence tests and a reproduction guide are given in the companion technical report~\cite{v12Report}.

\appendix
\predisplaypenalty=0\clubpenalty=150\widowpenalty=150\displaywidowpenalty=150%
\input{appendices/v13_aA_trace}
\input{appendices/v13_aB_casimir}
\input{appendices/v13_aC_shape}

\input{appendices/v13_aD_higgs_model}

\bibliography{refs}

\end{document}

%% file: dnec_macros.tex
\newcommand{\MPbar}{\bar M_{\mathrm{P}}}         %
\newcommand{\Tr}{\operatorname{tr}}
\newcommand{\NC}{N_C}                            %
\newcommand{\GeV}{\,\mathrm{GeV}}
\newcommand{\MSbar}{\overline{\mathrm{MS}}}
\newcommand{\mt}{m_t}                            %
\newcommand{\mtbar}{\bar m_t}                    %
\newcommand{\mub}{\mu_b}                         %
\newcommand{\Lb}{L_b}                            %
\newcommand{\lamc}{\lambda_c}                    %
\newcommand{\xiH}{\xi_H}                         %
\newcommand{\xic}{\xi_c}                         %
\newcommand{\Pone}{P_1}                          %
\newcommand{\Pbar}{\bar P_1}                     %
\newcommand{\fzm}{f_0}                           %
\newcommand{\dLSS}{d_{\mathrm{LSS}}}             %
\newcommand{\rT}{r_{\mathrm{T}}}                 %
\newcommand{\sfloor}{\sigma_{\mathrm{floor}}}

%% file: numbers.tex
\newcommand{\PbarOne}{0.73}        %
\newcommand{\PbarMSR}{0.64}        %
\newcommand{\PbarSigma}{0.86}      %
\newcommand{\mExclFCC}{173.7}      %
\newcommand{\xiHmin}{0.08}         %
\newcommand{\xiHcentral}{0.074}    %
\newcommand{\xiHhalf}{0.12}        %
\newcommand{\marginGeV}{0.36}      %
\newcommand{\marginLam}{0.0024}    %
\newcommand{\mcMinusMzero}{1.9}    %
\newcommand{\modelErrUp}{0.07}     %
\newcommand{\modelErrDn}{0.08}     %
\newcommand{\modelErrUpPnine}{0.04}     %
\newcommand{\modelErrDnPnine}{0.06}     %
\newcommand{\modelErrUpPone}{0.12}      %
\newcommand{\modelErrDnPone}{0.13}      %
\newcommand{\modelErrUpNoViPnine}{0.02} %
\newcommand{\modelErrDnNoViPnine}{0.04} %
\newcommand{\modelErrUpNoVi}{0.03}      %
\newcommand{\modelErrDnNoVi}{0.05}      %
\newcommand{\modelErrUpNoViPone}{0.04}  %
\newcommand{\modelErrDnNoViPone}{0.06}  %
\newcommand{\xiHminEW}{0.09}            %
\newcommand{\mSurvFCC}{172.7}         %

%% file: sections/v13_s01_intro.tex
\section{Introduction}
\label{sec:intro}

General relativity gives an expanding universe a singular beginning when matter obeys the standard energy conditions~\cite{HawkingPenrose1970}, and inflation does not remove the singularity, since a spacetime that expands on average is geodesically incomplete to the past~\cite{BordeGuthVilenkin2003}.
A spatially flat universe can avoid the singularity only if the null energy condition fails.
In the simplest case contraction turns into expansion, and the total energy density must become non-positive at the turning point~\cite{Molina-Paris:1998xmn,Novello:2008ra}.
In Ref.~\cite{Zhang:2026gva} we asked how much of the averaged condition must fail: a null-complete, homogeneous and anisotropic flat universe violates the averaged null energy condition along every complete light ray outside at most four directions, by an amount fixed by the expansion and shear that the ray crosses.
Quantum fields obey the averaged condition in Minkowski space~\cite{Faulkner:2016mzt,Hartman:2016lgu} but can violate it in curved spacetime~\cite{Graham:2007va,Kontou:2020bta}. They are expected to obey it along achronal rays~\cite{Graham:2007va,Kontou:2015yha}, so such a universe can be nonsingular only if space is compact and light rays circumnavigate it.
The present work grew out of that result.
It asks what a nonsingular flat universe looks like when its matter is the matter we know.

The first question is what can supply a violation of this size.
Classical fields that violate the null energy condition by that much tend to carry ghost or gradient instabilities~\cite{Dubovsky:2005xd}.
Kinetic gravity braiding avoids these instabilities while the condition is violated~\cite{Deffayet:2010qz,Easson:2011zy,Ijjas:2016tpn}, but in Horndeski theories, which include it, a nonsingular bounce cannot avoid them at all times~\cite{Kobayashi:2016xpl,Dobre:2017pnt}.
Quantum fields offer a source without new classical degrees of freedom: on a compact flat space, the vacuum energy density of untwisted massless bosons is negative and grows in magnitude as $L^{-4}$ when the space shrinks~\cite{DeWitt:1979dd}.
Together with a positive vacuum energy, this Casimir energy turns a contracting three-torus around along an exact, regular solution found by Zel'dovich and Starobinsky~\cite{Zeldovich:1984vk} and written in closed form by Guth and Vilenkin~\cite{Guth:2025dal}; toroidal universes that begin small were also studied in Refs.~\cite{Coule:1999wg,Linde:2004nz,Fornal:2026qfq}.
The compact space required by the null-energy argument thus supplies its own negative energy.
The light rays of this solution wind around the torus and are therefore not achronal, so the solution violates the averaged condition without contradicting the achronal one~\cite{Zhang:2026gva}.

Here we take the matter to be the Standard Model plus an inflaton on a plateau, and ask what follows in this scenario from the single requirement that the universe be nonsingular.
That requirement fixes much of what a cosmological model ordinarily assumes.
Without an $R^2$ term, the trace anomaly prevents conformal matter by itself from turning an anisotropic contraction around; the Casimir energy must therefore be accompanied by matter with $w<1/3$, most cheaply the positive potential of the inflaton (Sec.~\ref{sec:trace}).
With fermions antiperiodic around all three cycles, the Standard Model provides the negative energy, and the Friedmann constraint places the turning point at an edge of $722$ reduced Planck lengths. %
At a turning point where the torus is nearly cubic, as survival of both branches requires in the aspect directions (Sec.~\ref{sec:v13-meanshape}), the same constraint leaves room for only about five quanta of the lowest modes. %
The turn is therefore nearly empty whatever led to it: its entropy is at most a few tens, and entropy grows away from it along each branch (Sec.~\ref{sec:gate}). %
The turning point is the beginning of both branches, and its low entropy follows from the constraint rather than from a separate postulate.

The negative energy that turns the volume around also acts on the shape of the torus, and it cannot hold a shape, because a negative Casimir energy of massless fields has no shape minimum.
If the deformation runs away, one of the cycles collapses; a universe that stays nonsingular on both branches therefore turns with its edges nearly equal, and quantum mechanics restricts the two unstable shape modes there to a finite set of about $9\times10^2$ states. %
We take the state of the shape to be symmetric under the time reversal that exchanges the two branches (row~\ref{v13:assump-S5} of Table~\ref{v13:tab-assumptions}), and prove that a shape state can be time-symmetric about at most one instant.
With the survival region that we determine numerically, the shape then survives on both branches with probability $90\%$ only if that instant lies within $0.4/h$ of the turn, where $1/h$ is the Hubble time of the plateau. %
For a state that is symmetric about the turn itself, the uncertainty principle sets a floor under the typical late-time deformation.
As our benchmark we choose the cube, one of the two shapes at which the Casimir energy is stationary.
In the Gaussian state of minimal energy variance that is symmetric about the turn and centered on the cube, the torus freezes into a nearly rectangular box whose longest and shortest sides typically differ by $16$--$18\%$ (Sec.~\ref{sec:shape}). %
This frozen deformation is the shape fossil of the beginning.

The Standard Model itself puts such a beginning at risk.
The measured Higgs and top-quark masses place its vacuum near the boundary between stability and metastability~\cite{Degrassi:2012ry,Buttazzo:2013uya,Bednyakov:2015sca,Andreassen:2017rzq}: run to high energies, the Higgs quartic coupling $\lambda$ turns negative far below the Planck scale and is already negative at the scale of the turn.
Vacuum survival in the early universe has been studied in a spacetime that is already large and inflating, where Higgs fluctuations feel the curvature through the non-minimal coupling $\xiH$~\cite{Espinosa:2007qp,Herranen:2014cua,Kearney:2015vba,Espinosa:2015qea,Joti:2017fwe,Markkanen:2018pdo,Mantziris:2020rzh,Mantziris:2022fuu,Herranen:2015ima}, and across the singular bounces of a cyclic universe~\cite{Bars:2013vba}.
On a small torus that turns around regularly, the question takes a different form.
The relevant degree of freedom is the Higgs zero mode, the spatially constant part of the Higgs field, and a small compact volume speeds up its decay~\cite{Hassan:2026cmp,Brown:2015kgj,Antoniadis:2024ent}.
The gauge fields, however, have spatially constant modes of their own, whose zero-point energy grows linearly with the Higgs field and holds the Higgs zero mode in place down to a finite-volume threshold of the quartic coupling, established for a static box in the companion Letter~\cite{v12PRL}.
Across the turn this threshold is $\lamc=-0.0121$, below the value $-0.0098$ that the Standard Model gives at the measured masses. %
The electroweak vacuum therefore survives the turn on each branch with probability above one half for a top-quark pole mass below $172.96$\,GeV, for every curvature coupling $\xiH$ compatible with vacuum stability during and after inflation (Sec.~\ref{sec:higgs}). At FCC-ee, a measured pole mass above $\mExclFCC$\,GeV would exclude this beginning at $95\%$ confidence. %

Little of this beginning remains observable.
With the fields in their vacuum at the turn, inflation erases every local imprint, and the primordial spectra are those of the plateau.
The global shape of space survives, and the cosmic microwave background can read it if the shortest cycle of the torus lies between $0.985$ and about $1.04$ times the diameter of the last-scattering sphere (Sec.~\ref{sec:spectra}). %
The model does not select the present size of the torus, so this prediction is conditional; the top-quark criterion does not depend on the size.

Several ingredients of this picture have been studied separately.
Curvature fixes the trace of the stress tensor of conformal fields in every state~\cite{Duff:1993wm}.
Anomaly-driven cosmologies~\cite{Starobinsky:1980te,Fischetti:1979ue} and the damping of anisotropy by quantum effects~\cite{Hu:1978zd,Hartle:1979uf,Hartle:1980nn} have a long history, and earlier Casimir bounces used a single scalar field~\cite{Herdeiro:2005zj,Godlowski:2007gx}.
Standard Model Casimir energies on circles and tori were computed for compactifications to lower dimensions~\cite{Arkani-Hamed:2007ryu,Fornal:2011tw,Hamada:2017yji}, and the constant modes of gauge fields on a small torus give the leading order of the small-volume expansion of Yang--Mills theory~\cite{Luscher:1982uv,Luscher:1982ma,Luscher:1983gm,Weisz:1986vu,vanBaal:1988va,vanBaal:2000zc}.
Fornal proposed that our universe began as an empty, Planck-size three-torus with periodic boundary conditions, whose positive Casimir energy drives its expansion~\cite{Fornal:2026qfq}.
That scenario has no turning point and does not treat the Higgs zero mode, and its positive energy assumes trivial holonomy, which for periodic fermions is not the one-loop vacuum.
Guth and Vilenkin found that regularity at nucleation confines the Bogoliubov coefficients of the field modes to a narrow range near zero~\cite{Guth:2025dal}, and in the isotropic Casimir torus of Mithani and Vilenkin, excited radiation lowers the Casimir coefficient~\cite{Mithani:2014toa}.
The emptiness bound of Sec.~\ref{sec:gate} is a Lorentzian counterpart of the first observation and a quantized form of the second.
Finally, a torus smaller than the last-scattering sphere produces matched circles in the microwave background~\cite{Lachieze-Rey:1995qrb,Cornish:1997ab}, whose absence bounds its shortest translation from below~\cite{Vaudrevange:2012da,COMPACT:2022nsu}, whereas a somewhat larger torus still imprints correlations that are detectable in principle~\cite{COMPACT:2023rkp,Himeno:2026lgh}.

To our knowledge, three results here are new in kind: the entropy bound that makes the turning point a beginning; the theorem that bounds the instant of time symmetry of the shape state, with its consequence for the shape of space; and, together with the companion Letter, the finite-volume threshold for the Higgs zero mode and the top-mass criterion that follows from it.
We also derive the trace obstruction, the Standard Model Casimir weights and shape Hessian for non-periodic spin structures, and the extension of the shape no-go result to thermal states, zero modes and holonomies.

Section~\ref{v13:sec-turning-point} sets up the model and derives the turning point.
Section~\ref{sec:gate} shows why the turning point is a beginning, and Secs.~\ref{sec:shape} and~\ref{sec:higgs} follow the shape of space and the electroweak vacuum through it.
Section~\ref{sec:spectra} turns to observational tests, and Sec.~\ref{sec:disc} separates derived results from assumptions and lists the open questions, chiefly the present size of the torus and the origin of the plateau.
The companion technical report, Ref.~\cite{v12Report}, gives every derivation, with alternative methods and convergence checks, and a guide to reproducing the numerical results with Ref.~\cite{DNECv12code}.

%% file: sections/v13_s02_turning_point.tex
\makeatletter
\providecommand{\vxiiiTag}[1]{\def\@currentlabel{#1}}%
\makeatother

\section{A turning point with Standard Model fields}
\label{v13:sec-turning-point}

This section asks what can turn a contracting flat universe around and whether the Standard Model can pay for it.
Without an $R^2$ term conformal matter alone cannot turn an anisotropic universe around, so matter with $w<1/3$, most cheaply a positive potential, must supply a negative trace. With antiperiodic fermions the Casimir energy of the Standard Model supplies the negative energy, and the torus turns at an edge of 722 reduced Planck lengths. %

\subsection{Setup and assumptions}
\label{sec:setup}

The paper rests on one principle, that the universe is nonsingular.
Table~\ref{v13:tab-assumptions} lists it together with the scenario in which we evaluate its consequences, the approximations of the calculation and the measured inputs; this subsection fixes the geometry and the units.

We consider flat homogeneous cosmologies
\begin{equation}
ds^2=-dt^2+h_{ij}(t)\,dx^idx^j ,
\label{eq:pa-metric}
\end{equation}
with mean Hubble rate $H=\tfrac16\,d\ln\det h/dt$, shear $\sigma$ (the trace-free part of $\tfrac12h^{-1}\dot h$) and $S=\Tr\sigma^2$.
On a three-torus with unit coordinate periods we write $h_{ij}=L^2G_{ij}$ with $\det G=1$.
Then $L^3$ is the physical volume, and the five independent components of $G$ are the shape of the fundamental domain; a change of lattice basis acts on $G$ discretely, so a continuous change of $G$ is a physical change of shape.
We parametrize $G=\exp(2\sum_A\beta_AX_A)$ with symmetric trace-free matrices $X_A$, $\Tr X_AX_B=\delta_{AB}$.
About the cube the two aspect directions $X_d={\rm diag}(1,-1,0)/\sqrt2$ and $X_{d2}={\rm diag}(1,1,-2)/\sqrt6$ span the irreducible representation $E$ of the cubic group, and the three angle directions $X_{ij}=(E_{ij}+E_{ji})/\sqrt2$ span $T_2$.
For small amplitudes $\ln(L_1/L_2)=\sqrt2\,\beta_d$, and the angle between the first two edges is $\pi/2-\sqrt2\,\beta_{12}$.

Units are $\hbar=c=1$, and $\MPbar=(8\pi G)^{-1/2}=2.435\times10^{18}$\,GeV is the reduced Planck mass. %
The signature is $(-,+,+,+)$, and the trace of the stress tensor is $T=-\rho+3\bar p$, with $\bar p$ the mean pressure.
In Sec.~\ref{sec:trace} the symbols $a$ and $c$ are anomaly coefficients; elsewhere $a$ is a scale factor.
The benchmark plateau potential of the inflaton is
\begin{equation}
V(\phi)=3\MPbar^2h^2\big(1-e^{-\sqrt{2/3}\,\phi/\MPbar}\big)^2,
\label{eq:pa-plateau}
\end{equation}
the Einstein-frame potential of Starobinsky's model~\cite{Starobinsky:1980te} for a minimally coupled scalar, with plateau scale $h$.
We measure time $t$ from the turning point and write $\tau\equiv ht$.
The edge length at the turning point is $\Lb$, and $\mub\equiv1/\Lb$ is the matching scale.

\begin{table}[t]
\caption{The principle, the scenario, the approximations and the inputs of this work.
The text refers to the rows by their tags.}
\label{v13:tab-assumptions}
\begin{ruledtabular}
\begin{tabular}{@{}lp{0.89\columnwidth}@{}}
\multicolumn{2}{@{}l}{\textit{Principle}}\\[1pt]
P\vxiiiTag{P}\label{v13:assump-P} & The universe is nonsingular: its contraction turns around at an instant with $H=0$, and neither branch ends in a singularity.\\[3pt]
\multicolumn{2}{@{}l}{\textit{Scenario}}\\[1pt]
S1\vxiiiTag{S1}\label{assump:A1} & Flat three-torus; fermions antiperiodic on all three cycles (AAA); trivial gauge holonomies; no 't~Hooft flux.\\[2pt]
S2\vxiiiTag{S2}\label{v13:assump-S2} & Semiclassical Einstein gravity~\cite{Hartle:1981zt}; volume and inflaton classical, shape and zero modes quantum.\\[2pt]
S3\vxiiiTag{S3}\label{assump:A2}\label{assump:A6} & Standard Model with a curvature coupling $\xiH R\,\Phi^\dagger\Phi$, valid up to $2\pi\mub$; no further states that couple to the Higgs field.\\[2pt]
S4\vxiiiTag{S4}\label{assump:A3} & Plateau inflaton, Eq.~\eqref{eq:pa-plateau}, without direct coupling to the Higgs field.\\[2pt]
S5\vxiiiTag{S5}\label{v13:assump-S5} & State at the turn: non-zero modes and gauge zero modes in their ground state; shape and Higgs zero mode in the Gaussian state that is symmetric under time reversal about the turn and has minimal energy variance, centered on a stationary shape (the natural state, Sec.~\ref{sec:s-theorem}).\\[3pt]
\multicolumn{2}{@{}l}{\textit{Approximations}}\\[1pt]
X1\vxiiiTag{X1}\label{v13:assump-X1} & Zero-mode energy at leading order: a share $\fzm=0.30$ of the free Casimir energy (Sec.~\ref{sec:pa-zeromodes}).\\[2pt] %
X2\vxiiiTag{X2}\label{v13:assump-X2} & Free non-zero modes, matched to the zero modes at one loop at $\mub$.\\[2pt]
X3\vxiiiTag{X3}\label{assump:A5} & Reduced model for the Higgs zero mode.\\[3pt]
\multicolumn{2}{@{}l}{\textit{Inputs}}\\[1pt]
I1\vxiiiTag{I1}\label{assump:A7} & Top-quark pole mass $\mt$.\\[2pt]
I2\vxiiiTag{I2}\label{v13:assump-I2} & $\alpha_s(M_Z)=0.1180$~\cite{ParticleDataGroup:2026mpi}.\\[2pt] %
I3\vxiiiTag{I3}\label{v13:assump-I3} & $M_h=125.13$\,GeV~\cite{ParticleDataGroup:2026mpi}.\\[2pt] %
I4\vxiiiTag{I4}\label{v13:assump-I4} & Plateau scale $h$, fixed by the observed scalar amplitude for instant reheating.
\end{tabular}
\end{ruledtabular}
\end{table}

A turning point is an instant at which $H=0$ and the volume is minimal; we call it the turn, and the contracting and expanding halves of the history the two branches.
The Standard Model of row~\ref{assump:A2} is defined at $\mub$ and has the twelve gauge vectors, the Higgs doublet and 45 Weyl fermions (no right-handed neutrinos); in the Casimir energy the static graviton counts as one more Maxwell field. %
All its masses are negligible at the turn, where $1/\Lb$ exceeds the electroweak scale by thirteen orders of magnitude. %
The couplings $\lambda$, $g_3$, $g_2$, $g_Y$ and $y_t$ are running $\MSbar$ couplings at $\mub$, obtained by next-to-next-to-leading-order (NNLO) matching at the top-quark mass and three-loop running from the inputs \ref{assump:A7}--\ref{v13:assump-I3}; $g_Y$ has the Standard Model normalization, and the Higgs potential is $\lambda(\Phi^\dagger\Phi-v^2/2)^2$.
The pole mass $\mt$ follows from $\mtbar\equiv\mtbar(\mtbar)$ with the four-loop pole--$\MSbar$ relation; the directly measured mass, $172.60\pm0.27$\,GeV~\cite{ParticleDataGroup:2026mpi}, equals $\mt$ up to about $0.5$\,GeV~\cite{Hoang:2020iah}. %
Reference~\cite{v12Report}, Sec.~1, gives the matching and the mass definitions, including the MSR mass as a variant.

\subsection{What a turning point must supply}
\label{sec:trace}

What must matter supply for a contracting flat universe to turn around?
It must supply a non-positive total energy and a sufficiently negative trace.

Einstein's equations give $3\MPbar^2H^2=\rho+\tfrac12\MPbar^2S$ and $\dot H=-(\rho+\bar p)/2\MPbar^2-\tfrac12S$.
At a turning point $H=0$, so $\rho=-\tfrac12\MPbar^2S$ and $\dot H=(\rho-\bar p)/2\MPbar^2$, and a bounce, $\dot H>0$, requires~\cite{Solomons:2001ef}
\begin{equation}
\bar p<\rho=-\tfrac12\MPbar^2S\leq0\quad\Longleftrightarrow\quad T<-\MPbar^2S .
\label{eq:pa-trace-criterion}
\end{equation}

\label{sec:pa-square}%
\emph{The anomaly completes a square.}
Let the light fields form a conformal field theory (CFT) with anomaly coefficients $a,c>0$.
Its trace is fixed by the curvature in every state~\cite{Duff:1993wm},
\begin{equation}
\langle T^\mu{}_\mu\rangle=\frac{cW^2-aE_4}{16\pi^2}+b\,\Box R ,
\label{eq:pa-anomaly}
\end{equation}
where $W^2$ is the square of the Weyl tensor and $E_4$ the Euler density.
Let $P=\sigma^2-\tfrac13S\,\mathds{1}$ and $u=\dot\sigma$ in an orthonormal frame transported along the comoving observers.
At $H=0$ the curvature invariants of Appendix~\ref{app:curvature} combine, with $\varrho=a/c$, into
\begin{equation}
\begin{aligned}
cW^2-aE_4&=Q_0+4a\dot HS,\\
Q_0&=2c\Tr[u-2(\varrho-1)P]^2+\tfrac{4a}{3}(2-\varrho)S^2 .
\end{aligned}
\label{eq:pa-square}
\end{equation}
For $\varrho\leq2$ the Weyl square absorbs the indefinite Euler term, and $Q_0\geq0$.
A finite local $R^2$ counterterm shifts $b$, so $b=0$ is a choice of renormalization scheme, and we take the net $b$ to vanish.
With an extra trace $T_{\rm extra}$ from all other matter, the traced Einstein equation at $H=0$ is algebraic in $\dot H$:
\begin{equation}
\dot H=-\frac{\MPbar^2S+Q_0/16\pi^2+T_{\rm extra}}{6\MPbar^2+aS/4\pi^2}.
\label{eq:pa-hdot}
\end{equation}

\label{sec:pa-obstruction}%
\emph{Obstruction and rate bound.}
If $0<a\leq2c$ and all other matter has non-negative trace, $T_{\rm extra}\geq0$, a contracting universe cannot turn around.
Non-zero shear makes $\dot H<0$ at $H=0$, and the general-$H$ identity of Appendix~\ref{app:hidentity} excludes degenerate crossings.
The conformal collider bounds, which follow from positivity of the measured energy flux, restrict unitary, parity-preserving CFTs to $\tfrac13\leq a/c\leq\tfrac{31}{18}$~\cite{Hofman:2008ar,Hofman:2016awc}; the Standard Model at the turn has $a/c=1.17$ (Appendix~\ref{app:smcasimir}). %
Minimizing $Q_0$ over $u$ then gives, with $y=aS/24\pi^2\MPbar^2$,
\begin{equation}
\frac{\dot H}{S}\leq-\frac{1+2(2-\varrho)y}{6(1+y)}\leq-\frac5{54}-\frac{2}{27(1+y)},
\label{eq:pa-rate}
\end{equation}
so that $\dot H\leq-5S/54$ at every anisotropic zero of $H$, in any state and at any coupling. %
The bound is strict only at anisotropic turning points. In an isotropic history the anomaly terms of Eq.~\eqref{eq:pa-hdot} vanish at the turn, as in the anomaly-corrected Friedmann equation of a Casimir torus~\cite{Mithani:2014toa}.

\label{sec:pa-supply}%
\emph{The cheapest supplier.}
Suppose that a positive-energy component with $\bar p=w\rho$, $w<1/3$, supplies the trace and the CFT supplies the energy $\rho_{\rm CFT}=-\zeta\MPbar^2S$.
The energy constraint and Eq.~\eqref{eq:pa-hdot} then bound $\zeta$ from below.
At marginal turning the least negative CFT energy is $\zeta=\tfrac34+\tfrac{5}{36}y$ for vacuum energy and $\tfrac32+\tfrac59y$ for dust. %
Vacuum energy thus pays for the trace most cheaply but saturates the null energy condition, so the most economical turning point combines a positive potential with a separate source of negative energy.
With $T_{\rm extra}=-4\rho_\Lambda$ and $\rho_\Lambda=3\MPbar^2h^2$, Eq.~\eqref{eq:pa-hdot} becomes an exact criterion for every state of the conformal sector:
\begin{equation}
S_b<12h^2-\frac{Q_0}{16\pi^2\MPbar^2},
\label{eq:pa-exact-criterion}
\end{equation}
where $S_b$ is the shear at the turn; for a rolling inflaton $-T_{\rm extra}=4V-2K_\phi$, with $K_\phi=\tfrac12\dot\phi^2$.
Other routes break the conformal hypothesis through running couplings, masses or minimally coupled scalars~\cite{CesareSilva:2024nym,CesareSilva:2026zne}, or add a healthy $R^2$ term, which shifts $b$ and still requires negative matter energy~\cite{Anderson:1983nq,DeFelice:2010aj}.
We take the economical route: the plateau inflaton of row~\ref{assump:A3} and the negative Casimir energy of the Standard Model.
Reference~\cite{v12Report}, Sec.~2, gives the derivation, including rotating shear axes.

\subsection{The Standard Model pays}
\label{sec:casimir}

Can the Standard Model supply the negative energy?
It can when its fermions are antiperiodic. With all three cycles of a cubic torus antiperiodic, its signed Casimir weight, defined below, is $\NC=+25.515$, whereas periodic fermions give $-30$. %
Reference~\cite{v12Report}, Secs.~3 and~4, gives the derivations.

\label{sec:pa-image}%
\emph{Image sums and spin structures.}
On a flat torus the vacuum energy of free massless fields with flat twists $\bm\vartheta_f$ is the image sum~\cite{DeWitt:1979dd,Banach:1979iy,Bellucci:2009jr}
\begin{equation}
\begin{aligned}
E_C&=-\frac{1}{\pi^2L}\sum_{\bm n\neq0}\frac{w_{\bm n}}{(\bm n^{\mathsf T}G\bm n)^2},\\
w_{\bm n}&=\sum_fd_fs_f\cos(2\pi\bm\vartheta_f\cdot\bm n),
\end{aligned}
\label{eq:pa-imagesum}
\end{equation}
with $s_f=+1$ for bosons and $-1$ for fermions, and $d_f=1$ for a Maxwell field, a Weyl fermion or the static graviton and $\tfrac12$ for a real scalar.
Periodic fermions ($\bm\vartheta=0$) therefore raise the energy, and antiperiodic fermions lower it~\cite{Arkani-Hamed:2007ryu,Ibanez:2017kvh}; flat gauge holonomies twist charged fields in the same way.
At the cube we write $\rho_C=-\NC\gamma_C/L^4$, with $\gamma_C=Z_2(\mathds{1})/\pi^2=1.6750738$, $Z_s(G)=\sum'_{\bm n}(\bm n^{\mathsf T}G\bm n)^{-s}$ and the signed Casimir weight $\NC=\sum'_{\bm n}w_{\bm n}|\bm n|^{-4}/Z_2(\mathds{1})$. %
On the cubic three-torus with all three cycles antiperiodic (AAA) the alternating lattice sum is much weaker than the periodic sum,
\begin{equation}
\sum_{\bm n\neq0}\frac{(-1)^{n_1+n_2+n_3}}{|\bm n|^4}=-3.8631638, %
\label{eq:pa-aaa-sum}
\end{equation}
so an AAA Weyl fermion contributes $+0.233673$ to $\NC$, while a periodic one contributes $-1$. %
The gauge vectors, the Higgs doublet and the static graviton give $B=12+2+1=15$, and the Weyl fermions give $F=45$, hence %
\begin{equation}
\begin{aligned}
\NC^{\rm PPP}&=15-45=-30,\\ %
\NC^{\rm AAA}&=15+45\times0.233673=+25.515 . %
\end{aligned}
\label{eq:pa-NC}
\end{equation}

With one or two antiperiodic cycles the cube is not stationary, and the one-loop vacuum carries $SU(3)$ center holonomies. %
The AAA assignment keeps the cube stationary, gives the largest weight and leaves the holonomies trivial; it is our benchmark (row~\ref{assump:A1}).
Charged fields give flat holonomies a one-loop potential~\cite{Hosotani:1983xw}, and for AAA, trivial holonomy is its minimum at the shapes given below.
Where the twisted sum $Z_2(G,\bm\vartheta)$ is smallest at $\bm\vartheta=(\tfrac12,\tfrac12,\tfrac12)$, no flat connection of $SU(3)\times SU(2)\times U(1)$ lowers the energy below its value at trivial holonomy.
This holds for every rectangular torus (Appendix~\ref{app:smcasimir}), and numerically for shears $|\beta|\leq0.25$ of the cube and, at the body-centered cubic shape introduced below, for deformations $|\beta|\leq0.15$ and for its unstable shears up to $0.4$. These ranges contain the turning and fossil shapes of the benchmark state of row~\ref{v13:assump-S5}. The wider set of surviving shapes of Sec.~\ref{sec:v13-meanshape} extends beyond them in the angle directions, where the results assume trivial holonomy. %
Sectors with 't~Hooft flux need a hypercharge flux of at least $2\pi$, whose classical energy, $2\pi^2/g_Y^2L=100.7/L$, is $2.36$ times the free AAA Casimir energy, so they cannot turn around. %
The light inflaton fluctuation adds $\tfrac12$ to $\NC$, and three right-handed neutrinos lighter than $2\pi/\Lb$ would add $0.70$; we carry both as uncertainties. %

\begin{figure}[t]
\centering
\includegraphics[width=\columnwidth]{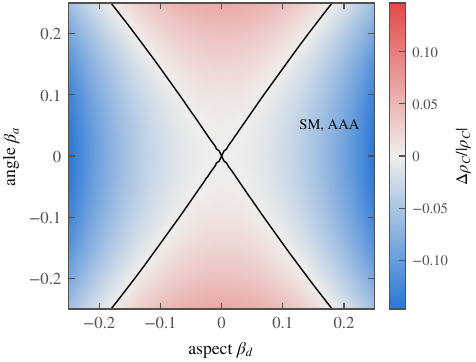}
\caption{Casimir energy near the cube. Fractional change $\Delta\rho_C/|\rho_C|$ of the AAA Casimir energy density of the Standard Model on the shape slice $G=\exp(2\beta_dX_d+2\beta_aX_{12})$; lines: no change. Aspect deformations lower the energy and angle deformations raise it.} %
\label{fig:v13-landscape}
\end{figure}

\emph{Shape dependence.}
At the cube the AAA energy is stationary, and its Hessian in the shape variables $\beta_A$ is diagonal in the representations $E$ and $T_2$.
Writing $E_C=-Z_{\rm eff}(G)/\pi^2L$, we measure the Hessian by the normalized stiffnesses $\kappa_A=3\lambda_A/Z_{\rm eff}$, where $\lambda_A$ are the Hessian eigenvalues of $Z_{\rm eff}$.
For free fields $\kappa_E=13.53$ and $\kappa_{T_2}=-6.35$, so aspect deformations lower the energy and angle deformations raise it [Fig.~\ref{fig:v13-landscape}]. %

The cube is not the only stationary shape of the AAA energy.
A numerical search, from $1200$ starting shapes with eigenvalue ratios of $G$ up to $12$ and along nine symmetric families of lattices, finds exactly one other: the body-centered cubic lattice with the three antiperiodic cycles along its primitive vectors, which also has the full cubic symmetry. %
There $\NC=22.25$, the two aspect deformations of the conventional cell are stable and the three shears are unstable with $\kappa=8.78$ for free fields. %
Away from a stationary shape the Casimir stress is anisotropic, and a torus at rest there starts to deform.
On the families of shapes that we followed, both branches keep their shape for time-symmetric data only near the cube: on the tetragonal axes one edge may be at most $9\%$ shorter than the other two, or up to twice as long on three thin spikes, and an edge angle may differ from $90^\circ$ by about $30^\circ$; this set contains the body-centered cubic shape (Sec.~\ref{sec:v13-meanshape}). %
At a stationary shape the stress is isotropic, so the Weyl curvature vanishes at a time-symmetric turn, as in Penrose's condition on the beginning~\cite{Penrose:1980ge}; we take the state at the turn to be centered there (row~\ref{v13:assump-S5}).
We use the cube, which has the larger Casimir weight, as the benchmark, and return to the second shape in Sec.~\ref{sec:disc}.

Antiperiodic cycles are natural in a quantum origin, because a cycle that caps off smoothly in a Euclidean geometry must be antiperiodic~\cite{GarciaEtxebarria:2020xsr}, and the smooth no-boundary fillings of $T^3$ cap one primitive cycle~\cite{Godet:2026gqw}.
Antiperiodic spinors and twisted fields were used before to make the Casimir energy of a created torus negative~\cite{Zeldovich:1984vk,Goncharov:1985gi,Linde:2004nz}, and the periodic Standard Model on $T^3$ has been studied for general shapes~\cite{Fornal:2011tw,Fornal:2026qfq}.

\label{sec:pa-zeromodes}%
\emph{Zero modes.}
The constant modes of the gauge and Higgs fields are not free.
Each of the twelve gauge fields has three of them and the Higgs doublet has four; their state is not fixed by the Hadamard condition~\cite{Negro:2026ooh}, and their nonlinear self-interaction gives their ground state a finite energy~\cite{Luscher:1982uv,Luscher:1982ma}.
For antiperiodic fermions the vacuum lies at trivial holonomy and is unique~\cite{vanBaal:1988va}.
We write the constant connections as $A^a_i=g\,x^a_i$ for each non-abelian factor, with $g=g_3$ or $g_2$, and $B_i=g_Y w_i$ for hypercharge, and the constant Higgs doublet as $\Phi/L$, with $\Phi=(\phi_0+i\phi_1,\phi_2+i\phi_3)/\sqrt2$ and $r=|\bm\phi|$.
At fixed volume their Hamiltonian is
\begin{equation}
\begin{aligned}
L\mathcal H_0={}&\sum_{\rm G}\Big[\tfrac12G_{ij}\,\pi_i\cdot\pi_j+\frac{g^2}{4}G^{ik}G^{jl}F_{ij}\cdot F_{kl}\Big]\\
&+\tfrac12G_{ij}\,\pi^w_i\pi^w_j+\tfrac12p_\Phi^2\\
&+G^{ij}\Phi^\dagger M_iM_j\Phi+\lambda(\Phi^\dagger\Phi)^2,
\end{aligned}
\label{eq:pa-H0}
\end{equation}
where $\pi_i$ and $\pi^w_i$ are the momenta conjugate to $x_i$ and $w_i$, $F^a_{ij}=f^{abc}x^b_ix^c_j$, $M_i=g_2\,x_i\cdot T+g_YYw_i$ acts on the doublet, and $p_\Phi^2$ sums over the four real components $\phi_a$.
Its spectrum does not depend on the shape. For $G=SS^{\mathsf T}$ with $\det S=1$, a linear canonical map removes $G$ from every term of Eq.~\eqref{eq:pa-H0} and from the Gauss law, with unit Jacobian (Appendix~\ref{app:zeromodes}).
The ground energy $Q/L$ is therefore a positive constant on the space of shapes at fixed volume; only the global identification of large holonomies and the matching to the non-zero modes depend on the shape.

At leading order of the small-volume expansion the pure-gauge part of $Q$ is $\epsilon_3g_3^{2/3}+\epsilon_2g_2^{2/3}$, with $\epsilon_2=4.1167$~\cite{Luscher:1983gm} and $\epsilon_3=12.587$~\cite{Weisz:1986vu,Pavel:2021mxn}, independent of the fermions at this order~\cite{vanBaal:1988va}. %
With $g_3\simeq0.54$, $g_2\simeq0.53$ and $g_Y\simeq0.44$ at $\mub$, this is $26\%$ of the free Casimir coefficient $C=\NC\gamma_C=42.74$, and it bounds $Q$ from below for $\lambda\geq0$. %
A diffusion Monte Carlo (DMC) computation of the coupled $SU(2)\times U(1)\times$Higgs ground state (Appendix~\ref{app:zeromodes}) gives $Q=12.8$, so the zero modes use up a fraction %
\begin{equation}
\fzm\equiv Q/C=0.30 %
\label{eq:pa-fzero}
\end{equation}
of the free Casimir energy, between the lower bound $0.26$ and the Gaussian variational bound $0.31$. %
We keep $\fzm=0.30$ as the benchmark and treat the next-order estimate, $\fzm\simeq0.28$ (Appendix~\ref{app:zeromodes}), as a one-sided uncertainty; higher orders are not estimated. %

\begin{table}[b]
\caption{Background quantities for free fields ($\fzm=0$) and with the leading-order zero-mode energy ($\fzm=0.30$, the benchmark). %
$\NC^{\rm eff}=(1-\fzm)\NC$; $\Lb$ and $h\Lb$ refer to the turning solution on the plateau; $\kappa_E$ and $\kappa_{T_2}$ are the normalized stiffnesses of the aspect and angle modes at the cube.}
\label{tab:pa-fzero}
\begin{ruledtabular}
\begin{tabular}{lcc}
 & $\fzm=0$ & $\fzm=0.30$\\ %
\hline
$\NC^{\rm eff}$ & $25.5$ & $17.9$\\ %
$\Lb\MPbar$ & $790$ & $722$\\ %
$h\Lb$ & $0.0048$ & $0.0044$\\ %
$\kappa_E$; $\kappa_{T_2}$ & $13.53$; $-6.35$ & $19.32$; $-9.07$ %
\end{tabular}
\end{ruledtabular}
\end{table}

At leading order $Q$ does not depend on the shape, so the zero modes leave the shape Hessian unchanged, lower the signed weight to $\NC^{\rm eff}=(1-\fzm)\NC=17.9$ and raise every normalized stiffness $\kappa_A$ to $\kappa_A/(1-\fzm)$ (Table~\ref{tab:pa-fzero}). %
This energy count does not settle the fate of the Higgs zero mode. At $\mub$ the quartic coupling is negative, so the Hamiltonian~\eqref{eq:pa-H0} has no normalizable ground state along the Higgs direction, and the zero mode becomes a narrow resonance, which Sec.~\ref{sec:higgs} treats.

\subsection{The turning solution and its null energy}
\label{sec:turning}

What does the turning point look like, and how does it escape the theorems that forbid a nonsingular flat universe?
The benchmark torus turns smoothly at the edge $\Lb=722\,\MPbar^{-1}$. %
The solution is null geodesically complete and violates the averaged null energy condition along every light ray; it remains consistent with the achronal version of that condition, because no light ray on the torus is achronal.

With a constant vacuum energy $3\MPbar^2h^2$ the negative Casimir energy turns a contracting torus around.
The leading solution~\cite{Zeldovich:1984vk,Guth:2025dal} is
\begin{equation}
L=\Lb\sqrt{\cosh2ht},\qquad \Lb^4=\frac{\NC^{\rm eff}\gamma_C}{3\MPbar^2h^2},
\label{eq:pa-turning-solution}
\end{equation}
with $H=h\tanh2ht$.
McInnes~\cite{McInnes:2005sa} discarded the contracting half of this solution as a redundant dual description; we keep both halves as the two branches of the turn.

On the plateau the turn sits at $\phi_b$ with $H=\dot\phi=0$, the background history is time symmetric, and $V(\phi_b)$ takes the place of $3\MPbar^2h^2$ in Eq.~\eqref{eq:pa-turning-solution}.
We write $V(\phi_b)=3\MPbar^2\sHeff^2$; at the reference turning point defined below, $\sHeff=0.98949\,h$, and the sizes quoted below use this value. %
We fix $h$ by the observed scalar amplitude, and the pivot $k_*=0.05\,{\rm Mpc}^{-1}$ by thermal matching.
With instant reheating this gives $N_*=55.722$ $e$-folds between pivot exit and the end of inflation, and $h/\MPbar=6.116\times10^{-6}$. %
The zero-mode energy lowers $\Lb$ by the factor $(1-\fzm)^{1/4}$, or $8.5\%$, so the benchmark turning point has %
\begin{equation}
\Lb\MPbar=722,\qquad h\Lb=0.0044, %
\label{eq:pa-benchmark}
\end{equation}
and the matching scale is $\mub=1/\Lb=\MPbar/722=3.37\times10^{15}$\,GeV. %
The torus is then about $230$ times smaller than the Hubble radius $1/h$. %
As a reference we take the turning point for which the present side of the torus is $L_0=6/H_0$, $0.96$ times the diameter $\dLSS$ of the last-scattering sphere. Reference~\cite{v12Report}, Sec.~5, gives its position on the plateau and the number of $e$-folds after it. Its present edge lies $2.5\%$ below the matched-circle bound of Sec.~\ref{sec:c-shortest}. This is immaterial for the quantities quoted at the reference turning point, because $\Lb$ changes by $0.6\%$ across the plateau. %

\emph{Null energy.}
Along a null geodesic $d\lambda=L\,dt/|\bm p|$, and $\int L\,dt$ diverges at both ends, so the solution is null geodesically complete.
By Corollary~3 of Ref.~\cite{Zhang:2026gva}, every null geodesic of a non-static Bianchi~I spacetime that obeys the null energy condition is incomplete in at least one affine direction.
The Casimir state must therefore violate that condition, and it does: $\rho+p=\tfrac43\rho_C<0$, while the potential contributes nothing to $\rho+p$.
With Eq.~\eqref{eq:pa-turning-solution} the averaged null curvature along every null geodesic is
\begin{equation}
\int R_{kk}\,d\lambda=-\frac{\sqrt2\,\Gamma(1/4)^2}{3\sqrt\pi}\,h\,\omega_b=-3.4961\,h\,\omega_b , %
\label{eq:pa-anec}
\end{equation}
where $\omega_b=|\bm p|/\Lb$ is the null frequency at the turning point and $\omega=|\bm p|/L$ the null frequency at time $t$.
Reference~\cite{v12Report}, Sec.~5, gives the integral and checks the sum rule below by quadrature.
The averaged null energy is $\int T_{kk}\,d\lambda=\MPbar^2\int R_{kk}\,d\lambda=-3.4961\,\MPbar^2h\,\omega_b$. %
The optical tensor of the congruence is $B_p=\omega H\,\mathds{1}_2$, so the optical bulk is $J_p=\int\Tr B_p^2\,d\lambda=2|\bm p|\int H^2L^{-1}dt$.
With $I_p=\int R_{kk}\,d\lambda$, Eq.~\eqref{eq:pa-anec} satisfies the sum rule $I_p=-J_p$ that Proposition~1 of Ref.~\cite{Zhang:2026gva} establishes for complete null geodesics. The Einstein-gravity form of the sum rule, Eq.~(27) of that reference, states $\int T_{kk}\,d\lambda=-\MPbar^2J_p$, and Sec.~V\,B of that reference obtains the same value for this solution.

At the turn $H=0$, and the optical tensor of every null congruence vanishes, so every null direction has a saturated instant in the language of Ref.~\cite{Zhang:2026gva}.
Lemma~1 of that reference states that the total energy density, potential included, is non-positive at a saturated instant; at an isotropic turn it vanishes, as it does here.
Away from the turn $B_p\neq0$, so no light ray is saturated~\cite{Zhang:2026gva}.

The violation is consistent with the achronal averaged null energy condition~\cite{Graham:2007va}, because no null geodesic of the torus is achronal.
During the whole history light crosses the torus $N_{\rm cross}=\int dt/L=c_\times/(h\Lb)$ times, with $c_\times=\sqrt\pi\,\Gamma(\tfrac14)/[2\Gamma(\tfrac34)]\simeq2.622$; here $N_{\rm cross}\simeq600$. %
Section~V\,B of Ref.~\cite{Zhang:2026gva} shows that no null geodesic is achronal for $h\Lb<2c_\times/\sqrt3\simeq3.03$, and the benchmark has $h\Lb=0.0044$. %

%% file: sections/v13_s03_beginning.tex
\section{The turning point is a beginning}
\label{sec:gate}

A turning point could be the middle of a bounce or the beginning of the universe.
We show that the Friedmann constraint decides between the two.
At the turning points from which both branches emerge with their shape intact, whose edges are nearly equal (Sec.~\ref{sec:v13-meanshape}), the light fields can carry only about five quanta of their lowest modes, whatever the history on either side. The entropy there is therefore at most a few tens. %
No branch can carry a larger entropy into the turning point, so the entropy that a branch has later was produced on the way out. The thermodynamic arrow of each branch points away from the turning point, and an observer on either branch finds it in the past.
A history that reaches the turn from a genuine past, with a coupling that reheats the universe, would have to carry more than $10^{94}$ times the Casimir energy in a correlated form and shed it before the turn. %

\subsection{The emptiness bound}
\label{sec:s-budget}

At a turning point the Friedmann constraint reads $\rho_\phi+\rho_C+\rho_{\rm exc}+\tfrac12\MPbar^2S=0$, where $\rho_{\rm exc}\geq0$ is the energy of the light fields above their ground state and $\rho_C$ is the ground-state (Casimir) energy, zero modes included.
Since $\rho_\phi\geq V>0$, the total excitation energy obeys $E_{\rm exc}<|E_C|=\sNeff\sgam/\Lb$, with $\sNeff=(1-\fzm)\NC$.
For free massless quanta, whose frequencies at the cube are $2\pi|\bm n-\bm\vartheta|/L$ with $\bm\vartheta$ the twist of the field ($|\bm n-\bm\vartheta|\geq\sqrt3/2$ for antiperiodic fermions), this becomes
\begin{equation}
\sum_{\rm quanta}|\bm n_i-\bm\vartheta_i|<\frac{\sNeff\sgam}{2\pi}=0.2666\,\sNeff .
\label{eq:s-emptiness}
\end{equation} %
For the Standard Model with $\fzm=0.30$ the right side is $4.76$, so %
the whole universe can hold about five quanta of its lowest non-zero bosonic modes, or $5.5$ of its lowest fermionic modes, all fields together. %
Across the allowed range of the zero-mode share, $0.258\leq\fzm\leq0.308$ (Sec.~\ref{sec:pa-zeromodes}), the bosonic bound runs from $4.7$ to $5.05$ ($6.80$ for free fields), and a single quantum with $|\bm n|\gtrsim5$ already forbids the turn. %
We call Eq.~\eqref{eq:s-emptiness} the emptiness bound.
It uses only the constraint at the instant where $H=0$, so at the cubic shape it holds for every state of the non-zero modes and for every history that turns there.
Away from the cube the right side depends on the shape, and it grows without limit as the torus degenerates into a slab or a rod. Over the time-symmetric turning points from which both branches survive (Sec.~\ref{sec:v13-meanshape}) it exceeds its value at the cube by at most about $10\%$ outside three thin spikes of elongated shapes, reaches $1.9$ times that value at their tips and is smaller in the angle directions. %

In the model of Mithani and Vilenkin, excited radiation lowers the Casimir coefficient by a tunable amount, which they note becomes discrete once the conformal fields are quantized on the compact space~\cite{Mithani:2014toa}. Equation~\eqref{eq:s-emptiness} is the quantized form of that statement.
Radiation and Casimir energy both dilute as $L^{-4}$, so the left side of Eq.~\eqref{eq:s-emptiness} is conserved by the adiabatic evolution of conformal fields~\cite{Parker:1969au}.
Excitation energy can therefore appear or disappear between the turn and a later time only through fields that are not conformal, through couplings that break conformal invariance or through the state itself.

\subsection{The entropy of a turning point}
\label{sec:v13-entropy}

The emptiness bound limits the number of states in which a turning point can be.
On the cube the non-zero modes of the Standard Model fields and of the graviton have $30$ bosonic states for each momentum $\bm n\neq0$, with energy $2\pi|\bm n|/L$, and $90$ fermionic states for each $\bm n$, with energy $2\pi|\bm n-\bm\vartheta|/L$.
Their Fock space contains $5.3\times10^{12}$ states with total energy below the bound~\eqref{eq:s-emptiness} at $\fzm=0.30$. %
The momentum constraint and Gauss's law remove some of these states, so this number is an upper limit. The entropy of the non-zero modes at a turning point, coarse grained in any way, is at most the logarithm of this number, $29$ in units of the Boltzmann constant. %
If the constraint is imposed on the mean energy only, as in semiclassical gravity, the largest entropy is that of the Gibbs state whose mean energy equals the bound, $S_{\max}=33$. %
Over the surviving shapes this value lies between $29$ and $55$, the largest at the tips of the spikes. %

The zero modes of the gauge and Higgs fields add levels below the same bound.
Their lowest levels lie $1.47/L$ and $1.24/L$ above the ground state for $SU(3)$ and $SU(2)$~\cite{vanBaal:1988va,Pavel:2021mxn}, and near $0.46/L$ for hypercharge (Ref.~\cite{v12Report}, Sec.~13), but their full spectrum is not known. %
If their $28$ physical coordinates are treated as oscillators with these gaps, with the Higgs zero mode in the well of Sec.~\ref{sec:higgs}, the largest entropy at a turning point becomes $S_{\max}\approx44$. %
It lies between $37$ and $55$ when all gaps are doubled or halved, and it depends on each gap only logarithmically. %
A turning point therefore has an entropy of at most a few tens.
The radiation in the same comoving volume carries an entropy of order $10^{90}$ today~\cite{Egan:2009yy}. %

The entropy bound turns a question of interpretation into a statement about solutions.
In every solution of the model whose two branches keep their shape, the entropy of the light fields, gravitational waves included, is at most a few tens at the turn.
Followed away from the turn, each branch expands, inflates and reheats, and its entropy is then larger by more than eighty orders of magnitude.
Neither branch can bring more than a few tens of units of entropy to the turning point, so on both branches the thermodynamic arrow of time points away from it.
An observer on either branch, whose records follow the growth of entropy, has the turning point in the past.
In this sense the turning point is the beginning of both branches.
In cosmologies with a singular beginning the low entropy of the early universe is a separate postulate about the initial state~\cite{Penrose:1980ge}. Here it follows from the constraint, for matter and for gravitational waves alike, given that the universe has a turning point from which both branches emerge regular.

Arrows of time that point away from a central surface appear in time-symmetric cosmologies~\cite{Carroll:2004pn,Barbour:2014bga}, in the bouncing histories of the no-boundary state~\cite{Hartle:2011rb}, in past-eternal inflation, where the surface is null~\cite{Aguirre:2001ks,Aguirre:2003ck}, and at a CPT-symmetric bang~\cite{Boyle:2018tzc}.
Nonsingular universes with two arrows of time have been argued to need special low-entropy conditions at the bounce~\cite{Vilenkin:2013rza}; in the present model the surface is a regular turning point, and the constraint enforces its low entropy.

\subsection{What a past would cost}
\label{sec:s-reheating}

The emptiness bound also measures what it would take for the contracting branch to be a genuine past, with entropy that decreases toward the turn.
The plateau history is time symmetric about the turning point (Sec.~\ref{sec:turning}), so its remote contracting branch is the time reverse of reheating, and any coupling that reheats the universe after inflation also acts there.
We take the visible coupling $-I^2(\phi)F_{\mu\nu}F^{\mu\nu}/4$ with $I^2=1+c_v\phi f(|\phi|)/\MPbar$ and $c_v=0.01$, where the smooth window $f$ vanishes for $|\phi|\geq1.5\MPbar$ and leaves the turning point exactly conformal. This coupling reheats the universe after inflation through its perturbative width~\cite{Garcia:2023obw}. %
Photon modes that start in the instantaneous vacuum at the $j$th extremum of the inflaton oscillation before its final climb onto the plateau reach the turn with an energy that exceeds the Casimir energy, zero modes included, by a factor between $10^{94.0}$ ($j=1$) and $10^{96.5}$ ($j=85$). %
The produced energy scales as $c_v^2$, and a coupling weak enough to bring this factor below one would be far too weak to reheat.
A contracting past with a reheating coupling can therefore reach the turn only from a correlated state that carries more than $10^{94}$ times the Casimir energy until the coupling removes it. %

\begin{figure}[t]
\centering
\includegraphics[width=\columnwidth]{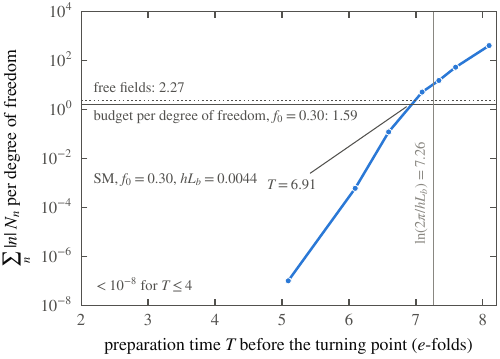}
\caption{Excitation of the non-conformal fields at the turning point, $\sum_{\bm n}|\bm n|N_{\bm n}$ per degree of freedom, for the instantaneous vacuum declared $T$ $e$-folds before the turn (Standard Model, $\fzm=0.30$, $h\Lb=0.0044$). It reaches the emptiness bound~\eqref{eq:s-emptiness} at $T=6.91$; vertical line: Hubble crossing of the lowest mode.} %
\label{fig:v13-past}
\end{figure}

Fields that are not conformal set a second limit.
Gravitons, any minimally coupled massless scalar and, up to slow-roll corrections, the inflaton fluctuation leave the Hubble radius when they are followed back in time, mode $\bm n$ at $\ln(2\pi|\bm n|/h\Lb)$ $e$-folds before the turning point.
A state declared $T$ $e$-folds before the turn can be a vacuum only for the modes that are still inside the Hubble radius, and the other modes return excited.
With three non-conformal degrees of freedom the excitation exhausts the bound at $T=6.91$ for the Standard Model with $\fzm=0.30$ and $h\Lb=0.0044$, and beyond that it grows roughly as $e^{4T}$ [Fig.~\ref{fig:v13-past}]. %
A state that is not arranged mode by mode must therefore be set within $6.9$ $e$-folds of the turning point, shortly after the lowest mode of the torus has entered the Hubble radius at $\ln(2\pi/h\Lb)=7.26$ $e$-folds before the turn. %
Classically, too, the shear grows as $L^{-6}$ on a contracting branch, so a genuine contracting past would also need finely tuned anisotropy~\cite{Battefeld:2014uga}.
Reference~\cite{v12Report}, Sec.~13, gives both calculations and an independent check by time reversal.

\subsection{The state at the beginning}
\label{sec:s-beginning}

The constraint fixes most of the state at the turning point and leaves a small remainder.
The non-zero modes are in their vacuum up to at most five quanta.
The emptiness bound does not select the vacuum among these states (Ref.~\cite{v12Report}, Sec.~13). In what follows we take all non-zero modes in their adiabatic vacuum at the turning point and the gauge zero modes in their ground state.
Two degrees of freedom are not settled by an energy bound, because near the turn their energy has no lower bound. They are the shape of the torus, which the Casimir energy destabilizes, and the Higgs zero mode, whose quartic coupling is negative there.
The next two sections show how the requirement of nonsingularity constrains their states.

Two remarks relate this beginning to earlier results.
The turning solution is geodesically complete, but its completeness is not an eternal past.
Both branches are futures of the turning point, and the past-incompleteness theorem for expanding spacetimes~\cite{BordeGuthVilenkin2003} does not apply, because $H$ averages to zero along every complete geodesic.
For a torus created from nothing, Guth and Vilenkin concluded that creation is either impossible or governed by Planck-scale physics~\cite{Guth:2025dal}. In a closed isotropic universe the conformal anomaly and the vacuum energy of conformal fields restrict a Euclidean density-matrix origin to a bounded range of the cosmological constant~\cite{Barvinsky:2006uh}.
The emptiness bound is a Lorentzian counterpart of these restrictions for the flat torus, and it does not refer to a creation event.

%% file: sections/v13_s04_shape.tex
\section{The shape of space}
\label{sec:shape}

The negative energy that turns the volume around also acts on the shape of the torus, and this section asks what it does there.
It cannot hold the shape, because a negative Casimir energy of massless fields has no shape minimum. Through the turning point the aspect modes grow by a factor $674$ for the Standard Model with its zero-mode energy. %
A shape that keeps growing squeezes one cycle of the torus to zero, which is a singular end of the branch on which it happens.
A nonsingular spacetime therefore needs a shape that survives on both branches, and this requirement confines the outgoing shapes of the two branches to a bounded region.
The two outgoing shapes are conjugate variables, so the region holds a finite number of quantum states, about $9\times10^2$ for the two unstable modes, and none of them leaves an exactly cubic torus. %
We take the state to be symmetric under the time reversal that exchanges the two branches (row~\ref{v13:assump-S5}).
We prove that a state is time-symmetric about at most one instant $t_x$. With nonlinear evolution we find that survival on both branches with probability $90\%$ ($99\%$) requires $h|t_x|\leq0.41$ ($0.31$) for every state whose survival on a branch is decided by its outgoing amplitude, a property we tested for amplitudes on the other branch of up to three times the uncertainty floor defined below. %
For every state that is time-symmetric about the turn, the uncertainty principle sets a floor on the outgoing shape.
The Gaussian state that is symmetric about the turn, centered on the cube and of minimal energy variance lies within $0.05\%$ of it in variance, and the torus then freezes into an almost rectangular box whose longest and shortest sides typically differ by $16$--$18\%$. %
Appendix~\ref{app:shapedyn} gives the proofs, and Ref.~\cite{v12Report}, Secs.~14--15, gives the calculations; Sec.~\ref{sec:sky} asks whether the microwave background can read the shape.

\subsection{No shape minimum for negative Casimir energy}
\label{sec:s-nominimum}

Can the Casimir energy that turns the volume around hold the shape of the torus at a minimum?
It cannot. The trace of its shape Hessian is a positive multiple of the energy itself, so a negative energy has a direction of instability at every shape.

We use the physical shape metric $ds^2_{\rm sh}=\tfrac14\Tr(G^{-1}dG)^2$ on the five-dimensional space of unimodular metrics $G$ of Sec.~\ref{sec:setup}.
For every image term of the Casimir energy, completeness of the basis $X_A$ gives (Appendix~\ref{app:shapelaplacian})
\begin{equation}
\Tr_{\rm sh}{\rm Hess}\,(\bm n^{\mathsf T}G\bm n)^{-s}=\frac{4s}{3}(2s-3)(\bm n^{\mathsf T}G\bm n)^{-s},
\end{equation}
so that for any mixture of twisted massless free fields
\begin{equation}
\Tr_{\rm sh}{\rm Hess}\,E_C=\frac83E_C .
\label{eq:s-Hesstrace}
\end{equation} %
The trace of the shape Hessian is the Laplace--Beltrami operator on the space of unimodular metrics, and Eq.~\eqref{eq:s-Hesstrace} is the Laplace eigen-equation of the $SL(3,\mathbb Z)$ Eisenstein series~\cite{Obers:1999um}.
A negative Casimir energy therefore has no shape minimum, at any shape.
The same mechanism is known on a two-torus~\cite{Ponton:2001hq,Luo:2022tqy,Aparici:2025kjj}.
Casimir back-reaction makes a torus ever thinner in $2+1$ gravity~\cite{Seriu:1995hc} and more elongated in de Sitter space~\cite{COMPACT:2026vdj}, and a fully compactified Standard Model on $T^3$ is unstable against decompactification, crunching or anisotropic evolution~\cite{Arkani-Hamed:2007ryu}.
The AAA Standard Model obeys Eq.~\eqref{eq:s-Hesstrace} to $10^{-8}$; aspect deformations lower its energy and angle deformations raise it [Fig.~\ref{fig:v13-landscape}]. %

The identity extends from the vacuum to excited and thermal states and to the zero modes, whose energy does not depend on the shape at leading order (Sec.~\ref{sec:pa-zeromodes}).
Thermal response and positive zero-mode energies make the instability stronger, and static classical holonomies cannot create a minimum either (Appendix~\ref{app:shapelaplacian}).
Massive fields, interactions beyond the zero-mode Hamiltonian, compact abelian flux sectors and holonomies pinned only by charged loops lie outside these statements. %

\subsection{Amplification through the turn}
\label{sec:s-modes}

How strongly does the turning point amplify a small deformation of the shape?
Every stationary-shape turning point has a mode that grows by at least a factor $4.50$ for free fields and $6.72$ at $\fzm=0.30$, and for the Standard Model the aspect modes grow by $674$. %

At a stationary shape the shape modes decouple at linear order.
With kinetic term $\tfrac12\MPbar^2L^3\dot\beta_A^2$, on the turning-point solution~\eqref{eq:pa-turning-solution} they obey
\begin{equation}
\ddot\beta_A+3H\dot\beta_A-\kappa_Ah^2\,{\rm sech}^2(2ht)\,\beta_A=0,\qquad \kappa_A=\frac{3\lambda_A}{Z_{\rm eff}},
\label{eq:s-shapeode}
\end{equation}
where $\lambda_A$ are the Hessian eigenvalues of $Z_{\rm eff}=-\pi^2LE_C$, so that Eq.~\eqref{eq:s-Hesstrace} gives $\sum_A\kappa_A=8/(1-\fzm)$.
A mode that starts constant with zero incoming canonical momentum leaves with $\beta_{A,+}=\sGain(\kappa_A)\beta_{A,-}$, where
\begin{equation}
\sGain(\kappa)=\sqrt2\cos\!\Big(\frac\pi4\sqrt{1-4\kappa}\Big)=\sqrt2\cosh\!\Big(\frac\pi4\sqrt{4\kappa-1}\Big),
\label{eq:s-gain}
\end{equation}
the second form holding for $\kappa>1/4$ (Appendix~\ref{app:theorem}).

For free fields the trace identity forces some $\kappa_A\geq8/5$.
Every stationary-shape turning point of this kind therefore has a mode with $\sGain\geq\sqrt2\cosh(\tfrac\pi4\sqrt{27/5})=4.500465$, whatever $h$, $\MPbar$ and $\NC$. %
A shape-independent zero-mode energy raises every $\kappa_A$ to $\kappa_A/(1-\fzm)$, and the guaranteed gain to $6.72$ at $\fzm=0.30$. %
For the cubic-symmetric twist of all fermions considered here the cube is a saddle of the lattice energy~\cite{Ennola:1964epstein,Betermin:2016cubic}.
For the AAA Standard Model with the stiffnesses of Table~\ref{tab:pa-fzero} the aspect gain is $216.5$ for free fields and $674$ with the zero-mode energy. %
The next-order terms of the zero-mode energy lower it to $529$, with $\kappa_E=18.0$. %
At the body-centered cubic shape of Sec.~\ref{sec:pa-image} the unstable modes are the three shears, with $\kappa=13.4$ and gain $210$ when the zero-mode energy is included. %
Two further effects, the finite-volume shape inertia of periodic Maxwell fields and pair creation by the shape motion, leave these numbers unchanged at the quoted precision (Ref.~\cite{v12Report}, Sec.~14).

\subsection{The uncertainty floor}
\label{sec:quantum}
\label{sec:s-floor}

How narrow can the outgoing shape be?
A gain of several hundred makes quantum fluctuations of the shape at the turn macroscopic afterward. The uncertainty principle sets a floor on the outgoing aspect variance of every state that is time-symmetric about the turn.

The canonical momentum of mode $A$ is $P_A=\MPbar^2L^3\dot\beta_A$.
With $\Pi_A=P_A/p_0$ and $p_0=3\MPbar^2\sHeff\Lb^3/2^{3/2}$,
\begin{equation}
[\beta_A,\Pi_B]=2i\delta\,\delta_{AB},\qquad
\delta=\frac{\sqrt2}{3X},\qquad X\equiv\MPbar^2\Lb^3\sHeff ,
\label{eq:s-delta}
\end{equation}
where, as in Sec.~\ref{sec:turning}, $\sHeff=[V(\phi_b)/3]^{1/2}/\MPbar=0.98949\,h$ is the rate of the turning-point solution on the $A_s$-calibrated plateau. %
On that plateau it replaces $h$ in Eq.~\eqref{eq:s-shapeode}; in this section we keep the symbol $h$ for it and measure times in units of $1/\sHeff$.
For the Standard Model with $\fzm=0.30$, $X=2282.5$ and $\delta=2.065\times10^{-4}$. %

A linear mode maps its incoming amplitude $\beta_-$ and momentum to the outgoing ones by a symplectic transfer matrix $\sTm$.
With $M$ the transfer from the turning point to the far future, time symmetry of the background, $t\to-t$ with $\Pi\to-\Pi$, makes the full transfer
\begin{equation}
\sTm=M\mathsf RM^{-1}\mathsf R,\qquad \mathsf R={\rm diag}(1,-1),
\label{eq:s-Tsym}
\end{equation}
so $\sTm_{12}=2m_{11}m_{12}$.
In terms of the data $\beta_0$, $\Pi_0$ at the turn the two outgoing amplitudes, one on each branch, are
\begin{equation}
\beta_\pm=m_{11}\beta_0\pm m_{12}\Pi_0 ,
\label{eq:v13-betapm}
\end{equation}
and Eq.~\eqref{eq:s-delta} gives $[\beta_-,\beta_+]=2i\delta\sTm_{12}$.

We call the shape state time-symmetric about $t_x$ if, at that instant, it is invariant under the reversal of all shape momenta, the antiunitary map $\beta\to\beta$, $\Pi\to-\Pi$.
For a pure state this is the statement that the wave function $\chi(\beta,t_x)$ is real up to a constant phase.
For $t_x=0$ the reversal of momenta at the turn, combined with the time symmetry of the background, maps the history $t>0$ onto the history $t<0$.
Equation~\eqref{eq:v13-betapm} then shows that $\beta_+$ and $\beta_-$ have the same distribution, and in particular $\sVar\,\beta_-=\sVar\,\beta_+$, for any state time-symmetric about the turn, pure or mixed, Gaussian or not.
Robertson's inequality, $\sVar\,\beta_-\,\sVar\,\beta_+\geq\tfrac14|\langle[\beta_-,\beta_+]\rangle|^2=\delta^2\sTm_{12}^2$, then gives for each aspect mode
\begin{equation}
\sVar\,\beta_\pm\geq\delta|\sTm_{12}|\equiv\sfloor^2 .
\label{eq:v13-symfloor}
\end{equation}
The uncertainty floor $\sfloor$ is therefore a lower bound on the outgoing shape of every state that is time-symmetric about the turn.
Without the symmetry the product bound remains, $\sVar\,\beta_-\,\sVar\,\beta_+\geq\sfloor^4$, so a state can leave one branch closer to the cube than the floor only by leaving the other farther from it.
Without any symmetry the floor also bounds the whole history: for an anti-restoring mode the maximum principle makes the larger endpoint variance an all-time bound, $\max_t\sVar\,\beta(t)\geq\delta|\sTm_{12}|$, so the shape cannot stay narrower than $\sfloor$ throughout, however it is prepared (Appendix~\ref{app:theorem}).

With $z=(1+\tanh2ht)/2$ the mode equation~\eqref{eq:s-shapeode} becomes Gauss's hypergeometric equation, and the connection formulas give $\sTm_{11}=\sGain(\kappa)$ and
\begin{equation}
\sTm_{12}=\frac{\Gamma(\tfrac74)\Gamma(\tfrac34)}{|\Gamma(\tfrac54+i\mu)|^2},\qquad
\mu=\frac{\sqrt{4\kappa-1}}{4}.
\label{eq:s-T12}
\end{equation}
The floor therefore has the closed form
\begin{equation}
\sfloor^2=\frac{\sqrt2\,\Gamma(\tfrac74)\Gamma(\tfrac34)}{3\,|\Gamma(\tfrac54+i\mu_E)|^2\,\MPbar^2\Lb^3\sHeff},
\label{eq:s-floor}
\end{equation}
where $\mu_E$ is the value of $\mu$ at $\kappa_E$.
For the Standard Model with $\fzm=0.30$, $\kappa_E=19.32$ and $\sTm_{12}=52.1$, so $\sfloor=0.1037$. %
At next order, $\kappa_E=18.0$ and $\fzm\simeq0.28$, the floor is $0.0934$. %

The bound~\eqref{eq:v13-symfloor} is attained.
A Gaussian with real wave function at the turn and $\sVar\,\beta_0=s$ evolves to
\begin{equation}
\sVar\,\beta_\pm=m_{11}^2s+\frac{m_{12}^2\delta^2}{s}=\delta|\sTm_{12}|\,\frac{x+x^{-1}}2,\qquad x=\frac{s}{s_*},
\label{eq:s-family}
\end{equation}
with $s_*=|m_{12}/m_{11}|\,\delta$, so the Gaussian of optimal width $s=s_*$ reaches the floor on both branches, and it is also the all-time optimum.
The natural local width of the inverted oscillator at the turn, $s_{\rm nat}=1/(2m_{\rm sh}|\omega|)$ with $m_{\rm sh}=\MPbar^2\Lb^3$, is the width whose spreading matches the instability.
For the Standard Model aspect modes the Gaussian of natural width lies $0.05\%$ above the floor in variance, with $\beta_{\rm rms}=0.0071$ at the turn (Ref.~\cite{v12Report}, Sec.~15). %

\subsection{States that keep both branches regular}
\label{sec:s-theorem}

Which states of the shape keep both branches regular, and about which instant can such a state be time-symmetric?

The state at the turn cannot be inherited from a vacuum of the contracting branch.
The shape modes freeze from $h|t|=\tfrac12{\rm arsinh}\sqrt{|\kappa_A|}$, which is $1.093$ for the aspect modes. %
On the plateau the anisotropic stiffness of every sector of our matter content is suppressed by the slow-roll parameter, so the modes have no oscillating asymptotic vacuum in the far past (Appendix~\ref{app:theorem}).
What the state can be is fixed instead by the symmetrized correlation $C(t)\equiv\tfrac12\langle\{\delta\beta,\delta\Pi\}\rangle$ of an aspect mode.
The reversal of momenta changes the sign of $C$, so a state time-symmetric about $t_x$ has $C(t_x)=0$.

\begin{stheorem}[Instant of time symmetry]\label{thm:s-real}
For the linearized dynamics of Eq.~\eqref{eq:s-shapeode}, for an aspect mode ($\kappa_E>0$) in any state, pure or mixed, Gaussian or not, $C(t)$ increases strictly with $t$.
The shape state is therefore time-symmetric about at most one instant $t_x$, and every state belongs to one of three classes: $C$ vanishes once [$C(-\infty)<0<C(+\infty)$]; $C\geq0$ throughout (outgoing); or $C\leq0$ throughout (incoming).
If $C(t_x)=0$, and in particular if the state is time-symmetric about $t_x$, the larger of the two outgoing variances obeys
\begin{equation}
\begin{gathered}
\max_\pm\sVar\,\beta_\pm\geq G(|t_x|)\,\delta|\sTm_{12}|,\\
G(|t_x|)=\frac{|M_{11}M_{12}|_{-|t_x|\to\infty}}{|M_{11}M_{12}|_{0\to\infty}},
\end{gathered}
\label{eq:s-Gtx}
\end{equation}
where the subscripts give the interval of the transfer $M$.
A state whose correlation never vanishes has $\max_\pm\sVar\,\beta_\pm\geq2\sGain(\kappa_E)\,\delta|\sTm_{12}|$ and $\sVar\,\beta_E(0)\geq2m_{12}m_{22}\delta$.
\end{stheorem}

The proof (Appendix~\ref{app:theorem}) uses only the linear equations of motion: by Ehrenfest's relations $\dot C$ is a sum of two positive terms, and the Robertson--Schr\"odinger inequality with $C(t_x)=0$ bounds the transfer from $t_x$.
Such growth of the correlation (squeezing) of inverted-oscillator modes is familiar from inflation~\cite{Polarski:1995jg}; new here are the location of the instant of time symmetry near the turn and, with the nonlinear survival below, the bound that follows from survival on both branches.
For the Standard Model, $G=2.395$ and $8.263$ at $h|t_x|=0.1$ and $0.25$. %
It grows to $45.10$ and $313.8$ at $h|t_x|=0.5$ and $1$, and to $2\sGain(\kappa_E)=1348$ as $|t_x|\to\infty$ (Ref.~\cite{v12Report}, Sec.~15). %
Because the background is time symmetric, an instant of time symmetry before the turn bounds the future branch, and one after the turn bounds the past branch.
A state that is time-symmetric after the turn can leave the future branch below the floor, by a factor $0.120$ in variance at $ht_x=0.25$, while the other branch then carries at least $G(|t_x|)\sfloor^2$; the floor holds on both branches only for a state that is symmetric about the turn itself. %

\begin{figure}[t]
\centering
\includegraphics[width=\columnwidth]{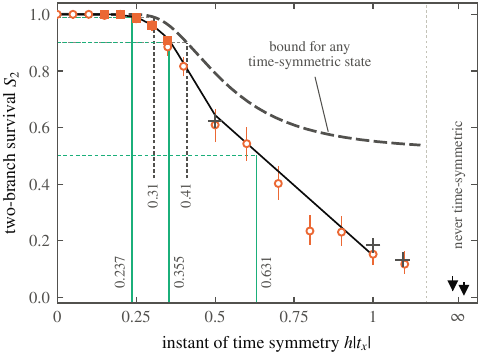}
\caption{Survival $S_2$ on both branches of the time-symmetric Gaussian of optimal width against its instant of time symmetry, for the Standard Model with $\fzm=0.30$: open circles, $256$--$512$ draws; squares, $2048$ draws (Wilson $95\%$ intervals); line, independent implementation; crosses, plateau background; at $\infty$, states whose correlation never vanishes. Thick dashed: the bound~\eqref{eq:v13-lpbound} for every time-symmetric state, with $D$ enlarged by $5\%$. Solid verticals: the Gaussian survival crosses $99\%$, $90\%$ and $50\%$; dashed verticals: the bound crosses $99\%$ and $90\%$.} %
\label{fig:v13-timesym}
\end{figure}

\label{sec:s-conditional}
The linear bound becomes a constraint on $t_x$ once nonlinear evolution is included, because large outgoing deformations squeeze a cycle to zero.
We say that a history survives on a branch if the volume turns, all three cycles end up inflating and no cycle shrinks to zero.
We prepared at $t_x$ the time-symmetric Gaussian of optimal width, the best case among time-symmetric Gaussians at linear order, and drew phase-space points from its Wigner function.
Each draw was evolved through the full equations for $L$ and the five-dimensional $G$, with twisted Ewald forces, on the turning-point solution. %
This truncated Wigner evolution is exact at linear order and omits the Moyal terms, and the volume and inflaton are classical.
The survival $S_2$ on both branches, the joint frequency of histories that survive on both branches in this classical (truncated Wigner) ensemble, falls to $0.609$ at $h|t_x|=0.5$ and to $0.152$ at $1$ [Fig.~\ref{fig:v13-timesym}]. %
The quantum bound derived below constrains the probability of survival on a single branch.
For Gaussian states the survival on both branches crosses the $99\%$ level at $h|t_*|=0.237$ and the $90\%$ level at $0.355$; at next order the $99\%$ level moves to $0.265$. %
The numbers of this subsection are reproduced by an independent implementation that was given the method sections of the derivations but not their results. It gives the thresholds $0.243\pm0.009$ and $0.360\pm0.004$ (Ref.~\cite{v12Report}, Sec.~15). %

\begin{figure}[t]
\centering
\includegraphics[width=\columnwidth]{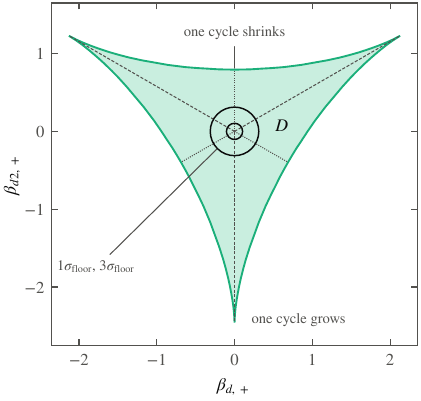}
\caption{Region $D$ of outgoing aspect amplitudes $(\beta_{d,+},\beta_{d2,+})$ for which the shape survives a Standard Model turning point ($\fzm=0.30$), with circles of radius $\sfloor$ and $3\sfloor$. Its area fixes the number of states that survive on both branches, Eq.~\eqref{eq:v13-count}.} %
\label{fig:v13-region}
\end{figure}

The restriction to Gaussian states can be removed, because the fate of a history is decided by its outgoing linear amplitude. The nonlinear terms act after the growing solution has come to dominate each aspect mode.
We evolved single histories from the turn with prescribed outgoing amplitudes $\bm b=(\beta_{d,+},\beta_{d2,+})$ of the two aspect modes and found that, when the amplitude on the other branch is at most a few floors, the shape survives if and only if $\bm b$ lies in a region $D$ of the aspect plane.
The region has the threefold symmetry of the cube: its boundary lies at $|\bm b|=0.79$ in the three directions in which one cycle shrinks and at $0.94$ between them, and it reaches $2.46$ at three cusps, in the directions in which one cycle grows [Fig.~\ref{fig:v13-region}]. %
Its area is about $3.7\times10^2\,\sfloor^2$. %
Survival on a branch is thus the event that the two commuting outgoing amplitudes of that branch lie in $D$, which has a probability in every state, Gaussian or not.
Both branches survive if $\bm\beta_+$ and $\bm\beta_-$ lie in $D$.
These are conjugate, $[\beta_{-,A},\beta_{+,B}]=2i\sfloor^2\delta_{AB}$, so the surviving set has the phase-space volume $({\rm Area}\,D)^2$, a quantum state occupies $(4\pi\sfloor^2)^2$, and the set holds
\begin{equation}
N_D=\Big(\frac{{\rm Area}\,D}{4\pi\sfloor^2}\Big)^{2}\approx9\times10^{2}
\label{eq:v13-count}
\end{equation} %
orthogonal states of the two aspect modes; $N_D$ is the trace of the product of the projectors on $\bm\beta_+\in D$ and on $\bm\beta_-\in D$ (Appendix~\ref{app:v13-lp}).
Since the two projectors do not commute, no state survives on both branches with certainty; the count estimates the number of states that survive with probability close to one.
Nonsingularity on both branches leaves the two unstable modes this finite set of states, whatever their symmetry.
The count is one of phase-space cells; it does not include the three angle modes, whose surviving range, edge angles within about $30^\circ$ of $90^\circ$, is some sixty times their quantum width. Nor does it include thin slivers along the rod axes (one edge longer than the other two, which stay equal), where histories with large and nearly equal amplitudes on both branches survive outside $D$ (below). %
For Gaussian states the probability that $\bm b$ lies in $D$ reproduces the surviving fractions above, $0.9082$ against $0.9087$ at $h|t_x|=0.35$. %
Now let the state be time-symmetric about $t_x$, let $X_A=M_{11}\beta_A+M_{12}\Pi_A$ be the outgoing amplitudes on the branch that crosses the turn, with $M$ the transfer from $t_x$ and $\beta_A$, $\Pi_A$ taken at $t_x$, and let $Y_A=M_{11}\beta_A-M_{12}\Pi_A$.
Time symmetry gives $\bm X$ and $\bm Y$ the same distribution, while
\begin{equation}
[Y_A,X_B]=2i\,G(|t_x|)\,\sfloor^2\,\delta_{AB},
\label{eq:v13-XYcomm}
\end{equation}
so $\bm X$ and $\bm Y$ are conjugate variables whose effective Planck constant grows with $|t_x|$.
Two conjugate variables cannot both be confined to a bounded region with certainty, and the sharp form of this statement, due to Landau and Pollak~\cite{LandauPollak1961}, gives (Appendix~\ref{app:v13-lp})
\begin{equation}
P(\bm X\in D)\leq\tfrac12\bigl[1+\sqrt{\lambda_0(G)}\,\bigr]
\label{eq:v13-lpbound}
\end{equation}
for every state that is time-symmetric about $t_x$, pure or mixed, where $\lambda_0$ is the largest eigenvalue of the operator that projects on $\bm X\in D$, then on $\bm Y\in D$ and again on $\bm X\in D$.
The right side is $0.9988$, $0.930$, $0.773$ and $0.547$ at $G=8.26$, $20$, $45.1$ and $313.8$. %
Survival on both branches with probability $99\%$ ($90\%$) therefore requires $h|t_x|\leq0.31$ ($0.41$) for every time-symmetric state whose survival is decided by its outgoing amplitude, where we have enlarged $D$ by $5\%$ to cover the weak dependence of the fate on the amplitude of each mode on the other branch; Gaussian states reach $0.24$ ($0.36$) [Fig.~\ref{fig:v13-timesym}]. %
The bound never falls below $1/2$: a superposition of a state and its time reverse is time-symmetric and survives on a given branch with probability one half, however large $|t_x|$ is.
States whose correlation never vanishes fare worse.
The extreme case is a mirror pair of states with $1.2$ and $2$ times the minimum width that such states need, and with extremal correlation. %
These survive on both branches in $4.3\%$ and $2.7\%$ of the trajectories. %

Combining Theorem~\ref{thm:s-real} with these results gives the consequences of nonsingularity for the shape.
\emph{The two aspect modes have about $9\times10^2$ phase-space cells of states that keep both branches regular. A shape state can be time-symmetric about at most one instant $t_x$. If it is, and if it survives on both branches with probability of at least $90\%$ ($99\%$), then $h|t_x|\leq0.41$ ($0.31$) on either side of the turning point, for every state whose survival is decided by $D$, and $h|t_x|\lesssim0.36$ ($0.24$) if the state is Gaussian.} %
The uniqueness, the linear bound, the count~\eqref{eq:v13-count} and the inequality~\eqref{eq:v13-lpbound} are proved for the linear amplitudes; the region $D$ and the Gaussian thresholds are numerical.
Survival does not force time symmetry: a displaced Gaussian, a state that carries angular momentum in the aspect plane and a state whose two aspect modes are symmetric about different instants are time-symmetric about no instant and can survive on both branches.
A state whose correlation never vanishes has, by Theorem~\ref{thm:s-real}, an outgoing variance of at least $1348\,\sfloor^2$ on one branch, and the most favorable Gaussian states of that class survive on both branches only in the few percent of histories found above. %
The classical counterpart is the tuning that a genuine contracting past would need (Sec.~\ref{sec:s-reheating}): for the shape to survive the turn, the incoming shape of a classical history must agree with the cube to about one part in a thousand, $|\beta_-|\lesssim0.79/\sGain=1.2\times10^{-3}$, and its incoming shear must be about twenty times smaller than the turning criterion~\eqref{eq:pa-exact-criterion} allows. %

The state that attains the floor has a second property.
Among pure centered Gaussians of an inverted oscillator, the unique time-symmetric state of minimal energy variance has the natural width (Appendix~\ref{app:theorem}), which lies $0.05\%$ above the floor in variance for the aspect modes. %
The energy variance of a real Gaussian also contains the square of the force at its center times its variance, so the minimum of the energy variance lies at a stationary shape; the cubic anharmonicity of the Casimir energy shifts it by far less than the width of the state.
We call the resulting Gaussian, time-symmetric about the turn and centered on a stationary shape, the natural state; we center it on the cube, our benchmark among the two stationary shapes (Sec.~\ref{sec:casimir}).

\label{sec:v13-meanshape}
Time symmetry by itself leaves the mean shape at the turn free.
Every shape with negative vacuum energy admits an instant of vanishing extrinsic curvature, and a shape that is not stationary starts to deform from rest.
We evolved such beginnings with the full equations.
On the families of shapes that we followed, both branches keep their shape for time-symmetric data only near the cube: on the tetragonal axes one edge may be at most $9\%$ shorter than the other two, or up to twice as long on three thin spikes, and an edge angle may differ from $90^\circ$ by about $30^\circ$; this set contains the body-centered cubic shape. %
On its boundary one cycle stays at $755\,\MPbar^{-1}$ while the other two inflate, and at the tips of the spikes two cycles stay at $676\,\MPbar^{-1}$. These histories end on the product of a static circle or two-torus with de Sitter space of lower dimension, which is nonsingular for a constant potential but unstable. %
A time-symmetric Gaussian of natural width survives on both branches with probability $90\%$ only if its mean aspect offset is below $0.06$--$0.085$, or $0.18$ along the three rod axes. %
Nonsingularity therefore confines the mean shape without fixing it, and $\sfloor$ is a lower bound on the outgoing spread of every state that is time-symmetric about the turn.
The floor is the whole fossil when the mean shape is stationary, that is, when the Casimir stress is isotropic and the Weyl curvature vanishes at the turn, which in our search is the case at the cube and at the body-centered cubic shape only (row~\ref{v13:assump-S5}).

\label{sec:s-equivalent}
We take the state at the turn to be symmetric under the time reversal that exchanges the two branches (row~\ref{v13:assump-S5} of Table~\ref{v13:tab-assumptions}), because both branches are futures of the turning point (Sec.~\ref{sec:gate}), the background is symmetric about it and the data at the turn do not distinguish the branches.
Time symmetry of the state is a statement about the wave function of the universe, where it has been discussed as a postulate~\cite{Hawking:1985af,Page:1985ei,Kiefer:1994gp}.
With the volume semiclassical and the shape in the Born--Oppenheimer approximation, in the allowed region $L>\Lb$
\begin{equation}
\Psi(L,\beta)=A(L)\big[e^{iS(L)}\chi(\beta,t(L))+e^{-iS(L)}\chi(\beta,-t(L))\big],
\label{eq:s-WDW}
\end{equation}
where $t(L)>0$ is the time at which the classical solution reaches $L$, and $\chi(\beta,t)$, with $i\partial_t\chi=H_s(t)\chi$, is the shape state carried along a single classical history through the turn.
Since $H_s$ is real in the $\beta$ representation and even in $t$, $\Psi$ is real if and only if $\chi(\beta,0)$ is real up to a phase (Appendix~\ref{app:theorem}).
A real $\Psi$ is a standing wave with equal weight on the two branches; with CP violation, the reality condition becomes invariance under the antiunitary $\Theta=CPT$.
Instantaneous-ground-state rules and the Hartle--Hawking proposal~\cite{Hartle:1983ai} with a $\Theta$-symmetric saddle and contour give such a state, while the tunneling proposal~\cite{Vilenkin:1984wp} does not. The CPT-symmetric proposal of Ref.~\cite{Boyle:2018tzc} is an analog.

The Euclidean continuation of the turning point does not select the shape state.
Continuing $t=-i\tau_E$, Eq.~\eqref{eq:s-shapeode} reaches the singular end $L_E=\Lb\sqrt{\cos2h\tau_E}\to0$ of the Euclidean solution, which Ref.~\cite{Guth:2025dal} rejects as an instanton. For $\kappa>1/4$ both solutions of the mode equation oscillate log-periodically there with the same amplitude, and the on-shell action oscillates without a limit, so no regularity condition selects a state.
The no-boundary proposal has been applied to flat compact three-manifolds~\cite{Louko:1989up}, and the complex no-boundary saddles of a three-torus~\cite{Godet:2026gqw} cap one cycle and have no turning point.
Within the ansatz of a disk times a square two-torus, the only smooth real cap belongs to an elongated torus: with the Casimir stress of $\mathbb R^2\times T^2$, a hemisphere of radius $\MPbar(2V)^{-1/2}$ times a two-torus of side $676\,\MPbar^{-1}$ solves the Euclidean equations, and on its surface of vanishing extrinsic curvature one cycle is $627$ times longer than the other two. %
Its Lorentzian continuation is the product solution above, which on the plateau collapses on both branches, and capping a cycle of length $\Lb$ smoothly would need about $3\times10^5$ times the curvature that the stress supplies (Ref.~\cite{v12Report}, Secs.~15 and~21). %
Nonsingularity constrains the alternatives to the natural state: a state that is time-symmetric about another instant keeps both branches with probability $90\%$ only if that instant lies within $0.41/h$ of the turn, and a time-symmetric Gaussian centered on the cube keeps them with probability $99\%$ only if its outgoing width is below about three times the floor (Sec.~\ref{sec:s-fossil}). %

\subsection{The shape fossil}
\label{sec:s-fossil}

What does the turning point leave of the shape at late times?
For the natural state it leaves an almost rectangular box whose longest and shortest sides typically differ by $16$--$18\%$, with right angles to half a degree. %

After a few Hubble times the Casimir force, which decays as $L^{-4}$ relative to the inertia, switches off and the shape freezes~\cite{Fornal:2026qfq}.
Its outgoing distribution is the shape fossil of the turning point (Table~\ref{tab:s-fossil}).

\begin{table}[t]
\caption{Rms outgoing shape per mode for the natural state: the uncertainty floor of the aspect ($E$) modes, Eq.~\eqref{eq:s-floor}, and the natural width of the angle ($T_2$) modes, with the aspect stiffness and gain and the typical difference between the longest and shortest sides, for the AAA Standard Model on the $A_s$-calibrated plateau. The natural aspect variances exceed the floor by $0.05\%$ ($\fzm=0.30$) and $0.12\%$ (free fields).} %
\label{tab:s-fossil}
\begin{ruledtabular}
\begin{tabular}{lccccc}
 & $\kappa_E$ & $\sGain$ & $\sfloor$ & $\sigma_{T_2}$ & sides\\
\hline
$\fzm=0.30$ & 19.32 & 674 & 0.104 & 0.0065 & $18\%$\\ %
next order & 18.0 & 529 & 0.0934 & --- & $16\%$\\ %
free fields & 13.53 & 216.5 & 0.0587 & 0.0066 & $10\%$ %
\end{tabular}
\end{ruledtabular}
\end{table}

\begin{figure}[t]
\centering
\includegraphics[width=\columnwidth]{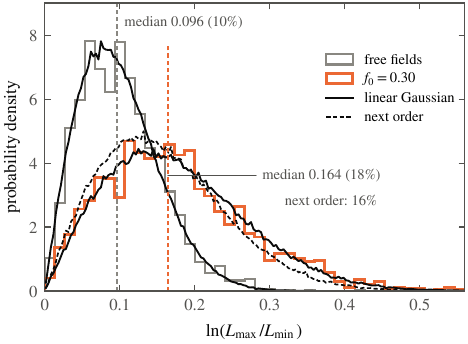}
\caption{Shape fossil: distribution of $\ln(L_{\max}/L_{\min})$ for $2048$ natural turning-point states evolved through the full nonlinear equations (steps) and for the linear Gaussian (lines; dashed: next order), for free fields and for $\fzm=0.30$. Vertical lines: medians.} %
\label{fig:v13-fossil}
\end{figure}

For the Standard Model the fundamental domain is an almost rectangular box.
With $\fzm=0.30$ each aspect mode freezes with rms $\sigma_E$, the rms of $\ln(L_i/L_j)$ is $\sqrt2\,\sigma_E=0.147$, and the median of $\ln(L_{\max}/L_{\min})$ is $1.588\,\sigma_E$. %
The longest and shortest sides therefore typically differ by $18\%$ at leading order and by $16\%$ with the next-order corrections [Fig.~\ref{fig:v13-fossil}]. %
Each edge angle deviates from $90^\circ$ by $\sqrt2\,\sigma_{T_2}=0.53^\circ$ rms. %
The net change of the shape between the far past and the far future is the fossil anisotropy of Ref.~\cite{Zhang:2026gva}, with rms $1.39\,\sigma_E$ per aspect mode. %

The aspect fossil obeys $\sigma_E^2=\delta|\sTm_{12}(\kappa_E)|$, so fewer light fields and a stronger instability give a larger fossil.
For plateau inflation $\sHeff/\MPbar\propto(A_s\rT)^{1/2}$, so that
\begin{equation}
\sigma_E\propto(A_s\rT)^{1/8}\,(\sNeff)^{-3/8}\,\sTm_{12}(\kappa_E)^{1/2}.
\label{eq:s-scaling}
\end{equation}
We quote leading-order values and treat the next-order shift as a one-sided uncertainty, with higher orders not estimated. The next-order shift lowers the floor by the factor $0.900$ and moves the $99\%$ threshold outward to $h|t_*|=0.265$. %
The remaining corrections, from the numerical error, classical nonlinear evolution, the inflaton fluctuation and the weak-decay calibration of $h$, are listed in Ref.~\cite{v12Report}, Sec.~15.
A shift of $0.02$ in $\fzm$ moves $\sigma_E$ by $5\%$. %
A classical mean offset of the shape grows by $m_{11}=10.6$ through the turn, so it must stay below about $0.01$, or $1.4$ quantum widths, for the quantum fossil to dominate. %

\label{sec:s-nonlinear}
Nonlinear shape dynamics leave the fossil of the natural state intact, but they end the shape when the outgoing deformation is large.
We evolved the natural state, and the same state scaled in width by a factor $s$, with the truncated Wigner method of Sec.~\ref{sec:s-theorem}.
For the natural state with $\fzm=0.30$, two sets of $2048$ draws all survive. %
Larger states fail in one way: the volume turns, and then one cycle is squeezed to zero within $0.3$--$4$ Hubble times. %
Of the trajectories, $99\%$ survive up to $s_{99}\simeq2.8$--$2.95$, that is, below an outgoing linear amplitude $\sigma_{\rm lin}\simeq0.29$--$0.31$ per aspect mode (Ref.~\cite{v12Report}, Sec.~15). %
Survival is mainly a function of $\sigma_{\rm lin}$: for $\sigma_{\rm lin}\lesssim1.5$ the amplitude scans, the time-symmetric states of Sec.~\ref{sec:s-theorem} and the states whose correlation never vanishes lie on one curve to within $\pm5$ percentage points. %
The Moyal term changes the aspect variance by $-0.020\%$ for the natural state. %
An independent implementation reproduces the amplitude scan, and small-amplitude runs reproduce the linear transfer to $2\times10^{-9}$ (Ref.~\cite{v12Report}, Sec.~15). %

%% file: sections/v13_s05_higgs.tex
\section{The electroweak vacuum}
\label{sec:higgs}

The turning point fixes the scale at which the Standard Model couplings act on the zero modes, $\mub=1/\Lb\approx3.4\times10^{15}\GeV$, and at this scale the Higgs quartic coupling is negative. %
A spatially constant Higgs field with $\lambda<0$ has no ground state, so we ask whether the Higgs zero mode stays near the origin while the universe turns around and inflates.
It does, with a probability set by the top-quark mass: on one branch it survives with probability $\Pone>1/2$ for a top-quark pole mass $m_t<172.96\GeV$, for any Higgs curvature coupling that vacuum stability during and after inflation allows (Sec.~\ref{sec:h-curvature}). %
The companion Letter~\cite{v12PRL} establishes the mechanism in a static box.
Appendix~\ref{app:h-matching} summarizes the zero-mode calculation, and Ref.~\cite{v12Report}, Secs.~6--12, gives every derivation.

\subsection{The zero mode at the turn}
\label{sec:h-model}

At the turn the Higgs zero mode feels a negative quartic coupling, yet it is held near the origin by the zero-point energy of the gauge zero modes, which grows linearly with the Higgs field.
For $m_t=172.60\GeV$ (the direct measurement), $\alpha_s(M_Z)=0.1180$ and $M_h=125.13\GeV$ we find $\lambda(\mub)=-0.0098$. %

We use units $L=1$, the Higgs radius $r=|\bm\phi|$ (with $\Phi^\dagger\Phi=r^2/2$) and $\MSbar$ couplings $g_2$, $g_Y$ and $\lambda$ at $\mu=1/L$.
The electroweak part of the zero-mode Hamiltonian, Eq.~(\ref{eq:pa-H0}), couples the Higgs zero mode to the SU(2) and hypercharge connections of Sec.~\ref{sec:pa-zeromodes}~\cite{Luscher:1982ma,Koller:1987fq,vanBaal:2000zc}, and the Letter~\cite{v12PRL} writes it out in these variables.
The gluon zero modes do not couple to $\Phi$ at tree level, and the negative quartic leaves the zero-mode share $\fzm$, and hence the turning solution, unchanged at the quoted precision.

At fixed $r$ the $W$ and $Z$ zero modes are oscillators, and for $r\gtrsim1$ their zero-point energy grows linearly with the Higgs radius~\cite{v12PRL},
\begin{equation}
E_{\rm g}(r)\simeq E_{\rm g}^{(0)}+\sigma_0 r,\qquad \sigma_0=\tfrac32 g_2+\tfrac34 g_Z\simeq1.31, %
\label{eq:h-linear}
\end{equation}
where $g_Z=(g_2^2+g_Y^2)^{1/2}$.
Against the negative quartic this energy builds a barrier, and the electroweak state becomes a resonance with complex energy $E=\varepsilon-i\Gamma/2$, where $\Gamma$ is the full width in conformal time. The Letter derives the barrier and the exponential fall of the width with decreasing $|\lambda|$ for a static box~\cite{v12PRL}.

Without the one-loop energy of the non-zero modes the zero mode would survive the turn almost surely at the central inputs ($\Pone\approx0.998$ in a one-dimensional version of the model); with them the outcome depends on the top-quark mass. %

The half-life of the zero mode in a static box at $\mub$, the origin of the barrier in the gauge zero modes and the exact reduction of the abelian Higgs model to two coordinates are given in the Letter~\cite{v12PRL}.

We solve for the resonance in $r$ and in the modulus $|\bm w|$ of the hypercharge zero mode, with the SU(2) zero modes in a conditional Gaussian corrected by diffusion Monte Carlo (DMC), and with the non-zero modes of all Standard Model fields in a one-loop finite-volume matching at $\mu=1/L$.
The width is that of a Siegert state~\cite{Siegert:1939zz}, outgoing in $r$ (Appendix~\ref{app:h-matching}).

\subsection{Survival along the turn}
\label{sec:h-hazard}
\label{sec:h-budget}

\begin{figure}[t]
\centering
\includegraphics[width=\columnwidth]{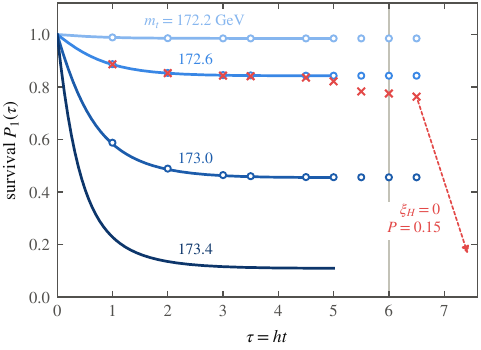}
\caption{Survival of the Higgs zero mode along the turning solution: $\Pone(\tau)=\exp[-I(\tau)]$, with $I(\tau)$ the decay exponent accumulated up to $\tau=ht$, for $m_t=172.2$--$173.4\GeV$ (lines), and the norm of the real-time evolution for $\xiH=1/6$ (circles) and $\xiH=0$ (crosses; arrow: final value after the classical stage from $\tau=6$).} %
\label{fig:v13-hazard}
\end{figure}

Along the turning solution the scale $1/L$ falls, $|\lambda|$ decreases and the decay slows down; the survival probability on one branch is the exponential of minus the integrated width [Fig.~\ref{fig:v13-hazard}].
At the central inputs it is $\Pone=0.84$, and $\Pone>1/2$ whenever %
\begin{equation}
\lambda(\mub)>\lamc(\mub)=-0.0121, %
\label{eq:h-threshold}
\end{equation}
where $\lamc(\mub)\equiv\lambda_{\rm SM}(\mub;m_c)$ is the Standard Model quartic at the critical top mass.
The central Standard Model quartic exceeds $\lamc$ by $\marginLam$, equivalent to $\marginGeV\GeV$ in $m_t$. %

Along the turning solution~(\ref{eq:pa-turning-solution}), with edge $L(\tau)=L_b(\cosh2\tau)^{1/2}$ and $\tau=ht$, the zero mode sees couplings at
\begin{equation}
\mu(\tau)=\frac{\mub}{(\cosh2\tau)^{1/2}},\qquad d\eta=\frac{d\tau}{hL_b(\cosh2\tau)^{1/2}},
\label{eq:h-node}
\end{equation}
where $hL_b$ is evaluated with the rate of the turning solution, $\sHeff\Lb=0.00437$, so that one branch lasts the conformal time $\eta_\infty=1.31103/(hL_b)=299.8$. %
At eleven nodes $\tau=0,0.5,\ldots,5$ we recompute the reduced model, with its matching and cross terms, for the couplings at $\mu(\tau)$ and solve directly for the resonance.
The decay exponent and the single-branch survival probability are
\begin{equation}
I(m_t)=\int_0^5\frac{\Gamma(m_t,\tau)\,d\tau}{hL_b(\cosh2\tau)^{1/2}},\qquad \Pone=e^{-I}.
\label{eq:h-hazard}
\end{equation}
The Letter writes this survival as a race between an equivalent duration of the turn in a static box at $\mub$ and the half-life of that box, which depends steeply on the top-quark mass~\cite{v12PRL}.

Under the principle the universe has a turning point, whereas a universe that joins the same solution at a larger size traverses less of it and survives with a higher probability~\cite{v12PRL}.

The decay exponent gives the survival of the resonant component of the state.
For a time-symmetric state $\psi_0$ at $\tau=0$ with weight $w_0$ on that component we find (Appendix~\ref{app:h-budget})
\begin{equation}
P_\infty=w_0\,\Pone\,(1+\delta_{\rm rt}),\qquad w_0=|(\varphi_{\rm res},\psi_0)|^2,
\label{eq:h-w0}
\end{equation}
with $|\delta_{\rm rt}|\le0.3\%$, where $(\cdot,\cdot)$ is the c-product with the resonance $\varphi_{\rm res}$. %
Section~\ref{sec:s-theorem} specified the shape state at the turn by two conditions: time symmetry about the turn and minimal energy variance among Gaussian states (row~\ref{v13:assump-S5}). This prescription centers the state on a stationary shape and places its width at the uncertainty floor.
We apply the same prescription to the Higgs zero mode, a different degree of freedom, and take the central Gaussian $\exp(-r^2/2\sigma_r^2-|\bm w|^2/2\sigma_s^2)$ of minimal energy variance. %
It gives $w_0=0.991$ at $m_t=172.6\GeV$, and $w_0$ lies between $0.980$ and $0.995$ for $172.2\le m_t\le173.4\GeV$. %
The criterion barely depends on this weight: it shifts the critical mass by at most $-0.016\GeV$ (Table~\ref{tab:budget}), and even $w_0=0.8$ would lower the critical mass for probability $1/2$ by only $0.114\GeV$. %

With $\xiH=1/6$ the zero-mode Hamiltonian is real and even in $\eta$, so a time-symmetric state survives on each branch with the same probability $\Pone$.
A branch on which the zero mode escapes does not stay regular, so the principle asks for survival on both branches, and quantum mechanics grants it only with a probability.
We quote single-branch probabilities, because the projectors onto the well at late times $\pm T$ on the two branches do not commute on $\psi_0$. Joint survival is therefore not a probability in the decoherent-histories sense~\cite{Gell-Mann:1991kdm,Goldstein:1994cp}, and any joint probability $P_2$ obeys only the Fr\'echet bounds $\max(0,2\Pone-1)\le P_2\le\Pone$ (Ref.~\cite{v12Report}, Sec.~9).
If the two branches decayed independently, $P_2=\Pone^2$ would equal one half at $m_t=172.76\GeV$. %

The Standard Model maps $\lambda(\mub)$ to the top-quark mass through three-loop running and next-to-next-to-leading-order matching.
Near the threshold $\lambda(\mub)$ is nearly linear in $m_t$, with $d\lambda(\mub)/dm_t\approx-0.0066\GeV^{-1}$, so the threshold becomes the steep criterion %
\begin{equation}
P_1>\tfrac12 \;\Longleftrightarrow\; m_t<172.96~\mathrm{GeV}
\quad\bigl[\,\bar m_t<163.16~\mathrm{GeV}\,\bigr], %
\label{eq:criterion}
\end{equation}
for $\alpha_s(M_Z)=0.1180$, where $\mtbar\equiv\mtbar(\mtbar)$ is the $\MSbar$ mass.

\begin{table}[t]
\caption{Critical top-quark pole mass $m_c$ at which $\Pone$ takes the value in the column header, and its uncertainties (GeV), for $\alpha_s(M_Z)=0.1180$ and $M_h=125.13\GeV$. %
Model items are combined in quadrature on each side; one-sided items carry their sign.
For the $\MSbar$ form the Standard Model mapping and $\alpha_s$ rows are $\pm0.07$ and $\pm0.25$; the other shifts differ from their pole-mass values by about $5\%$. %
Reference~\cite{v12Report}, Sec.~11, lists every item.}
\label{tab:budget}
\begin{ruledtabular}
\begin{tabular}{lccc}
 & $\Pone=0.9$ & $\Pone=0.5$ & $\Pone=0.1$ \\
\hline
$m_c$ & $172.50$ & $172.96$ & $173.42$ \\ %
$\bar m_c$ & $162.73$ & $163.16$ & $163.60$ \\[2pt] %
Zero-mode model\footnotemark[1] & $^{+\modelErrUpNoViPnine}_{-\modelErrDnNoViPnine}$ & $^{+\modelErrUpNoVi}_{-\modelErrDnNoVi}$ & $^{+\modelErrUpNoViPone}_{-\modelErrDnNoViPone}$ \\[2pt] %
\quad of which preparation weight & $-0.016$ & $-0.006$ & $-0.004$ \\ %
Inflation scale\footnotemark[2] & $\pm0.035$ & $\pm0.067$ & $\pm0.112$ \\ %
Model total & $^{+\modelErrUpPnine}_{-\modelErrDnPnine}$ & $^{+\modelErrUp}_{-\modelErrDn}$ & $^{+\modelErrUpPone}_{-\modelErrDnPone}$ \\[2pt] %
\hline
Standard Model mapping & $\pm0.10$ & $\pm0.10$ & $\pm0.10$ \\ %
$\alpha_s(M_Z)\pm0.0009$ & $\pm0.35$ & $\pm0.35$ & $\pm0.35$ \\ %
$M_h\pm0.11\GeV$ & $\pm0.053$ & $\pm0.053$ & $\pm0.053$ \\ %
Measured $m_t$ (direct) & $\pm0.27$ & $\pm0.27$ & $\pm0.27$ \\ %
Interpretation of $m_t$\footnotemark[3] & $\pm0.5$ & $\pm0.5$ & $\pm0.5$ \\ %
\end{tabular}
\end{ruledtabular}
\footnotetext[1]{One-loop matching, reduction to two coordinates, cross terms, numerics, and preparation weight (one-sided).}
\footnotetext[2]{$\rT\in[10^{-3},10^{-2}]$ in the plateau family; computed without the last one-loop cross terms, which change it by less than $0.001\GeV$.} %
\footnotetext[3]{Monte Carlo mass read as the pole mass.}
\end{table}

In Table~\ref{tab:budget} the model error at $\Pone=1/2$, $^{+\modelErrUp}_{-\modelErrDn}\GeV$, is dominated by the inflation scale, which enters through the duration of the turn. %
Without that item the model error is $^{+\modelErrUpNoVi}_{-\modelErrDnNoVi}\GeV$, comparable to the Standard Model theory error. %
The parametric error from $\alpha_s$ exceeds the theory and model errors combined, and only the interpretation of the measured mass as a pole mass contributes more.

The survival probability given current data, $\Pbar$, is the average of $\Pone$ over Gaussian distributions of the measured masses, $\alpha_s$ and the Standard Model error, with the asymmetric model error carried by one coherent nuisance parameter.
For the direct mass we find $\Pbar=\PbarOne$. %
Top-mass definitions differ by about $0.5\GeV$~\cite{Butenschoen:2016lpz,Hoang:2020iah}. %
Across them $\Pbar$ lies between $\PbarMSR$, for the MSR mass at $1.3\GeV$~\cite{Hoang:2017suc}, and $\PbarSigma$, for the cross-section pole mass~\cite{ParticleDataGroup:2026mpi} (Ref.~\cite{v12Report}, Sec.~11). %

\subsection{The curvature coupling}
\label{sec:h-curvature}

\begin{figure}[t]
\centering
\includegraphics[width=\columnwidth]{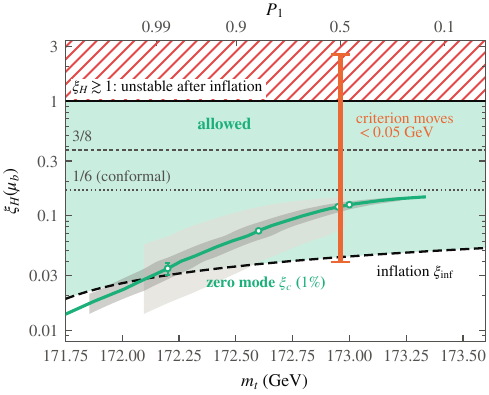}
\caption{Curvature coupling and the criterion. Shaded: values of $\xiH(\mub)$ allowed by vacuum stability during plateau inflation (above $\xi_{\rm inf}$, dashed) and after it ($\xiH\lesssim1$~\cite{Herranen:2015ima}; hatched above). Solid line and circles: zero-mode threshold $\xic$ for a $1\%$ loss of survival (universal curve and direct calculations), with the Standard Model mapping (dark) and $\alpha_s$ (light) bands. Bar: range in which the critical mass at $\Pone=1/2$ moves by less than $0.05\GeV$. Dotted: $1/6$; thin dashed: $3/8$, the onset of tachyonic resonance; upper axis: $\Pone$.} %
\label{fig:v13-xi}
\end{figure}

The survival probability $\Pone$ refers to the conformal coupling $\xiH=1/6$, for which the curvature term of the zero-mode equation vanishes; for smaller $\xiH$ the late turn can destroy the zero mode, which sets a lower bound on $\xiH$.
At the measured top mass $\Pone$ is unchanged to $1\%$ for $\xiH(\mub)\gtrsim\xiHcentral$ in the zero-mode calculation ($0.075$ once the non-zero modes are included) and for $\xiH(\mub)\gtrsim\xiHmin$ in all prescription variants. %
The criterion itself needs only a looser tolerance (Sec.~\ref{sec:h-universality}): the critical mass at $\Pone=1/2$ moves by less than $0.05\GeV$ for $0.04\lesssim\xiH(\mub)\lesssim2.6$ [Fig.~\ref{fig:v13-xi}]. %

For $\xiH<1/6$ the curvature term adds to the zero-mode Hamiltonian in the conformal frame the inverted harmonic potential $-(1-6\xiH)(hL)^2r^2$, with $hL=hL_b(\cosh2\tau)^{1/2}$. This potential grows like $\cosh2\tau$ and dominates at $\tau\gtrsim4$.
A lens transformation maps the conformal-frame equation exactly to the physical field, whose late dynamics is classical, so from $\tau=6$ we continue with Wigner-sampled classical trajectories with Hubble friction. The result does not depend on this switch (Appendix~\ref{app:h-budget}).

We define the threshold $\xic$ by
\begin{equation}
P_{\rm surv}(\xic)=0.99\,P_{\rm surv}(\tfrac16), %
\label{eq:h-xic}
\end{equation}
with $P_{\rm surv}$ from the real-time evolution, so that for $\xiH\ge\xic$ the survival probability is $\Pone$ to $1\%$ [Fig.~\ref{fig:v13-xi}].
The main uncertainty is the late-time running prescription, $\lambda$ at the field value or at $1/L$, and we quote the midpoint and half-difference.
In the first prescription a corrected classification of trajectories that linger near the barrier raises $\xic$ by $0.0001$ at the measured mass and by $0.0017$ at $172.2\GeV$ (Ref.~\cite{v12Report}, Sec.~12). The values quoted here are those of the main calculation. %
Below $\xic$ the survival decreases gradually, and for $\xiH=0$ the zero mode runs away: only $0.10$--$0.15$ of the probability has not yet run away after 60 $e$-folds, almost all of it on undecided trajectories [Fig.~\ref{fig:v13-hazard}]. %

The threshold $\xic$ is close to a universal function of $\Pone$ (Ref.~\cite{v12Report}, Sec.~12).
In the zero-mode calculation both versions of the model (with and without the last one-loop cross terms) and the variants of the Standard Model inputs and $\alpha_s$ whose $\Pone$ lies in the tested range follow one curve $\xic(\Pone)$ to within $0.002$. The non-zero modes, discussed at the end of this subsection, move this curve up by $0.006$, $0.0017$ and $0.0003$ at $\Pone=0.985$, $0.84$ and $0.50$. %
Stated through $\Pone$, or through $\lambda(\mub)$, the condition is therefore free of the Standard Model and $\alpha_s$ errors at first order; at fixed $m_t=172.6\GeV$ these errors move $\xic$ along the curve by $+0.010/{-0.017}$ and $+0.045/{-0.037}$. %
At the point $\Pone=1/2$ the threshold is $\xiHhalf$. %
The coupling runs as $\xi(\mu)-\tfrac16=(\xi_{\rm EW}-\tfrac16)\,\Xi(\mu)$ with $\Xi(\mub)=1.155$~\cite{Herranen:2014cua,Mantziris:2020rzh}, so $\xiHmin$ at $\mub$ corresponds to $\approx\xiHminEW$ at the electroweak scale, and the conformal value is a fixed point of the one-loop running. %
The coupling $\xiH$ is required by renormalization in curved spacetime~\cite{Birrell:1982ix}, so this bound is a condition on a Standard Model parameter.

After the turn the universe inflates on the plateau, and long-wavelength Higgs fluctuations can nucleate the true vacuum~\cite{Espinosa:2007qp,Hook:2014uia,Kearney:2015vba,Espinosa:2015qea,Joti:2017fwe,Markkanen:2018pdo}.
Because the plateau inflaton is minimally coupled and its potential has no $R^2$ term, we use the field-theory criterion of Ref.~\cite{Mantziris:2020rzh}, not the $R^2$ bound of Ref.~\cite{Mantziris:2022fuu}. This criterion combines the RG-improved one-loop curved-space effective potential~\cite{Markkanen:2018bfx} and the Hawking--Moss rate~\cite{Hawking:1981fz}, and it requires less than one bubble in the past light cone (Ref.~\cite{v12Report}, Sec.~12).
For $m_t=172.2$, $172.6$ and $173.0\GeV$ the bound is $\xi_{\rm inf}(\mub)=0.030$, $0.038$ and $0.044$. %
At $\mub$ the two conditions coincide near $m_t=172.1\GeV$ in the zero-mode calculation and at a lower mass with the non-zero modes, and above this mass the zero-mode condition is the stronger one. %

The two calculations cover adjacent epochs, the zero-mode calculation until the non-zero modes leave the Hubble radius and the light-cone count afterward, and one effect lies between them.
Once the lowest non-zero modes have left the Hubble radius, $7.3$ $e$-folds after the turn, a region of the torus is lost when its coarse-grained field crosses the curvature barrier, even if the mean field does not. %
This affects only zero-mode trajectories that linger at the top of the barrier, where the non-zero modes have the mass squared $-24\xiH h^2$ and grow at the same rate as the displacement of the zero mode from the top. %
At $m_t=172.6\GeV$ and $\xiH=0.07$ it removes a further $(5$--$8)\times10^{-4}$ of the probability, and it raises $\xic$ by $0.0017(2)$. The condition $\xiH\gtrsim\xiHmin$, which covers all prescription variants, is unchanged, with a margin of $0.001$ (Ref.~\cite{v12Report}, Sec.~12). %

After inflation the inflaton oscillates with $H_{\rm end}\approx7\times10^{12}\GeV$, far above the instability scale, so the analysis of Ref.~\cite{Herranen:2015ima} applies. %
Tachyonic resonance then sets in for $\xi>3/8$, but it destabilizes the vacuum only for $\xi$ of order one or larger~\cite{Herranen:2015ima}.
The allowed window is therefore $\max(\xic,\xi_{\rm inf})\le\xiH\lesssim1$, and the conformal value $\xiH=1/6$ lies inside it at both scales.
Over the range $0.044\lesssim\xiH(\mub)\lesssim1$ that vacuum stability during and after inflation allows at the criterion, the critical mass moves by less than $0.05\GeV$. %

\subsection{Robustness}
\label{sec:h-universality}

The threshold on $\lambda(\mub)$ depends on the duration of the turn only logarithmically and on the couplings with logarithmic derivatives below $0.7$, so its size, $\lamc\approx-0.01$, does not depend on the details of one point in parameter space. %
A change of $10\%$ in any one coupling changes $|\lamc|$ by at most $6.6\%$, so the magnitude $\lamc\approx-0.01$ is fixed to within a factor of about $1.2$. %

A finite threshold exists in every sector that can host a turning point and keeps a Higgs zero mode, because there the zero-point motion of the constant gauge modes still confines the zero mode.
In the PPA and PAA spin structures, which also turn around (Sec.~\ref{sec:pa-image}), the hypercharge curvature on the periodic edges is negative near the well. An indicative estimate shifts $m_c$ by at most $+0.08\GeV$ relative to AAA, and the exact value needs an anisotropic model (Ref.~\cite{v12Report}, Sec.~12). %
Periodic fermions on all three cycles would reverse the sign of the linear term, because the constant modes of the top quark would outweigh the gauge zero-point energy~\cite{v12PRL}. Sectors with 't~Hooft flux cannot turn around (Sec.~\ref{sec:pa-image}).

The criterion assumes that the inflaton does not couple directly to the Higgs field and that no further states couple to it below $2\pi\mub$ (rows~\ref{assump:A3} and~\ref{assump:A2}); both assumptions can be given a tolerance.
Since $R=12h^2$ along the turning solution, a portal coupling $\tfrac12\lambda_{h\phi}\phi^2\Phi^\dagger\Phi$ acts on the zero mode as a shift $\Delta\xiH=\lambda_{h\phi}\phi_b^2/24h^2\approx3.5\times10^{10}\,\lambda_{h\phi}$ of the curvature coupling, and a mixed term $\zeta(\phi^2/\MPbar^2)R\,\Phi^\dagger\Phi$ as a shift $31\,\zeta$. %
Measured from the conformal value at the reference turning point, the critical mass at $\Pone=1/2$ moves by less than $0.05\GeV$ for $-3.6\times10^{-12}<\lambda_{h\phi}<6.8\times10^{-11}$ and for $-0.004<\zeta<0.08$. %
The windows are asymmetric because a negative shift acts like a smaller curvature coupling: the lower edge corresponds to an effective $\xiH\approx0.04$. The $1\%$ condition of Sec.~\ref{sec:h-curvature} needs $\lambda_{h\phi}\gtrsim-2.6\times10^{-12}$ at the measured mass and $\gtrsim-1.3\times10^{-12}$ at the criterion. %

Right-handed neutrinos lower $\lambda(\mub)$ through the running above their mass (their effect on the Casimir coefficient is not included). For one neutrino with $m_\nu=0.05$\,eV the critical mass changes by less than $0.05\GeV$ if $M_N<5\times10^{13}\GeV$. %
A real singlet scalar with portal coupling $\lambda_{HS}=0.1$ and mass $10^{9}\GeV$ raises the critical mass by $0.07\GeV$ through the running and by about $0.16\GeV$, an estimate good to a factor of two, through its constant mode, which adds $(\lambda_{HS}/8)^{1/2}$ to the slope $\sigma_0$ (Ref.~\cite{v12Report}, Sec.~12). %

\subsection{What the Standard Model mapping adds}
\label{sec:h-criterion}

The Standard Model mapping turns the threshold on $\lambda(\mub)$ into the criterion~(\ref{eq:criterion}) on the top-quark mass; here we give its inputs, its uncertainties and what the threshold adds to the known stability bounds.

Our own code starts at $\mu_0=m_t$ from the SMDR reference model~\cite{Martin:2019lqd}, in the tadpole-free pure $\MSbar$ scheme, and shifts it to $(m_t,\alpha_s,M_h)$ with the slopes of Ref.~\cite{Buttazzo:2013uya}.
The $\beta$ functions are the complete three-loop Standard Model ones~\cite{Mihaila:2012fm,Bednyakov:2012rb,Bednyakov:2012en,Bednyakov:2013eba,Chetyrkin:2012rz,Chetyrkin:2013wya} with four-loop gauge~\cite{Bednyakov:2015ooa,Zoller:2015tha,Davies:2019onf} and leading four- and five-loop QCD terms~\cite{Martin:2015eia,Chetyrkin:2016ruf,Baikov:2014qja,Baikov:2016tgj,Herzog:2017ohr}, and the pole and $\MSbar$ masses are related through four loops~\cite{Marquard:2015qpa}.

The all-orders pole mass lies $0.27\GeV$ above our four-loop truncation, and the renormalon ambiguity is $0.07$--$0.25\GeV$~\cite{Beneke:2016cbu,Hoang:2020iah}; the $\mtbar$ form of Eq.~(\ref{eq:criterion}) is free of both. %
Reference~\cite{v12Report}, Sec.~10, gives the benchmarks of the mapping and a tadpole-including variant.

The new ingredient is the threshold.
It places the criterion $m_c-m_0=\mcMinusMzero\GeV$ above the line $\lambda(\mub)=0$ at $m_0=171.09\GeV$, which with our inputs lies just above the absolute-stability bound~\cite{Buttazzo:2013uya,Bednyakov:2015sca}, and far below the metastability bound, $177.7\GeV$~\cite{Andreassen:2017rzq}. %
In this relative form the criterion carries mainly the model error, because a shift of $\lambda(\mub)$ from the Standard Model mapping moves $m_c$ and $m_0$ together.

%% file: sections/v13_s06_observables.tex
\section{What can be observed}
\label{sec:spectra}

If the fields are in their vacuum at the turn, inflation after the turn removes every trace of the turning point from the primordial spectra.
Two traces of the beginning remain: the shape fossil of Sec.~\ref{sec:s-fossil}, which the microwave background can read only in a narrow window of present sizes, and the top-quark mass, which decides whether the Higgs zero mode survived the turn.

\subsection{Primordial spectra}
\label{sec:c-nofeature}

For fields in their adiabatic vacuum at the turn, the turning point leaves no feature in the primordial spectra.
Their tilt and tensor-to-scalar ratio are those of the plateau potential, evaluated at the number of $e$-folds that reheating fixes.

All modes of the torus are inside the Hubble radius at the turn by at least a factor $2\pi/(h\Lb)\simeq1400$ (1300 for free fields). %
The emptiness bound of Sec.~\ref{sec:gate} allows them at most five quanta there, and we take them in their adiabatic vacuum (row~\ref{v13:assump-S5}).
This vacuum evolves into the Bunch--Davies state of the subsequent plateau, with Bogoliubov coefficients suppressed by powers of $h\Lb/2\pi|\bm n|$.
The scalar and tensor spectra on the torus are therefore those of plateau inflation, sampled on the reciprocal lattice (Ref.~\cite{v12Report}, Sec.~16).

For instant reheating, $N_*=55.72$, we find $n_s=0.96534$ and $\rT=0.00346$. %
For the weak visible decay through the reheating coupling of Sec.~\ref{sec:s-reheating}, $N_*=50.33$, we find $n_s=0.96174$ and $\rT=0.00420$. %

Against the CMB-only combination of SPT-3G, ACT and Planck (CMB-SPA), $n_s=0.9679\pm0.0033$~\cite{SPT-3G:2025bzu}, the tilt lies $0.8\sigma$ below the measured value for instant reheating and $1.9\sigma$ below it for the weak decay. %
When DESI DR2 baryon acoustic oscillations~\cite{DESI:2025zgx} are added, the measured tilt rises to $0.9726\pm0.0028$ (CMB-SPA, Ref.~\cite{SPT-3G:2025bzu}) or $0.9752\pm0.0030$ (P-ACT-LB2, Ref.~\cite{AtacamaCosmologyTelescope:2025blo}). The one-dimensional offsets then grow to $2.6$--$3.3\sigma$ for instant reheating and $3.9$--$4.5\sigma$ for the weak decay. %
The tension belongs to the plateau potential and not to the turning point. A potential with a larger tilt at the same $N_*$ changes these numbers and leaves the turn intact, since the turn needs only $V(\phi_b)>0$ and the criterion~\eqref{eq:pa-exact-criterion}.
The tensor ratio is an order of magnitude below the BICEP/Keck bound $\rT<0.036$~\cite{BICEP:2021xfz} and three to four times the LiteBIRD target $\delta\rT=10^{-3}$~\cite{LiteBIRD:2022cnt}. %

The scale of inflation enters the two other observables only weakly.
At fixed $A_s$ the width of the shape fossil grows as $\rT^{1/8}$, Eq.~\eqref{eq:s-scaling}, so a measurement $\rT\simeq0.004\pm10^{-3}$ fixes $\sigma_E$ to $3.1\%$. %
Over the plateau class, $10^{-3}\leq\rT\leq10^{-2}$, the critical top mass at $\Pone=1/2$ moves by $\pm0.067\GeV$ (Table~\ref{tab:budget}). %

\subsection{The fossil in the sky}
\label{sec:sky}

The shape fossil survives to the present unchanged, and the microwave background can read it only if the shortest cycle of the torus lies between $0.985$ and about $1.04\,\dLSS$. %
The fundamental domain today therefore has the frozen shape of Sec.~\ref{sec:s-fossil}, scaled to a present size $L_0$ defined by the volume $L_0^3$.
The model fixes the width of the distribution of present sizes but not its center, so it does not predict whether the fossil is readable.

\label{sec:c-shortest}
The absence of matched circles requires every translation of the torus to exceed $0.985\,\dLSS$~\cite{Vaudrevange:2012da,COMPACT:2022nsu}. %
A somewhat larger torus still correlates harmonic coefficients of different $\ell$ and $m$~\cite{COMPACT:2023rkp,Himeno:2026lgh}, and these correlations carry its shape.
For $55$ deformed tori that cover the predicted shapes we computed the exact covariance of the temperature and $E$-mode coefficients for $2\leq\ell\leq80$ on the full sky, and its Kullback--Leibler divergence $D$ (not the region of Sec.~\ref{sec:s-theorem}) from the closest circle-free cubic torus. %
The covariance uses transfer functions from CAMB~\cite{Lewis:1999bs} for the Planck 2018 $\Lambda$CDM parameters~\cite{Planck:2018vyg}; Ref.~\cite{v12Report}, Sec.~17, gives the method, the convergence tests and an independent reproduction (code: Ref.~\cite{DNECv12code}).
We call a fossil readable if $D\geq4.5$; for nested Gaussian models $2D=9$ corresponds to $3\sigma$ for a deformation of known direction. %

\begin{figure*}[t]
\centering
\includegraphics[width=\textwidth]{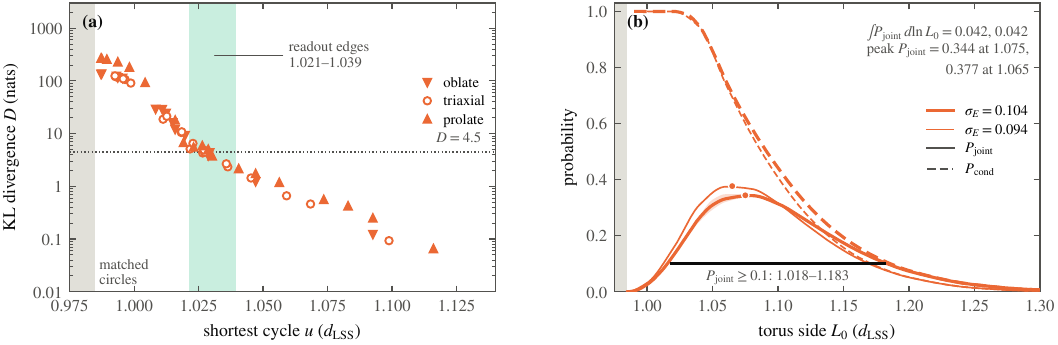}
\caption{Microwave-background readability of the shape fossil, set by the shortest cycle. (a)~Kullback--Leibler divergence of $55$ computed fossils from the closest circle-free cube, for temperature and $E$ modes with $2\leq\ell\leq80$ on the full sky, versus their shortest cycle $u$; band: readout edges of the 15 interpolated shapes; dotted: the readability criterion $D=4.5$; gray: excluded by matched circles. (b)~Probability that a predicted fossil is readable with no matched circles (solid) and given that circles are absent (dashed), versus the present size $L_0$, for $\sigma_E=0.104$ (thick; light band: edge-interpolation bounds) and $0.094$ (thin); bar: $P_{\rm joint}\geq0.1$. Below $L_0\simeq1.03\,\dLSS$ the dashed curve depends on the readout edge for $A<0.25$, where no fossil was computed; $P_{\rm cond}$ must vanish as $L_0\to0.985\,\dLSS$ (Ref.~\cite{v12Report}, Sec.~17).} %
\label{fig:c-sky}
\end{figure*}

The aspect part of the fossil lies in the $E$ plane,
\begin{equation}
\beta=A\,\sigma_E\big(\cos\theta\,X_d+\sin\theta\,X_{d2}\big),
\label{eq:c-Eplane}
\end{equation}
with $A$ Rayleigh distributed (median $1.177$, rms $\sqrt2$) and $\theta$ uniform for the Gaussian of Sec.~\ref{sec:s-fossil}. %
We write $u$ for the shortest side and neglect the angle modes, which change $D$ only at second order, by about $0.4\%$. %
The divergence is a steep function of $u$ and depends on the amplitude and orientation of the fossil almost only through it [Fig.~\ref{fig:c-sky}(a)].
It falls to about $4.5$ at $u\simeq1.027\,\dLSS$, to $1$ at $1.05\,\dLSS$ and to $0.1$ at $1.10\,\dLSS$. %
Over the $15$ computed shapes the readout edge $u_{\max}$, the shortest cycle at which $D=4.5$, lies between $1.021$ and $1.039\,\dLSS$. %

\label{sec:c-ensemble}
At a given present size the fossil is readable if $0.985\,\dLSS\leq u\leq u_{\max}$. %
Averaging over the Gaussian distribution of Eq.~\eqref{eq:c-Eplane} gives the joint probability $P_{\rm joint}(L_0)$ that the fossil is readable and matched circles are absent, and the conditional probability $P_{\rm cond}$ given that circles are absent [Fig.~\ref{fig:c-sky}(b)].
For temperature and $E$ modes to $\ell=80$, $P_{\rm joint}$ peaks at $0.344$ near $L_0=1.075\,\dLSS$, with bounds $0.339$--$0.351$ from the interpolation of the readout edges. %
It exceeds $0.1$ for $1.018<L_0/\dLSS<1.183$, and given that circles are absent the fossil is readable with probability above one half up to $L_0=1.090\,\dLSS$. %
Averaging over shapes spreads the readable sizes without adding to them: for each shape the readable range of $\ln L_0$ has the width $\ln(u_{\max}/0.985\,\dLSS)$, so that
\begin{equation}
\int P_{\rm joint}\,d\ln L_0=\Big\langle\ln\frac{u_{\max}}{0.985\,\dLSS}\Big\rangle=0.042,
\label{eq:c-measure}
\end{equation} %
almost exactly the width, $0.0415$, of the window of a single typical fossil, $|\beta_E|=\sqrt2\,\sigma_E$ along $X_d$. %
With the next-order fossil, $\sigma_E=0.094$ (with the asymptotic plateau rate; $0.0934$ at the rate of the turning point, Sec.~\ref{sec:s-fossil}), the peak rises to $0.377$, while the integrated measure does not change. %

\label{sec:c-size}
The present size of the torus is not selected by the model.
The constraint at the turn, $V(\phi_b)\Lb^4=\NC^{\rm eff}\gamma_C$, leaves a one-parameter family of time-symmetric turning points labeled by the inflaton value $\phi_b$ at the turn.
As $\phi_b$ runs from $4.9$ to $6.3\,\MPbar$, the number of $e$-folds after the turn grows from $38$ to $125$, while $\Lb$ changes by $0.6\%$. %
This number sets $\ln L_0$, and nothing in the model selects $\phi_b$.

The width of the distribution, in contrast, is bounded from below once the inflaton zero mode is treated quantum mechanically, beyond row~\ref{v13:assump-S2}.
To linear order $N_{\rm post}-\bar N=\mathcal A\,\delta\phi_b+\mathcal B\,\delta P_b$, where $P_b$ is the momentum of the inflaton zero mode at the turn.
Time reversal maps $Q=\mathcal A\phi_b+\mathcal BP_b$ to $\bar Q=\mathcal A\phi_b-\mathcal BP_b$, so in a time-symmetric state $Q$ and $\bar Q$ have the same distribution, while $[Q,\bar Q]=-2i\mathcal A\mathcal B$; hence
\begin{equation}
\sigma_N^2\geq|\mathcal A\mathcal B|,
\label{eq:c-sizebound}
\end{equation}
provided the other sectors are uncorrelated with $(\phi_b,P_b)$ at linear order.
At the reference turning point this gives $\sigma_N\geq0.935$ $e$-fold. %

A probability for the center needs a measure on $\phi_b$.
For an amplitude independent of $\phi_b$, the Liouville measure on the turning surface and the norms of the expanding branch all give $p(N_{\rm post})\simeq1.17/N_{\rm post}$ for $38\leq N_{\rm post}\leq125$, which cannot be normalized on the plateau. %
For any prior on $\ln L_0$ that is broad on the scale of the window, the probability of a readable fossil is proportional to the integrated measure, Eq.~\eqref{eq:c-measure}.
A Gaussian prior centered near the window, with the width $\sigma_N\simeq0.96$ of the time-symmetric minimal-variance state of the inflaton zero mode, gives $P_{\rm cond}\approx3.3\%$. %
The width $1.9$ of an inflaton packet that neglects the slope of the plateau gives $P_{\rm cond}\approx1.7\%$, and a prior $\propto L_0^{-\alpha}$ reaches $P_{\rm cond}=1/2$ only for $\alpha\simeq17$. %
Two extensions of the model do not change this.

A wave function that decays under the Casimir barrier, $L<\Lb$, weights the turning points by $e^{2S_E(\phi_b)}$, where $S_E\simeq1.2\times10^4$ is the barrier action. %
This weight grows with the Casimir coefficient and therefore does not favor the cube, in conflict with row~\ref{v13:assump-S5}.
Near the window this is a prior $\propto L_0^{-1.8}$, for which $P_{\rm cond}\approx7\%$, and without the condition on the circles it favors a torus far smaller than the horizon. %

The rule of minimal energy variance, which fixes the centers of the other homogeneous modes at stationary points of their finite-volume energy, has no solution for the inflaton, because $V-\NC^{\rm eff}\gamma_C/L^4$ has no stationary point on the plateau.
If matter couples to the inflaton as in the Jordan frame of $R+R^2$ gravity, the running of the zero-mode energy makes the Casimir coefficient depend on $\phi_b$ and creates one, about $10^3$ $e$-folds before the end of inflation, and the matching scale of the Higgs criterion would then change as well. On a hilltop the observed tilt requires more than $250$ $e$-folds. %
In both cases the torus is far larger than the horizon.
A readable fossil therefore needs a prior on the size sharply concentrated near the circle bound, which neither the model nor these extensions provide. The present size of the torus remains an open problem (Ref.~\cite{v12Report}, Sec.~17, gives the measures, the priors and the extensions).

\label{sec:c-tests}
For a cubic torus the Planck likelihoods require $L>0.92\,\dLSS$ (2013)~\cite{Planck:2013okc} and $L>0.97\,\dLSS$ (2015)~\cite{Planck:2015gmu}, weaker than the circle bound~\cite{COMPACT:2022nsu}. %
A shortest cycle between $0.985$ and about $1.04\,\dLSS$ is therefore allowed by circle and likelihood searches. It lies within the range $L\lesssim1.2\,\dLSS$ where off-diagonal correlations remain detectable in principle~\cite{COMPACT:2023rkp}, and no published analysis has excluded it. %

If a torus is found with its shortest cycle in the readable window, its aspect amplitudes should be compatible with an isotropic Gaussian of width $0.104$ per mode ($0.094$ at next order), truncated at large elongation by the absence of circles. %
Its angles should lie within about a degree of $90^\circ$; angle deviations above about $20^\circ$, detectable at $3\sigma$ for $L_0=1.10\,\dLSS$, would exclude the natural state centered on the cube (Sec.~\ref{sec:s-fossil}). %
A single sky is one draw from the predicted distribution. The Fisher information on $\ln\sigma_E$ from the two aspect amplitudes of one torus is $4$, so even a perfect measurement tests the predicted width only to about $50\%$. %
This suffices to separate the uncertainty floor from a shape state that is time symmetric about an instant later than about $0.25/h$ after the turn. %
A null topology search out to a shortest cycle of about $1.05\,\dLSS$ would close the window.

\subsection{The top-quark mass}
\label{v13:sec:obs-top}

The second trace of the beginning is the top-quark pole mass, which Sec.~\ref{sec:higgs} ties to the survival of the Higgs zero mode along the turn.
FCC-ee~\cite{Freitas:2019bre,FCC:2025lpp} can decide the question in both directions: a measured pole mass below $\mSurvFCC\GeV$ would establish $\Pone>1/2$, and one above $\mExclFCC\GeV$ would establish $\Pone<0.1$, each at $95\%$ confidence. %

With present data the survival probability, marginalized over the measured masses, $\alpha_s$ and the theory and model errors, is $\Pbar=\PbarOne$ for the direct top mass (Sec.~\ref{sec:h-budget}), and current data neither favor nor disfavor this beginning. %
The Letter gives the FCC-ee thresholds, the marginalized probabilities across top-mass definitions and the Bayes factors~\cite{v12PRL}.
The reach of the HL-LHC is limited by the error of $\alpha_s$ (Ref.~\cite{v12Report}, Sec.~11).
The test addresses the scenario as a whole, a beginning at the Standard Model turning point with a curvature coupling in the range that inflation allows, and none of its ingredients separately. %

%% file: sections/v13_s07_discussion.tex
\section{Discussion}
\label{sec:disc}

We have asked what follows from the single requirement that a flat universe filled with Standard Model fields be nonsingular.
Within the scenario of Table~\ref{v13:tab-assumptions}, the following results are derived from this principle.
Without an $R^2$ term the trace anomaly forbids conformal matter by itself to turn an anisotropic universe around, so the negative Casimir energy needs a partner with $w<1/3$, most cheaply a positive potential (Sec.~\ref{sec:trace}).
With antiperiodic fermions the Friedmann constraint fixes the turning point at an edge of $722$ reduced Planck lengths (Sec.~\ref{sec:turning}). %
Survival of both branches requires the torus to be nearly cubic there in the aspect directions (Sec.~\ref{sec:v13-meanshape}), and at such a turning point the same constraint lets the light fields hold only about five quanta, which bounds the entropy by a few tens and makes the turning point the beginning of both branches (Sec.~\ref{sec:gate}). %
The negative energy cannot hold the shape of the torus, and survival of the shape on both branches leaves its two unstable modes about $9\times10^2$ quantum states and bounds the instant about which a time-symmetric state can be symmetric (Sec.~\ref{sec:shape}).
Time symmetry about the turn (row~\ref{v13:assump-S5}) and the uncertainty principle then set a floor under the shape fossil, and in the natural state centered on the cube the torus freezes into a nearly rectangular box whose sides typically differ by $16$--$18\%$. %
Although the Higgs quartic coupling is negative at the turn, the zero-point energy of the gauge zero modes holds the Higgs zero mode down to the threshold set by its finite-volume Hamiltonian, and the zero mode survives each branch with probability above one half for a top-quark pole mass below $172.96\GeV$, for any curvature coupling that vacuum stability during and after inflation allows (Sec.~\ref{sec:higgs}). %

The low entropy of the beginning, a separate postulate in cosmologies with a singular beginning~\cite{Penrose:1980ge}, is thus a consequence of the principle here.
The time symmetry of the state of the universe, also discussed as a postulate~\cite{Hawking:1985af,Page:1985ei,Kiefer:1994gp}, remains one; the principle bounds the instant about which it can hold, the mean shape and the width of the state.

The rest is assumed.
The principle itself is the postulate of this work (row~\ref{v13:assump-P}).
Space is a flat three-torus with fermions antiperiodic around all three cycles (row~\ref{assump:A1}).
Compactness is not a choice, because Ref.~\cite{Zhang:2026gva} shows that a nonsingular flat universe requires it if quantum fields obey the achronal averaged null energy condition.
The torus, among the compact flat spaces, and its spin structure are choices.
At the turn the state is the natural state, the Gaussian of minimal energy variance that is symmetric about the turn and centered on a stationary shape, with the non-zero modes and the gauge zero modes in their ground state (row~\ref{v13:assump-S5}); of the two stationary shapes we take the cube.
The inflaton is a plateau field without a direct coupling to the Higgs field (row~\ref{assump:A3}), and the Standard Model, with its curvature coupling $\xiH$, holds up to the scale $2\pi\mub$ of the turning point (row~\ref{assump:A2}), within semiclassical Einstein gravity (row~\ref{v13:assump-S2}).

Four problems remain open; we list them in order of weight.
First, the model leaves the present size of the torus undetermined: the number of $e$-folds after the turn is set by the inflaton value at the turn, and nothing in the model selects that value.
The size decides whether the shape fossil can be seen, because the microwave background reads it only if the shortest cycle lies between $0.985$ and about $1.04\,\dLSS$ (Sec.~\ref{sec:sky}). %
The inflaton is the one homogeneous mode of the model that does not sit at a stationary point of its finite-volume energy, and the extensions that give it one place the torus far beyond the horizon (Sec.~\ref{sec:c-size}).
By contrast, the top-quark criterion does not depend on the size.

Second, we do not derive the spin structure, the centering of the state on a stationary shape or the choice between the two such shapes.
For the fully antiperiodic spin structure a numerical search finds exactly two stationary shapes of the Casimir energy, the cube and the body-centered cubic torus (Sec.~\ref{sec:pa-image}).
At the second, with Casimir weight $\NC=22.25$, the three shear modes are unstable and grow by a factor $210$ through the turn. %
The fossil of this shape therefore lies in its shears: the angles between the edges of its conventional cubic cell deviate from right angles by $5.8^\circ$ rms while the edges stay nearly equal. %
Because the shortest translations of that torus meet at $70.5^\circ$ and $109.5^\circ$, the angle test of Sec.~\ref{sec:c-tests} applies only to the cube. %
With one or two periodic cycles the stationary shape is tetragonal (Appendix~\ref{app:smcasimir}), and the fossil would be centered on that shape, which we have not evaluated; the fully periodic sector, whose one-loop vacuum has a nontrivial holonomy, is not analyzed.
A measured shape would tell these cases apart, and it would also tell the quantum fossil from the deformation that a few quanta at the turn can leave.
Each excited quantum of a non-abelian gauge zero mode keeps the cube stationary but raises the fossil width $\sigma_E$ by $8$--$9\%$. %
A pair of hypercharge photons with opposite momenta along one edge costs $42\%$ of the emptiness bound and, when the axis of the pair is definite, freezes a cubic torus into an elongated box with one side $4.96$ times as long as the other two. %

Third, the inflaton sector is assumed.
The model does not explain the origin of the plateau, and for instant reheating the scalar tilt of our benchmark potential lies $0.8\sigma$ below the CMB-only value and $2.6$--$3.3\sigma$ below the values that include baryon acoustic oscillations (Sec.~\ref{sec:c-nofeature}). %
This tension belongs to the plateau potential, not to the turning mechanism, which needs only a positive potential energy at the turn, $V(\phi_b)>0$, together with the criterion~\eqref{eq:pa-exact-criterion}.

Fourth, the principle itself is not explained.
We take nonsingularity as the starting point and derive its consequences; why the universe should be nonsingular lies outside the model.

The calculation has the following limits.
The dynamics of the Higgs zero mode is computed in a reduced model, and full-dimensional checks bound the reduction error in the critical mass at $\Pone=1/2$ by $+0.020\GeV$. %
Two-loop finite-volume matching is not computed; the variation of the matching scale, $+0.012/{-0.049}\GeV$, which estimates its logarithmic part, is included in Table~\ref{tab:budget}. %
Taking the time-symmetric Gaussian of minimal energy variance as the state of the Higgs zero mode at the turn lowers the critical masses by at most $0.016\GeV$. %
The complete model error, $+\modelErrUp/{-\modelErrDn}\GeV$, is dominated by the unknown scale of inflation. %
The non-zero Higgs modes that leave the Hubble radius while the zero mode lingers at the top of the curvature barrier are treated in linear theory with nonlinear and light-cone corrections, not on a lattice; at the measured mass they raise $\xic$ by $0.0017(2)$, and they leave the condition $\xiH\gtrsim\xiHmin$ unchanged (Sec.~\ref{sec:h-curvature}). %
The interpretation of the measured top mass as a pole mass, uncertain by about $0.5\GeV$, dominates the present uncertainty: across mass definitions $\Pbar$ ranges from $\PbarMSR$ to $\PbarSigma$. %

In the shape sector the zero-mode energy is taken at leading order; the next order, a one-sided shift, lowers the uncertainty floor by the factor $0.900$. %
Shape evolution is computed with classical nonlinear dynamics and small quantum corrections; quantum corrections to the survival tail, the dynamical response of the minimally coupled graviton and the joint quantum state of volume, inflaton and shape remain to be computed.
The sky forecast assumes idealized full-sky maps; masks and foregrounds can only lower the readability probabilities, and a likelihood on masked Planck maps is the next step.
None of these limits changes the mechanism, and those that move the critical top mass move it by no more than about the present $\alpha_s$ uncertainty, $\pm0.35\GeV$. %

Two measurements can test this picture.
At FCC-ee a top-quark pole mass below $\mSurvFCC\GeV$ would establish $\Pone>1/2$ at $95\%$ confidence, and one above $\mExclFCC\GeV$ would establish $\Pone<0.1$ at the same confidence and exclude this beginning. %
A topology search that finds the torus with its shortest cycle in the readable window would measure its shape: the aspect amplitudes should then be drawn from an isotropic Gaussian of width $0.104$ per mode, and the edge angles should lie within about a degree of $90^\circ$. %
A Casimir turning point begins almost empty, with the Higgs zero mode held in place by the zero-point motion of its own gauge fields, and what it passes on to later times is the shape of space.

%% file: appendices/v13_aA_trace.tex
\section{Trace identities and the zero-expansion boundary}
\label{app:geometry}

\subsection{Curvature invariants of Bianchi~I}
\label{app:curvature}

For $ds^2=-dt^2+h_{ij}(t)dx^idx^j$ let $K_{ij}=\tfrac12\dot h_{ij}$, $\mathsf K^i{}_j=h^{ik}K_{kj}$ and $\mathcal V=(\det h)^{1/2}$.
Flat homogeneous slices give $R_{ijkl}=K_{ik}K_{jl}-K_{il}K_{jk}$, $R_{0i0j}=-(\dot K_{ij}-K_{ik}\mathsf K^k{}_j)$ and $R_{0ijk}=0$.
In an orthonormal frame transported along the comoving geodesics we write $\mathsf K=H\mathds{1}+\sigma$, with $u=\dot\sigma$ and $P=\sigma^2-\tfrac13S\,\mathds{1}$ as in Sec.~\ref{sec:trace}.
The magnetic part of the Weyl tensor vanishes, and
\begin{equation}
\begin{aligned}
W^2&=2\Tr(u+H\sigma+2P)^2,\\
E_4&=24H^4+24H^2\dot H-(12H^2+4\dot H)S\\
&\quad-8H\Tr(u\sigma)+8\Tr(uP)+8H\Tr\sigma^3,
\end{aligned}
\label{eq:pa-invariants}
\end{equation}
with $E_4=8\mathcal V^{-1}d(\mathcal V\det\mathsf K)/dt$.
Neither a diagonal $\mathsf K$ nor $[\sigma,u]=0$ is assumed.
The characteristic equation of a trace-free $3\times3$ matrix gives $\Tr\sigma^4=S^2/2$ and $\Tr P^2=S^2/6$, and at $H=0$ Eq.~\eqref{eq:pa-invariants} reduces to the expressions used in Eq.~\eqref{eq:pa-square}.
We checked these identities symbolically for unrestricted trace-free symmetric matrices, and numerically from the coordinate Riemann tensor of 500 random metric jets with noncommuting shear; the largest scaled error is $4.9\times10^{-14}$. %

\subsection{The zero-expansion boundary}
\label{app:hidentity}

For general $H$ the combination of Eq.~\eqref{eq:pa-square} becomes
\begin{equation}
\begin{aligned}
cW^2-aE_4&=Q_0+4a\dot H(S-6H^2)+H\mathcal P,\\
\mathcal P&=4(c+2a)\Tr(u\sigma)+8(c-a)\Tr\sigma^3\\
&\quad+(2c+12a)HS-24aH^3,
\end{aligned}
\label{eq:pa-genH}
\end{equation}
and the traced Einstein equation with $b=0$ reads
\begin{equation}
\begin{aligned}
\Big[6\MPbar^2+\frac{a(S-6H^2)}{4\pi^2}\Big]\dot H={}&-\MPbar^2(12H^2+S)\\
&-\frac{Q_0+H\mathcal P}{16\pi^2}-T_{\rm extra}.
\end{aligned}
\label{eq:pa-genH-trace}
\end{equation}
Near a regular zero of $H$ the bracket exceeds $3\MPbar^2$.
With $\varrho\leq2$ and $T_{\rm extra}\geq0$ every term on the right except $-H\mathcal P/16\pi^2$ is non-positive, so $\dot H\leq k|H|$ for a constant $k$ on a compact neighborhood.
If $H$ became positive after a zero at $t_*$, then $d(e^{-kt}H)/dt\leq0$ wherever $H>0$, which contradicts $H(t_*)=0$.
A regular contracting solution thus cannot cross $H=0$, even at a degenerate zero with $\dot H(t_*)=0$, which completes the obstruction of Sec.~\ref{sec:trace}.

%% file: appendices/v13_aB_casimir.tex
\section{Casimir sums, spin structures and zero-mode energy}
\label{app:casimirzm}

\subsection{Twisted image sums and spin structures}
\label{app:smcasimir}

We evaluate the twisted Epstein sums $Z_2(G,\bm\vartheta)=\sum'_{\bm n}\cos(2\pi\bm\vartheta\cdot\bm n)(\bm n^{\mathsf T}G\bm n)^{-2}$ by an Ewald split into rapidly convergent sums over the lattice and its dual, and check them by direct summation (Ref.~\cite{v12Report}, Sec.~3).
At the cube this gives $Z_2(\mathds{1},0)=16.5323159598$ and $Z_2(\mathds{1},(\tfrac12,\tfrac12,\tfrac12))=-3.8631638072$. %
For the Standard Model plus the static graviton, $Z_{\rm eff}=15Z_2(G,0)-45Z_2(G,\bm\vartheta_F)$, so that $E_C=-Z_{\rm eff}/\pi^2L$ in Eq.~\eqref{eq:pa-imagesum}; Table~\ref{tab:pa-spin} lists the results at the cube. %
For AAA the normalized stiffnesses $\kappa_A=3\lambda_A/Z_{\rm eff}$ are $\kappa_E=13.5286$ and $\kappa_{T_2}=-6.3524$ for free fields, with $\sum_A\kappa_A=8.0000$. %
The Standard Model has anomaly coefficients $c=283/120$ and $a=995.5/360$, so $a/c=1.17$, inside the collider bound $31/18$ of Sec.~\ref{sec:trace}. %

\begin{table}[!htb]
\caption{Standard Model plus the static graviton on the cubic three-torus at trivial holonomy, for fermion spin structures with periodic (P) and antiperiodic (A) cycles: signed Casimir weight, stationarity of the cube and the Hessian eigenvalues of $Z_{\rm eff}$ in the aspect ($E$) and angle ($T_2$) directions, with their multiplicities.}
\label{tab:pa-spin}
\begin{ruledtabular}
\begin{tabular}{lcccc}
spin structure & $\NC$ & stationary & $\lambda_E$ ($\times2$) & $\lambda_{T_2}$ ($\times3$)\\
\hline
PPP & $-30.00$ & yes & $-991.56$ & $+220.18$\\ %
PPA & $+13.12$ & no & & \\ %
PAA & $+20.87$ & no & & \\ %
AAA & $+25.515$ & yes & $+1902.24$ & $-893.20$ %
\end{tabular}
\end{ruledtabular}
\end{table}

Flat holonomies enter Eq.~\eqref{eq:pa-imagesum} as additional twists, $\vartheta_i=Yg_i$ for hypercharge $Y$ and $\vartheta_i=\pm k_i/3$ on (anti)triplets for the $SU(3)$ center.
A twist can only raise the energy of a boson, since $Z_2(G,\bm\vartheta)\leq Z_2(G,\bm0)$ term by term, and it can lower the energy of an antiperiodic fermion only if $Z_2(G,\bm\vartheta)$ is not smallest at $\bm\vartheta=(\tfrac12,\tfrac12,\tfrac12)$.
For a rectangular torus it is smallest there, because $Z_2$ is the integral over $t$ of $t$ times the heat trace $\sum_{\bm n}e^{-t\,\bm n^{\mathsf T}G\bm n}\cos(2\pi\bm n\cdot\bm\vartheta)$ without its $\bm n=0$ term, and for diagonal $G$ this trace is a product of three theta functions $\vartheta_3(\pi\vartheta_i,e^{-tL_i^2})$, each positive and smallest at $\vartheta_i=\tfrac12$. %

For one or two antiperiodic edges (PPA, PAA) the one-loop vacuum on each periodic edge is an $SU(3)$ center holonomy, which twists the quarks by $\pm1/3$ along that edge.
The energy is then stationary at a tetragonal shape, with $\NC=19.72$ for PPA and $22.84$ for PAA, and both aspect directions are unstable, as they are for AAA at the cube. %
With periodic fermions the hypercharge holonomy $\bm g=g(1,1,1)$ raises $\NC$ to at most $17.79$. %
Reference~\cite{v12Report}, Sec.~3, gives the shapes, the center-holonomy weights and the flux sectors.

\subsection{Zero modes}
\label{app:zeromodes}

\emph{Isospectrality.}
In temporal gauge, with ${\rm tr}\,T_aT_b=\tfrac12\delta_{ab}$ and the constant modes of Sec.~\ref{sec:casimir}, the Legendre transform of the reduced action gives Eq.~\eqref{eq:pa-H0}; the Gauss law is imposed on states, and the AAA fermions have no constant modes.
For $G=SS^{\mathsf T}$ with $\det S=1$, the canonical map $x_i=S_{ij}\tilde x_j$, $w=S\tilde w$, $\pi=S^{-\mathsf T}\tilde\pi$, $\pi^w=S^{-\mathsf T}\tilde\pi^w$, with $\Phi$ unchanged, gives $G_{ij}\pi_i\cdot\pi_j=\tilde\pi^2$ and $G^{ik}G^{jl}F_{ij}\cdot F_{kl}=F[\tilde x]^2$.
Since $M_i(x,w)=S_{ij}M_j(\tilde x,\tilde w)$ and $S^{\mathsf T}G^{-1}S=\mathds1$, also $G^{ij}\Phi^\dagger M_iM_j\Phi=\sum_j|M_j(\tilde x,\tilde w)\Phi|^2$.
The Gauss generators are unchanged and the Jacobian is $\det(\mathds1\otimes S)\det S=1$, so the map carries $\mathcal H_0(G)$ unitarily into $\mathcal H_0(\mathds1)$ on the local domain, and the whole local spectrum is shape independent.

\emph{Ground energy.}
For $\lambda\geq0$ the hypercharge and Higgs terms are non-negative, so $Q\geq\epsilon_2g_2^{2/3}+\epsilon_3g_3^{2/3}$, where $\epsilon_N$ is the ground energy of the pure-gauge Hamiltonian for $SU(N)$ in the variables $y=g^{1/3}x$.
Because $\epsilon_3$ is a variational value~\cite{Weisz:1986vu,Pavel:2021mxn}, this bound is numerical rather than rigorous.
A Gaussian trial state gives the upper bound.
For $\lambda\geq0$ the ground state has no nodes, so diffusion Monte Carlo guided by the optimal Gaussian gives the exact local ground energy up to statistical and time-step errors; it reproduces $\epsilon_2$ and $\epsilon_3$ within these errors.
The coupled electroweak sector gives $Q=12.790$ at $\lambda=0$ and $12.818$ at $\lambda=0.01$. %
Hence $\fzm=0.2993$ and $0.2999$, between the pure-gauge bound $\fzm\geq0.2580$ and the Gaussian bound $\fzm\leq0.3083$. %
At the Standard Model value $\lambda(\mub)=-0.0098$ the real part of the resonance energy gives $\fzm=0.298$. %

\emph{Next order.}
At $\mathcal O(g^{4/3})$ the zero modes feel the curvature of the one-loop potential that the non-zero modes of charged fields create for constant connections~\cite{Luscher:1982ma,vanBaal:1988va}.
With Gaussian spreads of the zero modes the first-order shift is $\Delta Q=-0.60$, or $\Delta\fzm=-0.014$; the one-loop mass of the Higgs zero mode lowers $\fzm$ by a further $0.003$, so that $\fzm\simeq0.28$. %
These terms depend on the shape through twisted Epstein sums at $s=1$, whose Hessians at the cube give $\kappa_E=18.0$ and $\kappa_{T_2}=-8.34$ at $\fzm\simeq0.28$. %
The expansion parameter $g^{2/3}\simeq0.66$ is not small, but at $g=0.53$ the corrections of the published $SU(2)$ series are $-11\%$, $-0.4\%$ and $-0.15\%$ of the leading term. %
Reference~\cite{v12Report}, Sec.~4, gives the reduced action, the curvature formula, the checks of isospectrality and the Monte Carlo benchmarks.

%% file: appendices/v13_aC_shape.tex
\section{Shape Laplacian, transfer and the time-symmetry theorem}
\label{app:shapedyn}

\subsection{The shape Laplacian}
\label{app:shapelaplacian}

Along $G_A(\beta)=G^{1/2}e^{2\beta X_A}G^{1/2}$ a single image term has
\begin{equation}
\frac{d^2}{d\beta^2}(\bm n^{\mathsf T}G_A\bm n)^{-s}\Big|_0
=4s(s+1)\frac{(\bm y^{\mathsf T}X_A\bm y)^2}{|\bm y|^{2s+4}}-4s\frac{\bm y^{\mathsf T}X_A^2\bm y}{|\bm y|^{2s+2}},
\end{equation}
with $\bm y=G^{1/2}\bm n$.
Completeness of the basis gives $\sum_A(\bm y^{\mathsf T}X_A\bm y)^2=\tfrac23|\bm y|^4$ and $\sum_AX_A^2=\tfrac53\mathds{1}$, hence Eq.~\eqref{eq:s-Hesstrace} for $s=2$. With the normalized stiffnesses $\kappa_A=3\lambda_A/Z_{\rm eff}$ of the turning-point solution, the trace identity gives $\sum_A\kappa_A=8/(1-\fzm)$.
For $q=\bm k^{\mathsf T}G^{-1}\bm k$ with fixed $\bm k$ the same completeness relations give
\begin{equation}
\Delta_{\rm sh}\,q^p=\frac{4p(2p+3)}{3}\,q^p .
\end{equation}
For $p=\tfrac12$ the eigenvalue is $\tfrac83$.
Every mode frequency $2\pi|G^{-1/2}(\bm k+\bm\vartheta)|/L$, with any fixed twist, and hence every energy of fixed occupations, therefore has the eigenvalue of the Casimir energy.

\emph{Excited and thermal states.}
Let the light fields have levels $E_s(G)=e_s(G)+q_s/L$ at fixed volume and couplings, where the free parts obey $\Delta_{\rm sh}e_s=\tfrac83e_s$ and the zero-mode energies $q_s\geq0$ do not depend on the shape (Sec.~\ref{sec:pa-zeromodes}).
The free energy has $\partial_AF=\langle f_A\rangle$, with $f_A=\partial_AE_s$ the shape forces on the levels, and covariant Hessian $\langle\nabla_A\nabla_BE\rangle-\sCov(f_A,f_B)/T$.
For a Gibbs state at temperature $T$ its trace is
\begin{equation}
\Delta_{\rm sh}F=\frac83\Big(U-\frac{\langle q\rangle}{L}\Big)-\frac1T\sum_A\sVar\,f_A ,
\label{eq:s-nogo}
\end{equation}
where $U$ is the internal energy.
The shape Laplacian is thus at most $\tfrac83(U-\langle q\rangle/L)$, and matter with negative internal energy has no static shape minimum.
A local minimum $G_0$ of $E_{\rm eff}(G)=\min_\theta E(G,\theta)$ over static holonomies with $E_{\rm eff}(G_0)<0$, attained at $\theta_0$, would be a local minimum of $E(\cdot,\theta_0)$, which the identity forbids.

\emph{First-order zero-mode energies.}
The matching energy $E_{\rm m}$ of the zero modes to the non-zero modes and the induced Higgs-mass energy $E_H$ are $s=1$ Epstein sums with eigenvalue $-\tfrac43$.
They preserve the absence of a static shape minimum as long as $Q\geq-\tfrac32L(E_{\rm m}+E_H)$.
With the estimates of Appendix~\ref{app:zeromodes} the right-hand side is $1.1$, about one twelfth of the zero-mode energy $Q$. %

\subsection{Transfer, time-symmetry theorem and the wave function}
\label{app:theorem}

Here $L$ is in units of $\Lb$ and $t$ in units of $1/\sHeff$, and we write $h$ for $\sHeff$.
A shape mode then obeys $\dot\beta=c\,\Pi/L^3$ and $\dot\Pi=(\kappa/c)\,L^3{\rm sech}^2(2t)\,\beta$, with $c=3/2^{3/2}$ and the normalization of Eq.~\eqref{eq:s-delta}.
With $z=(1+\tanh2t)/2$ the mode equation~\eqref{eq:s-shapeode} becomes
\begin{equation}
z(1-z)\beta_{zz}+\big(\tfrac14-\tfrac z2\big)\beta_z-\tfrac\kappa4\beta=0,
\end{equation}
Gauss's equation with parameters $a,b=-\tfrac14\pm i\mu$, $\mu=\tfrac14\sqrt{4\kappa-1}$.
The solution with $\beta\to1$ and $\Pi\to0$ in the far past is ${}_2F_1(a,b;\tfrac14;z)$, and Gauss's summation formula at $z\to1$ gives $\sTm_{11}=\Gamma(\tfrac14)\Gamma(\tfrac34)/|\Gamma(\tfrac12+i\mu)|^2=\sqrt2\cosh\pi\mu$, which is Eq.~\eqref{eq:s-gain}.
The solution with $\beta\to0$ and $\Pi\to1$ is $z^{3/4}{}_2F_1(\tfrac12+i\mu,\tfrac12-i\mu;\tfrac74;z)$, which gives Eq.~\eqref{eq:s-T12}.
With $\fzm=0.30$ ($\kappa_E=19.3237$) the transfer from the turn to the far future is $m_{11}=10.554$, $m_{12}=2.4685$, $m_{21}=136.37$, $m_{22}=31.99$. %
The mode equation is invariant under $t\to-t$, $\Pi\to-\Pi$, so the transfer from $-\infty$ to $0$ is $\mathsf RM^{-1}\mathsf R$, which gives Eq.~\eqref{eq:s-Tsym}, and its inverse $\mathsf RM\mathsf R$ gives the incoming amplitude in Eq.~\eqref{eq:v13-betapm}.

\emph{All-time bound.}
Let $u,v$ solve the anti-restoring equation $(m\dot\beta)^{\boldsymbol\cdot}=k\beta$, $m>0$, $k\geq0$, with $u(t_-)=1$, $u(t_+)=0$, $v(t_-)=0$ and $v(t_+)=1$.
The maximum principle gives $u,v\geq0$ and $u+v\leq1$.
The pure Gaussian with equal endpoint variances $\delta|\sTm_{12}|$ and zero symmetrized endpoint covariance therefore has $\sVar\,\beta(t)=\delta|\sTm_{12}|(u^2+v^2)\leq\delta|\sTm_{12}|$ at all times.

\emph{The far past.}
In the far past of the plateau, slow roll limits $\omega^2/H^2$ to $\mathcal O(\epsilon_V)\ll9/4$, so the modes have no oscillating vacuum there.
Sectors that would evade this, such as vector fields with an inflaton-dependent kinetic function~\cite{Watanabe:2009ct}, non-abelian gauge fields with a vacuum expectation value~\cite{Maleknejad:2011jw,Adshead:2012kp} or solid inflation~\cite{Endlich:2012pz}, are absent from the Standard Model with a plateau inflaton.

\emph{Proof of Theorem~\ref{thm:s-real}.}
For linear dynamics the Ehrenfest relations close on the second moments, and
\begin{equation}
\frac{dC}{dt}=\frac{c}{L^3}\,\sVar\,\Pi+\frac{\kappa_EL^3}{c}\,{\rm sech}^2(2t)\,\sVar\,\beta>0
\label{eq:s-dCdt}
\end{equation}
in any state, pure or mixed, Gaussian or not.
Both terms decay at least as $e^{-|t|}$, so $C(\pm\infty)$ are finite, and a strictly increasing $C$ vanishes at most once.
For the bound, let $M=M(t_{\rm ref}\to\infty)$; since $\kappa_E>0$, $M_{11}$ and $M_{12}$ are positive.
If $C(t_{\rm ref})\geq0$, then
\begin{equation}
\begin{split}
\sVar\,\beta_+&=M_{11}^2\sVar\,\beta+2M_{11}M_{12}C+M_{12}^2\sVar\,\Pi\\
&\geq2M_{11}M_{12}\,\delta,
\end{split}
\end{equation}
by the Robertson--Schr\"odinger inequality $\sVar\,\beta\,\sVar\,\Pi-C^2\geq\delta^2$.
With $t_{\rm ref}=t_x\leq0$ and $\delta|\sTm_{12}|=2m_{11}m_{12}\delta$ this is Eq.~\eqref{eq:s-Gtx}; for $t_x>0$ the reflection exchanges the branches.
As $t_{\rm ref}\to-\infty$, $M_{11}M_{12}\to\sTm_{11}\sTm_{12}$, which gives the bound $2\sGain\,\delta|\sTm_{12}|$ for an outgoing state whose correlation never vanishes, and the same bound on $\sVar\,\beta_-$ for an incoming one.
For the width condition write, with $M=M(0\to\infty)$,
\begin{equation}
\begin{split}
C(+\infty)={}&m_{11}m_{21}\sVar\,\beta+(m_{11}m_{22}+m_{12}m_{21})C\\
&+m_{12}m_{22}\sVar\,\Pi .
\end{split}
\end{equation}
Minimizing over $C$ and $\sVar\,\Pi\geq(\delta^2+C^2)/\sVar\,\beta$ with $\det M=1$ gives $\min C(+\infty)=m_{12}m_{22}\delta^2/\sVar\,\beta-\sVar\,\beta/(4m_{12}m_{22})$, and likewise for $C(-\infty)$.
States whose correlation never vanishes thus need $\sVar\,\beta_E(0)\geq2m_{12}m_{22}\delta=157.9\,\delta$ at $\fzm=0.30$. %
The bound uses only second moments and holds for mixed and non-Gaussian states.

\emph{Time symmetry and the Wheeler--DeWitt wave function.}
In Eq.~\eqref{eq:s-WDW} let $\tilde\chi(t)\equiv\chi(-t)^*$.
Since $H_s$ is real and $H_s(-t)=H_s(t)$, $i\partial_t\tilde\chi=H_s(t)\tilde\chi$ with $\tilde\chi(0)=\chi(0)^*$.
A real $\chi(0)$ therefore gives $\chi(-t)=\chi(t)^*$ at all times, and $\Psi=2A\,{\rm Re}[e^{iS}\chi(t)]$ is real.
Conversely, if $\Psi$ is real, the coefficients of the linearly independent factors $e^{\pm iS}$ are complex conjugates, $\chi(\beta,-t)=\chi(\beta,t)^*$, and continuity through the turn along the single history gives $\chi(0)=\chi(0)^*$.

\emph{Minimal energy variance.}
For $\psi\propto\exp[-(a-ib)x^2]$, $\langle x^2\rangle=1/4a$, and the inverted oscillator $H=p^2/2m-kx^2/2$ has
\begin{equation}
\begin{split}
\sVar\,H=\frac{1}{32a^2m^2}\big(&16a^4+32a^2b^2+8a^2km\\
&+16b^4-8b^2km+k^2m^2\big).
\end{split}
\end{equation}
At $b=0$ the minimum lies at $a=\sqrt{km}/2$, that is $\langle x^2\rangle=1/(2m|\omega|)$ with $|\omega|=(k/m)^{1/2}$, where $\sVar\,H=k/2m$.
For $|b|<\sqrt{km}/2$ the same value is reached at $a=(km-4b^2)^{1/2}/2$.
This one-parameter set is the orbit of the natural state under the inverted-oscillator evolution, and $b=0$ is its only time-symmetric member.

\subsection{Survival bound for all time-symmetric states}
\label{app:v13-lp}

Let $\mathsf K$ be complex conjugation in the $\beta$ representation at $t_x$, so that $\mathsf K\beta_A\mathsf K=\beta_A$ and $\mathsf K\Pi_A\mathsf K=-\Pi_A$. We call a state time-symmetric about $t_x$ if it is invariant under $\mathsf K$: a pure state with a real wave function at $t_x$, or a mixture with a real density matrix.
With $X_A=M_{11}\beta_A+M_{12}\Pi_A$ and $Y_A=\mathsf KX_A\mathsf K=M_{11}\beta_A-M_{12}\Pi_A$, Eq.~\eqref{eq:s-delta} gives $[Y_A,X_B]=4iM_{11}M_{12}\delta\,\delta_{AB}=2i\,G\,\sfloor^2\,\delta_{AB}$, which is Eq.~\eqref{eq:v13-XYcomm}.
Let $P$ and $Q$ be the projectors on $\bm X\in D$ and on $\bm Y\in D$; then $Q=\mathsf KP\mathsf K$.
For a unit vector $\psi$, the angles between $\psi$ and the ranges of two projectors obey~\cite{LandauPollak1961}
\begin{equation}
\arccos\|P\psi\|+\arccos\|Q\psi\|\geq\arccos\|PQ\| ,
\end{equation}
and $\|PQ\|^2=\lambda_0$ is the largest eigenvalue of $PQP$.
If $\mathsf K\psi=\psi$, then $\|Q\psi\|=\|P\psi\|=p^{1/2}$ with $p=P(\bm X\in D)$, so that $2\arccos p^{1/2}\geq\arccos\lambda_0^{1/2}$, which is Eq.~\eqref{eq:v13-lpbound}.
A real density matrix has real eigenvectors, and the bound, which is the same for each of them, holds for their mixture.
In the $\bm X$ representation $\bm Y=i\hbar_x\partial_{\bm X}+\bm f(\bm X)$ with $\hbar_x=2G\sfloor^2$, and $\bm f$ is removed by a phase factor that depends on $\bm X$ alone and commutes with $P$, so $PQP$ has the eigenvalues of the integral operator on $L^2(D)$ with kernel
\begin{equation}
K(\bm x,\bm x')=\frac{1}{(2\pi\hbar_x)^2}\int_D d^2y\;e^{i\bm y\cdot(\bm x-\bm x')/\hbar_x} .
\end{equation}
For a disk it is the concentration operator of the two-dimensional prolate functions~\cite{Slepian1964}, which we used to check the numerical eigenvalues to $10^{-6}$; for the region $D$ of Sec.~\ref{sec:s-theorem} they are converged to $10^{-5}$ in the discretization. The uncertainty of the boundary of $D$ changes the bound by at most $0.007$. %
For $t_x=0$, where $\bm X=\bm\beta_+$, $\bm Y=\bm\beta_-$ and $\hbar_x=2\sfloor^2$, the trace $\Tr PQ=\int_Dd^2x\,K(\bm x,\bm x)=({\rm Area}\,D)^2/(2\pi\hbar_x)^2$ is the number $N_D$ of Eq.~\eqref{eq:v13-count}. Up to boundary terms it is the number of eigenvalues of $PQP$ close to one, which estimates the number of orthogonal states of the two aspect modes that survive on both branches.

The identification of survival with $\bm X\in D$ is exact if the fate depends on the outgoing amplitude alone.
The fate depends on the second amplitude of each mode and on the angle modes only through a displacement of the boundary of $D$. For amplitudes on the other branch up to three floors and angle amplitudes up to three natural widths, the boundary moves by at most $3.5\%$ away from the cusps, and at the cusps by up to $5\%$ outward and $11\%$ inward. %
All surviving histories found with amplitudes on the other branch up to three floors lie inside the region enlarged by $5\%$, which is the region used for the thresholds quoted in the text. %
Along the rod axes, histories with large and nearly equal amplitudes on both branches survive outside it, up to $|\bm b|\approx6$ (Sec.~\ref{sec:v13-meanshape}). %

%% file: appendices/v13_aD_higgs_model.tex
\section{The zero-mode calculation in brief}
\label{app:h-matching}

We use units $L=1$ and $\MSbar$ couplings at $\mu=1/L$.
Reference~\cite{v12Report} gives the matching (Sec.~6), the reduced model (Sec.~7), the cross terms (Sec.~8), the survival computation (Sec.~9) and the checks (Secs.~18--20).

\subsection{Reduced Hamiltonian}
\label{app:h-reduced}
\label{sec:h-reduced}

We solve the radial and hypercharge motion jointly, with the SU(2) zero modes in a conditional Gaussian corrected by DMC. End Matter~A of the Letter~\cite{v12PRL} gives the reasons for this reduction and the model Hamiltonian, Eq.~(A1) of the Letter.
We write its matching energy $\Delta V$ as $\Delta V_{\rm MC}$.

\subsection{Matching and its scale}
\label{sec:hx-rg}

The non-zero modes see the zero mode as a constant background that gives squared masses proportional to $r^2$ to the $W$, $Z$, Goldstone, radial and top-quark modes.
For each degree of freedom, $\Delta V_{\rm MC}(r)$ contains the difference between the finite-volume sum and its continuum integral, plus the $\MSbar$ Coleman--Weinberg term~\cite{Coleman:1973jx}, which is analytic in the squared mass even when the mass is tachyonic.
With the Coulomb-gauge Higgs zero mode as the slow coordinate, the explicit $\mu$ dependence of $\Delta V_{\rm MC}$ cancels the one-loop running of $\lambda$, $g_2$, $g_Y$ and $y_t$ in the tree potential if and only if the Higgs field is renormalized with the anomalous dimension
\begin{equation}
\gamma_\phi^{C}=\frac{3y_t^2-\frac32g_2^2-\frac12g_Y^2}{16\pi^2},
\label{eq:h-gammaC}
\end{equation}
which equals the one-loop Feynman-gauge value; numerically the identity holds to $10^{-14}$. %
The choice $\mu=1/L$ is then a convention, and we take the residual scale dependence as the two-loop error. Varying $\mu$ over $[\frac12,2\pi]/L$ and the top-threshold scale over $[\frac12,2]/L$ moves $m_c$ at $\Pone=1/2$ by $+0.012$ and $-0.049\GeV$. %

\subsection{Cross terms}
\label{app:h-cross}
\label{sec:h-cross}

A constant connection along the charge direction of a field twists its boundary conditions, so the one-loop energy of the non-zero modes depends jointly on $r$ and on the holonomies $(\bm x,\bm w)$.
At $r>0$ these cross terms add a radial potential $E_X$ and a hypercharge curvature $\Omega_X^2$ to the reduced Hamiltonian. The curvature is $+0.137$ for $r\le1.5$ and is dominated by the 45 Weyl fermions, whose antiperiodic boundary conditions pin the hypercharge holonomy at its trivial value. %
The logarithms of all one-loop terms of this order, including the vev mixing, the Coulomb mismatch and the top-quark chiral loop, reproduce the running of the tree gauge--Higgs coupling with the anomalous dimension~(\ref{eq:h-gammaC}) to $97$--$99\%$ at $r=2$--$5$, so $\mu=1/L$ is a consistent matching scale for them as well. %
At $\Pone=1/2$ the cross terms shift the critical mass by $-0.08\GeV$ from the electroweak zero modes and $-0.04\GeV$ from the gluon zero modes through the top quark. %
In total the cross terms lower the critical mass by $0.118\pm0.010\GeV$ at $\Pone=1/2$, and by $0.080$ and $0.160\GeV$ at $\Pone=0.9$ and $0.1$; two-loop cross terms are not computed. %

\subsection{Checks}
\label{app:h-budget}

Full-dimensional DMC of the sixteen-coordinate zero-mode Hamiltonian gives $\Gamma_{\rm full}/\Gamma_{\rm 2D}=0.976$, which shifts $m_c$ by $+0.008\GeV$ at $\Pone=1/2$. A WKB estimate that keeps a spurious term under the barrier gives $0.937$, which Table~\ref{tab:budget} carries as a one-sided $+0.020\GeV$. A tail-amplitude estimate gives $0.92(16)$ and $1.03(7)$ and agrees with the DMC ratio within its statistical errors. %
Real-time evolution reproduces $P=w_0\Pone$; for $\xiH=1/6$ the real-time survival of the resonant component equals $\Pone$ to $0.22\%$. For $\xiH<1/6$ any switch to classical evolution at $\tau\ge5.5$ moves $\xic$ by at most $0.0005$. %
The full-dimensional DMC, the comparison with complex absorbing potentials and the blind reimplementations are described in End Matter~A of the Letter~\cite{v12PRL}.
These checks test the implementation, not the approximations it shares with the main calculation, which Table~\ref{tab:budget} covers.

%% file: main.bbl
\begin{thebibliography}{159}%
\makeatletter
\providecommand \@ifxundefined [1]{%
 \@ifx{#1\undefined}
}%
\providecommand \@ifnum [1]{%
 \ifnum #1\expandafter \@firstoftwo
 \else \expandafter \@secondoftwo
 \fi
}%
\providecommand \@ifx [1]{%
 \ifx #1\expandafter \@firstoftwo
 \else \expandafter \@secondoftwo
 \fi
}%
\providecommand \natexlab [1]{#1}%
\providecommand \enquote  [1]{``#1''}%
\providecommand \bibnamefont  [1]{#1}%
\providecommand \bibfnamefont [1]{#1}%
\providecommand \citenamefont [1]{#1}%
\providecommand \href@noop [0]{\@secondoftwo}%
\providecommand \href [0]{\begingroup \@sanitize@url \@href}%
\providecommand \@href[1]{\@@startlink{#1}\@@href}%
\providecommand \@@href[1]{\endgroup#1\@@endlink}%
\providecommand \@sanitize@url [0]{\catcode `\\12\catcode `\$12\catcode
  `\&12\catcode `\#12\catcode `\^12\catcode `\_12\catcode `\%12\relax}%
\providecommand \@@startlink[1]{}%
\providecommand \@@endlink[0]{}%
\providecommand \url  [0]{\begingroup\@sanitize@url \@url }%
\providecommand \@url [1]{\endgroup\@href {#1}{\urlprefix }}%
\providecommand \urlprefix  [0]{URL }%
\providecommand \Eprint [0]{\href }%
\providecommand \doibase [0]{https://doi.org/}%
\providecommand \selectlanguage [0]{\@gobble}%
\providecommand \bibinfo  [0]{\@secondoftwo}%
\providecommand \bibfield  [0]{\@secondoftwo}%
\providecommand \translation [1]{[#1]}%
\providecommand \BibitemOpen [0]{}%
\providecommand \bibitemStop [0]{}%
\providecommand \bibitemNoStop [0]{.\EOS\space}%
\providecommand \EOS [0]{\spacefactor3000\relax}%
\providecommand \BibitemShut  [1]{\csname bibitem#1\endcsname}%
\let\auto@bib@innerbib\@empty
\bibitem [{\citenamefont {Hawking}\ and\ \citenamefont
  {Penrose}(1970)}]{HawkingPenrose1970}%
  \BibitemOpen
  \bibfield  {author} {\bibinfo {author} {\bibfnamefont {S.~W.}\ \bibnamefont
  {Hawking}}\ and\ \bibinfo {author} {\bibfnamefont {R.}~\bibnamefont
  {Penrose}},\ }\bibfield  {title} {\bibinfo {title} {The singularities of
  gravitational collapse and cosmology},\ }\href
  {https://doi.org/10.1098/rspa.1970.0021} {\bibfield  {journal} {\bibinfo
  {journal} {Proc. R. Soc. Lond. A}\ }\textbf {\bibinfo {volume} {314}},\
  \bibinfo {pages} {529} (\bibinfo {year} {1970})}\BibitemShut {NoStop}%
\bibitem [{\citenamefont {Borde}\ \emph {et~al.}(2003)\citenamefont {Borde},
  \citenamefont {Guth},\ and\ \citenamefont
  {Vilenkin}}]{BordeGuthVilenkin2003}%
  \BibitemOpen
  \bibfield  {author} {\bibinfo {author} {\bibfnamefont {A.}~\bibnamefont
  {Borde}}, \bibinfo {author} {\bibfnamefont {A.~H.}\ \bibnamefont {Guth}},\
  and\ \bibinfo {author} {\bibfnamefont {A.}~\bibnamefont {Vilenkin}},\
  }\bibfield  {title} {\bibinfo {title} {Inflationary spacetimes are incomplete
  in past directions},\ }\href {https://doi.org/10.1103/PhysRevLett.90.151301}
  {\bibfield  {journal} {\bibinfo  {journal} {Phys. Rev. Lett.}\ }\textbf
  {\bibinfo {volume} {90}},\ \bibinfo {pages} {151301} (\bibinfo {year}
  {2003})}\BibitemShut {NoStop}%
\bibitem [{\citenamefont {Molina-Paris}\ and\ \citenamefont
  {Visser}(1999)}]{Molina-Paris:1998xmn}%
  \BibitemOpen
  \bibfield  {author} {\bibinfo {author} {\bibfnamefont {C.}~\bibnamefont
  {Molina-Paris}}\ and\ \bibinfo {author} {\bibfnamefont {M.}~\bibnamefont
  {Visser}},\ }\bibfield  {title} {\bibinfo {title} {{Minimal conditions for
  the creation of a Friedman-Robertson-Walker universe from a `bounce'}},\
  }\href {https://doi.org/10.1016/S0370-2693(99)00469-4} {\bibfield  {journal}
  {\bibinfo  {journal} {Phys. Lett. B}\ }\textbf {\bibinfo {volume} {455}},\
  \bibinfo {pages} {90} (\bibinfo {year} {1999})},\ \Eprint
  {https://arxiv.org/abs/gr-qc/9810023} {arXiv:gr-qc/9810023} \BibitemShut
  {NoStop}%
\bibitem [{\citenamefont {Novello}\ and\ \citenamefont
  {Perez~Bergliaffa}(2008)}]{Novello:2008ra}%
  \BibitemOpen
  \bibfield  {author} {\bibinfo {author} {\bibfnamefont {M.}~\bibnamefont
  {Novello}}\ and\ \bibinfo {author} {\bibfnamefont {S.~E.}\ \bibnamefont
  {Perez~Bergliaffa}},\ }\bibfield  {title} {\bibinfo {title} {{Bouncing
  Cosmologies}},\ }\href {https://doi.org/10.1016/j.physrep.2008.04.006}
  {\bibfield  {journal} {\bibinfo  {journal} {Phys. Rept.}\ }\textbf {\bibinfo
  {volume} {463}},\ \bibinfo {pages} {127} (\bibinfo {year} {2008})},\ \Eprint
  {https://arxiv.org/abs/0802.1634} {arXiv:0802.1634 [astro-ph]} \BibitemShut
  {NoStop}%
\bibitem [{\citenamefont {Zhang}\ and\ \citenamefont
  {Zhou}(2026)}]{Zhang:2026gva}%
  \BibitemOpen
  \bibfield  {author} {\bibinfo {author} {\bibfnamefont {A.}~\bibnamefont
  {Zhang}}\ and\ \bibinfo {author} {\bibfnamefont {Y.}~\bibnamefont {Zhou}},\
  }\href@noop {} {\bibinfo {title} {{How much averaged null energy must an
  anisotropic bounce violate?}}} (\bibinfo {year} {2026}),\ \Eprint
  {https://arxiv.org/abs/2608.29556} {arXiv:2608.29556 [gr-qc]} \BibitemShut
  {NoStop}%
\bibitem [{\citenamefont {Faulkner}\ \emph {et~al.}(2016)\citenamefont
  {Faulkner}, \citenamefont {Leigh}, \citenamefont {Parrikar},\ and\
  \citenamefont {Wang}}]{Faulkner:2016mzt}%
  \BibitemOpen
  \bibfield  {author} {\bibinfo {author} {\bibfnamefont {T.}~\bibnamefont
  {Faulkner}}, \bibinfo {author} {\bibfnamefont {R.~G.}\ \bibnamefont {Leigh}},
  \bibinfo {author} {\bibfnamefont {O.}~\bibnamefont {Parrikar}},\ and\
  \bibinfo {author} {\bibfnamefont {H.}~\bibnamefont {Wang}},\ }\bibfield
  {title} {\bibinfo {title} {{Modular Hamiltonians for Deformed Half-Spaces and
  the Averaged Null Energy Condition}},\ }\href
  {https://doi.org/10.1007/JHEP09(2016)038} {\bibfield  {journal} {\bibinfo
  {journal} {JHEP}\ }\textbf {\bibinfo {volume} {2016}}\bibfield  {number}
  {\bibinfo  {number} { (09)},\ \bibinfo {pages} {038}},\ }\Eprint
  {https://arxiv.org/abs/1605.08072} {arXiv:1605.08072 [hep-th]} \BibitemShut
  {NoStop}%
\bibitem [{\citenamefont {Hartman}\ \emph {et~al.}(2017)\citenamefont
  {Hartman}, \citenamefont {Kundu},\ and\ \citenamefont
  {Tajdini}}]{Hartman:2016lgu}%
  \BibitemOpen
  \bibfield  {author} {\bibinfo {author} {\bibfnamefont {T.}~\bibnamefont
  {Hartman}}, \bibinfo {author} {\bibfnamefont {S.}~\bibnamefont {Kundu}},\
  and\ \bibinfo {author} {\bibfnamefont {A.}~\bibnamefont {Tajdini}},\
  }\bibfield  {title} {\bibinfo {title} {{Averaged Null Energy Condition from
  Causality}},\ }\href {https://doi.org/10.1007/JHEP07(2017)066} {\bibfield
  {journal} {\bibinfo  {journal} {JHEP}\ }\textbf {\bibinfo {volume}
  {2017}}\bibfield  {number} {\bibinfo  {number} { (07)},\ \bibinfo {pages}
  {066}},\ }\Eprint {https://arxiv.org/abs/1610.05308} {arXiv:1610.05308
  [hep-th]} \BibitemShut {NoStop}%
\bibitem [{\citenamefont {Graham}\ and\ \citenamefont
  {Olum}(2007)}]{Graham:2007va}%
  \BibitemOpen
  \bibfield  {author} {\bibinfo {author} {\bibfnamefont {N.}~\bibnamefont
  {Graham}}\ and\ \bibinfo {author} {\bibfnamefont {K.~D.}\ \bibnamefont
  {Olum}},\ }\bibfield  {title} {\bibinfo {title} {{Achronal averaged null
  energy condition}},\ }\href {https://doi.org/10.1103/PhysRevD.76.064001}
  {\bibfield  {journal} {\bibinfo  {journal} {Phys. Rev. D}\ }\textbf {\bibinfo
  {volume} {76}},\ \bibinfo {pages} {064001} (\bibinfo {year} {2007})},\
  \Eprint {https://arxiv.org/abs/0705.3193} {arXiv:0705.3193 [gr-qc]}
  \BibitemShut {NoStop}%
\bibitem [{\citenamefont {Kontou}\ and\ \citenamefont
  {Sanders}(2020)}]{Kontou:2020bta}%
  \BibitemOpen
  \bibfield  {author} {\bibinfo {author} {\bibfnamefont {E.-A.}\ \bibnamefont
  {Kontou}}\ and\ \bibinfo {author} {\bibfnamefont {K.}~\bibnamefont
  {Sanders}},\ }\bibfield  {title} {\bibinfo {title} {{Energy conditions in
  general relativity and quantum field theory}},\ }\href
  {https://doi.org/10.1088/1361-6382/ab8fcf} {\bibfield  {journal} {\bibinfo
  {journal} {Class. Quant. Grav.}\ }\textbf {\bibinfo {volume} {37}},\ \bibinfo
  {pages} {193001} (\bibinfo {year} {2020})},\ \Eprint
  {https://arxiv.org/abs/2003.01815} {arXiv:2003.01815 [gr-qc]} \BibitemShut
  {NoStop}%
\bibitem [{\citenamefont {Kontou}\ and\ \citenamefont
  {Olum}(2015)}]{Kontou:2015yha}%
  \BibitemOpen
  \bibfield  {author} {\bibinfo {author} {\bibfnamefont {E.-A.}\ \bibnamefont
  {Kontou}}\ and\ \bibinfo {author} {\bibfnamefont {K.~D.}\ \bibnamefont
  {Olum}},\ }\bibfield  {title} {\bibinfo {title} {{Proof of the averaged null
  energy condition in a classical curved spacetime using a null-projected
  quantum inequality}},\ }\href {https://doi.org/10.1103/PhysRevD.92.124009}
  {\bibfield  {journal} {\bibinfo  {journal} {Phys. Rev. D}\ }\textbf {\bibinfo
  {volume} {92}},\ \bibinfo {pages} {124009} (\bibinfo {year} {2015})},\
  \Eprint {https://arxiv.org/abs/1507.00297} {arXiv:1507.00297 [gr-qc]}
  \BibitemShut {NoStop}%
\bibitem [{\citenamefont {Dubovsky}\ \emph {et~al.}(2006)\citenamefont
  {Dubovsky}, \citenamefont {Gregoire}, \citenamefont {Nicolis},\ and\
  \citenamefont {Rattazzi}}]{Dubovsky:2005xd}%
  \BibitemOpen
  \bibfield  {author} {\bibinfo {author} {\bibfnamefont {S.}~\bibnamefont
  {Dubovsky}}, \bibinfo {author} {\bibfnamefont {T.}~\bibnamefont {Gregoire}},
  \bibinfo {author} {\bibfnamefont {A.}~\bibnamefont {Nicolis}},\ and\ \bibinfo
  {author} {\bibfnamefont {R.}~\bibnamefont {Rattazzi}},\ }\bibfield  {title}
  {\bibinfo {title} {{Null energy condition and superluminal propagation}},\
  }\href {https://doi.org/10.1088/1126-6708/2006/03/025} {\bibfield  {journal}
  {\bibinfo  {journal} {JHEP}\ }\textbf {\bibinfo {volume} {2006}}\bibfield
  {number} {\bibinfo  {number} { (03)},\ \bibinfo {pages} {025}},\ }\Eprint
  {https://arxiv.org/abs/hep-th/0512260} {arXiv:hep-th/0512260} \BibitemShut
  {NoStop}%
\bibitem [{\citenamefont {Deffayet}\ \emph {et~al.}(2010)\citenamefont
  {Deffayet}, \citenamefont {Pujolas}, \citenamefont {Sawicki},\ and\
  \citenamefont {Vikman}}]{Deffayet:2010qz}%
  \BibitemOpen
  \bibfield  {author} {\bibinfo {author} {\bibfnamefont {C.}~\bibnamefont
  {Deffayet}}, \bibinfo {author} {\bibfnamefont {O.}~\bibnamefont {Pujolas}},
  \bibinfo {author} {\bibfnamefont {I.}~\bibnamefont {Sawicki}},\ and\ \bibinfo
  {author} {\bibfnamefont {A.}~\bibnamefont {Vikman}},\ }\bibfield  {title}
  {\bibinfo {title} {{Imperfect Dark Energy from Kinetic Gravity Braiding}},\
  }\href {https://doi.org/10.1088/1475-7516/2010/10/026} {\bibfield  {journal}
  {\bibinfo  {journal} {JCAP}\ }\textbf {\bibinfo {volume} {2010}}\bibfield
  {number} {\bibinfo  {number} { (10)},\ \bibinfo {pages} {026}},\ }\Eprint
  {https://arxiv.org/abs/1008.0048} {arXiv:1008.0048 [hep-th]} \BibitemShut
  {NoStop}%
\bibitem [{\citenamefont {Easson}\ \emph {et~al.}(2011)\citenamefont {Easson},
  \citenamefont {Sawicki},\ and\ \citenamefont {Vikman}}]{Easson:2011zy}%
  \BibitemOpen
  \bibfield  {author} {\bibinfo {author} {\bibfnamefont {D.~A.}\ \bibnamefont
  {Easson}}, \bibinfo {author} {\bibfnamefont {I.}~\bibnamefont {Sawicki}},\
  and\ \bibinfo {author} {\bibfnamefont {A.}~\bibnamefont {Vikman}},\
  }\bibfield  {title} {\bibinfo {title} {{G-Bounce}},\ }\href
  {https://doi.org/10.1088/1475-7516/2011/11/021} {\bibfield  {journal}
  {\bibinfo  {journal} {JCAP}\ }\textbf {\bibinfo {volume} {2011}}\bibfield
  {number} {\bibinfo  {number} { (11)},\ \bibinfo {pages} {021}},\ }\Eprint
  {https://arxiv.org/abs/1109.1047} {arXiv:1109.1047 [hep-th]} \BibitemShut
  {NoStop}%
\bibitem [{\citenamefont {Ijjas}\ and\ \citenamefont
  {Steinhardt}(2016)}]{Ijjas:2016tpn}%
  \BibitemOpen
  \bibfield  {author} {\bibinfo {author} {\bibfnamefont {A.}~\bibnamefont
  {Ijjas}}\ and\ \bibinfo {author} {\bibfnamefont {P.~J.}\ \bibnamefont
  {Steinhardt}},\ }\bibfield  {title} {\bibinfo {title} {{Classically stable
  nonsingular cosmological bounces}},\ }\href
  {https://doi.org/10.1103/PhysRevLett.117.121304} {\bibfield  {journal}
  {\bibinfo  {journal} {Phys. Rev. Lett.}\ }\textbf {\bibinfo {volume} {117}},\
  \bibinfo {pages} {121304} (\bibinfo {year} {2016})},\ \Eprint
  {https://arxiv.org/abs/1606.08880} {arXiv:1606.08880 [gr-qc]} \BibitemShut
  {NoStop}%
\bibitem [{\citenamefont {Kobayashi}(2016)}]{Kobayashi:2016xpl}%
  \BibitemOpen
  \bibfield  {author} {\bibinfo {author} {\bibfnamefont {T.}~\bibnamefont
  {Kobayashi}},\ }\bibfield  {title} {\bibinfo {title} {{Generic instabilities
  of nonsingular cosmologies in Horndeski theory: A no-go theorem}},\ }\href
  {https://doi.org/10.1103/PhysRevD.94.043511} {\bibfield  {journal} {\bibinfo
  {journal} {Phys. Rev. D}\ }\textbf {\bibinfo {volume} {94}},\ \bibinfo
  {pages} {043511} (\bibinfo {year} {2016})},\ \Eprint
  {https://arxiv.org/abs/1606.05831} {arXiv:1606.05831 [hep-th]} \BibitemShut
  {NoStop}%
\bibitem [{\citenamefont {Dobre}\ \emph {et~al.}(2018)\citenamefont {Dobre},
  \citenamefont {Frolov}, \citenamefont {G{\'a}lvez~Ghersi}, \citenamefont
  {Ramazanov},\ and\ \citenamefont {Vikman}}]{Dobre:2017pnt}%
  \BibitemOpen
  \bibfield  {author} {\bibinfo {author} {\bibfnamefont {D.~A.}\ \bibnamefont
  {Dobre}}, \bibinfo {author} {\bibfnamefont {A.~V.}\ \bibnamefont {Frolov}},
  \bibinfo {author} {\bibfnamefont {J.~T.}\ \bibnamefont {G{\'a}lvez~Ghersi}},
  \bibinfo {author} {\bibfnamefont {S.}~\bibnamefont {Ramazanov}},\ and\
  \bibinfo {author} {\bibfnamefont {A.}~\bibnamefont {Vikman}},\ }\bibfield
  {title} {\bibinfo {title} {{Unbraiding the Bounce: Superluminality around the
  Corner}},\ }\href {https://doi.org/10.1088/1475-7516/2018/03/020} {\bibfield
  {journal} {\bibinfo  {journal} {JCAP}\ }\textbf {\bibinfo {volume}
  {2018}}\bibfield  {number} {\bibinfo  {number} { (03)},\ \bibinfo {pages}
  {020}},\ }\Eprint {https://arxiv.org/abs/1712.10272} {arXiv:1712.10272
  [gr-qc]} \BibitemShut {NoStop}%
\bibitem [{\citenamefont {DeWitt}\ \emph {et~al.}(1979)\citenamefont {DeWitt},
  \citenamefont {Hart},\ and\ \citenamefont {Isham}}]{DeWitt:1979dd}%
  \BibitemOpen
  \bibfield  {author} {\bibinfo {author} {\bibfnamefont {B.~S.}\ \bibnamefont
  {DeWitt}}, \bibinfo {author} {\bibfnamefont {C.~F.}\ \bibnamefont {Hart}},\
  and\ \bibinfo {author} {\bibfnamefont {C.~J.}\ \bibnamefont {Isham}},\
  }\bibfield  {title} {\bibinfo {title} {{Topology and quantum field theory}},\
  }\href {https://doi.org/10.1016/0378-4371(79)90207-3} {\bibfield  {journal}
  {\bibinfo  {journal} {Physica A}\ }\textbf {\bibinfo {volume} {96}},\
  \bibinfo {pages} {197} (\bibinfo {year} {1979})}\BibitemShut {NoStop}%
\bibitem [{\citenamefont {Zeldovich}\ and\ \citenamefont
  {Starobinsky}(1984)}]{Zeldovich:1984vk}%
  \BibitemOpen
  \bibfield  {author} {\bibinfo {author} {\bibfnamefont {Y.~B.}\ \bibnamefont
  {Zeldovich}}\ and\ \bibinfo {author} {\bibfnamefont {A.~A.}\ \bibnamefont
  {Starobinsky}},\ }\bibfield  {title} {\bibinfo {title} {{Quantum creation of
  a universe in a nontrivial topology}},\ }\href@noop {} {\bibfield  {journal}
  {\bibinfo  {journal} {Sov. Astron. Lett.}\ }\textbf {\bibinfo {volume}
  {10}},\ \bibinfo {pages} {135} (\bibinfo {year} {1984})}\BibitemShut
  {NoStop}%
\bibitem [{\citenamefont {Guth}\ and\ \citenamefont
  {Vilenkin}(2025)}]{Guth:2025dal}%
  \BibitemOpen
  \bibfield  {author} {\bibinfo {author} {\bibfnamefont {A.~H.}\ \bibnamefont
  {Guth}}\ and\ \bibinfo {author} {\bibfnamefont {A.}~\bibnamefont
  {Vilenkin}},\ }\bibfield  {title} {\bibinfo {title} {{On quantum creation of
  a toroidal universe}}\ }(\bibinfo {year} {2025})\ \Eprint
  {https://arxiv.org/abs/2508.08747} {arXiv:2508.08747 [gr-qc]} \BibitemShut
  {NoStop}%
\bibitem [{\citenamefont {Coule}\ and\ \citenamefont
  {Martin}(2000)}]{Coule:1999wg}%
  \BibitemOpen
  \bibfield  {author} {\bibinfo {author} {\bibfnamefont {D.~H.}\ \bibnamefont
  {Coule}}\ and\ \bibinfo {author} {\bibfnamefont {J.}~\bibnamefont {Martin}},\
  }\bibfield  {title} {\bibinfo {title} {{Quantum cosmology and open
  universes}},\ }\href {https://doi.org/10.1103/PhysRevD.61.063501} {\bibfield
  {journal} {\bibinfo  {journal} {Phys. Rev. D}\ }\textbf {\bibinfo {volume}
  {61}},\ \bibinfo {pages} {063501} (\bibinfo {year} {2000})},\ \Eprint
  {https://arxiv.org/abs/gr-qc/9905056} {arXiv:gr-qc/9905056} \BibitemShut
  {NoStop}%
\bibitem [{\citenamefont {Linde}(2004)}]{Linde:2004nz}%
  \BibitemOpen
  \bibfield  {author} {\bibinfo {author} {\bibfnamefont {A.~D.}\ \bibnamefont
  {Linde}},\ }\bibfield  {title} {\bibinfo {title} {{Creation of a compact
  topologically nontrivial inflationary universe}},\ }\href
  {https://doi.org/10.1088/1475-7516/2004/10/004} {\bibfield  {journal}
  {\bibinfo  {journal} {JCAP}\ }\textbf {\bibinfo {volume} {2004}}\bibfield
  {number} {\bibinfo  {number} { (10)},\ \bibinfo {pages} {004}},\ }\Eprint
  {https://arxiv.org/abs/hep-th/0408164} {arXiv:hep-th/0408164} \BibitemShut
  {NoStop}%
\bibitem [{\citenamefont {Fornal}(2026)}]{Fornal:2026qfq}%
  \BibitemOpen
  \bibfield  {author} {\bibinfo {author} {\bibfnamefont {B.}~\bibnamefont
  {Fornal}},\ }\bibfield  {title} {\bibinfo {title} {{Universe originating from
  an empty Planck-size torus}},\ }\href {https://doi.org/10.1103/x381-qbm1}
  {\bibfield  {journal} {\bibinfo  {journal} {Phys. Rev. D}\ }\textbf {\bibinfo
  {volume} {114}},\ \bibinfo {pages} {L021303} (\bibinfo {year} {2026})},\
  \Eprint {https://arxiv.org/abs/2603.24669} {arXiv:2603.24669 [hep-th]}
  \BibitemShut {NoStop}%
\bibitem [{\citenamefont {Degrassi}\ \emph {et~al.}(2012)\citenamefont
  {Degrassi}, \citenamefont {Di~Vita}, \citenamefont {Elias-Miro},
  \citenamefont {Espinosa}, \citenamefont {Giudice}, \citenamefont {Isidori},\
  and\ \citenamefont {Strumia}}]{Degrassi:2012ry}%
  \BibitemOpen
  \bibfield  {author} {\bibinfo {author} {\bibfnamefont {G.}~\bibnamefont
  {Degrassi}}, \bibinfo {author} {\bibfnamefont {S.}~\bibnamefont {Di~Vita}},
  \bibinfo {author} {\bibfnamefont {J.}~\bibnamefont {Elias-Miro}}, \bibinfo
  {author} {\bibfnamefont {J.~R.}\ \bibnamefont {Espinosa}}, \bibinfo {author}
  {\bibfnamefont {G.~F.}\ \bibnamefont {Giudice}}, \bibinfo {author}
  {\bibfnamefont {G.}~\bibnamefont {Isidori}},\ and\ \bibinfo {author}
  {\bibfnamefont {A.}~\bibnamefont {Strumia}},\ }\bibfield  {title} {\bibinfo
  {title} {{Higgs mass and vacuum stability in the Standard Model at NNLO}},\
  }\href {https://doi.org/10.1007/JHEP08(2012)098} {\bibfield  {journal}
  {\bibinfo  {journal} {JHEP}\ }\textbf {\bibinfo {volume} {2012}}\bibfield
  {number} {\bibinfo  {number} { (08)},\ \bibinfo {pages} {098}},\ }\Eprint
  {https://arxiv.org/abs/1205.6497} {arXiv:1205.6497 [hep-ph]} \BibitemShut
  {NoStop}%
\bibitem [{\citenamefont {Buttazzo}\ \emph {et~al.}(2013)\citenamefont
  {Buttazzo}, \citenamefont {Degrassi}, \citenamefont {Giardino}, \citenamefont
  {Giudice}, \citenamefont {Sala}, \citenamefont {Salvio},\ and\ \citenamefont
  {Strumia}}]{Buttazzo:2013uya}%
  \BibitemOpen
  \bibfield  {author} {\bibinfo {author} {\bibfnamefont {D.}~\bibnamefont
  {Buttazzo}}, \bibinfo {author} {\bibfnamefont {G.}~\bibnamefont {Degrassi}},
  \bibinfo {author} {\bibfnamefont {P.~P.}\ \bibnamefont {Giardino}}, \bibinfo
  {author} {\bibfnamefont {G.~F.}\ \bibnamefont {Giudice}}, \bibinfo {author}
  {\bibfnamefont {F.}~\bibnamefont {Sala}}, \bibinfo {author} {\bibfnamefont
  {A.}~\bibnamefont {Salvio}},\ and\ \bibinfo {author} {\bibfnamefont
  {A.}~\bibnamefont {Strumia}},\ }\bibfield  {title} {\bibinfo {title}
  {{Investigating the near-criticality of the Higgs boson}},\ }\href
  {https://doi.org/10.1007/JHEP12(2013)089} {\bibfield  {journal} {\bibinfo
  {journal} {JHEP}\ }\textbf {\bibinfo {volume} {2013}}\bibfield  {number}
  {\bibinfo  {number} { (12)},\ \bibinfo {pages} {089}},\ }\Eprint
  {https://arxiv.org/abs/1307.3536} {arXiv:1307.3536 [hep-ph]} \BibitemShut
  {NoStop}%
\bibitem [{\citenamefont {Bednyakov}\ \emph {et~al.}(2015)\citenamefont
  {Bednyakov}, \citenamefont {Kniehl}, \citenamefont {Pikelner},\ and\
  \citenamefont {Veretin}}]{Bednyakov:2015sca}%
  \BibitemOpen
  \bibfield  {author} {\bibinfo {author} {\bibfnamefont {A.~V.}\ \bibnamefont
  {Bednyakov}}, \bibinfo {author} {\bibfnamefont {B.~A.}\ \bibnamefont
  {Kniehl}}, \bibinfo {author} {\bibfnamefont {A.~F.}\ \bibnamefont
  {Pikelner}},\ and\ \bibinfo {author} {\bibfnamefont {O.~L.}\ \bibnamefont
  {Veretin}},\ }\bibfield  {title} {\bibinfo {title} {Stability of the
  electroweak vacuum: Gauge independence and advanced precision},\ }\href
  {https://doi.org/10.1103/PhysRevLett.115.201802} {\bibfield  {journal}
  {\bibinfo  {journal} {Phys. Rev. Lett.}\ }\textbf {\bibinfo {volume} {115}},\
  \bibinfo {pages} {201802} (\bibinfo {year} {2015})},\ \Eprint
  {https://arxiv.org/abs/1507.08833} {arXiv:1507.08833} \BibitemShut {NoStop}%
\bibitem [{\citenamefont {Andreassen}\ \emph {et~al.}(2018)\citenamefont
  {Andreassen}, \citenamefont {Frost},\ and\ \citenamefont
  {Schwartz}}]{Andreassen:2017rzq}%
  \BibitemOpen
  \bibfield  {author} {\bibinfo {author} {\bibfnamefont {A.}~\bibnamefont
  {Andreassen}}, \bibinfo {author} {\bibfnamefont {W.}~\bibnamefont {Frost}},\
  and\ \bibinfo {author} {\bibfnamefont {M.~D.}\ \bibnamefont {Schwartz}},\
  }\bibfield  {title} {\bibinfo {title} {Scale invariant instantons and the
  complete lifetime of the standard model},\ }\href
  {https://doi.org/10.1103/PhysRevD.97.056006} {\bibfield  {journal} {\bibinfo
  {journal} {Phys. Rev. D}\ }\textbf {\bibinfo {volume} {97}},\ \bibinfo
  {pages} {056006} (\bibinfo {year} {2018})},\ \Eprint
  {https://arxiv.org/abs/1707.08124} {arXiv:1707.08124} \BibitemShut {NoStop}%
\bibitem [{\citenamefont {Espinosa}\ \emph {et~al.}(2008)\citenamefont
  {Espinosa}, \citenamefont {Giudice},\ and\ \citenamefont
  {Riotto}}]{Espinosa:2007qp}%
  \BibitemOpen
  \bibfield  {author} {\bibinfo {author} {\bibfnamefont {J.~R.}\ \bibnamefont
  {Espinosa}}, \bibinfo {author} {\bibfnamefont {G.~F.}\ \bibnamefont
  {Giudice}},\ and\ \bibinfo {author} {\bibfnamefont {A.}~\bibnamefont
  {Riotto}},\ }\bibfield  {title} {\bibinfo {title} {{Cosmological implications
  of the Higgs mass measurement}},\ }\href
  {https://doi.org/10.1088/1475-7516/2008/05/002} {\bibfield  {journal}
  {\bibinfo  {journal} {JCAP}\ }\textbf {\bibinfo {volume} {2008}}\bibfield
  {number} {\bibinfo  {number} { (05)},\ \bibinfo {pages} {002}},\ }\Eprint
  {https://arxiv.org/abs/0710.2484} {arXiv:0710.2484 [hep-ph]} \BibitemShut
  {NoStop}%
\bibitem [{\citenamefont {Herranen}\ \emph {et~al.}(2014)\citenamefont
  {Herranen}, \citenamefont {Markkanen}, \citenamefont {Nurmi},\ and\
  \citenamefont {Rajantie}}]{Herranen:2014cua}%
  \BibitemOpen
  \bibfield  {author} {\bibinfo {author} {\bibfnamefont {M.}~\bibnamefont
  {Herranen}}, \bibinfo {author} {\bibfnamefont {T.}~\bibnamefont {Markkanen}},
  \bibinfo {author} {\bibfnamefont {S.}~\bibnamefont {Nurmi}},\ and\ \bibinfo
  {author} {\bibfnamefont {A.}~\bibnamefont {Rajantie}},\ }\bibfield  {title}
  {\bibinfo {title} {Spacetime curvature and the {Higgs} stability during
  inflation},\ }\href {https://doi.org/10.1103/PhysRevLett.113.211102}
  {\bibfield  {journal} {\bibinfo  {journal} {Phys. Rev. Lett.}\ }\textbf
  {\bibinfo {volume} {113}},\ \bibinfo {pages} {211102} (\bibinfo {year}
  {2014})},\ \Eprint {https://arxiv.org/abs/1407.3141} {arXiv:1407.3141}
  \BibitemShut {NoStop}%
\bibitem [{\citenamefont {Kearney}\ \emph {et~al.}(2015)\citenamefont
  {Kearney}, \citenamefont {Yoo},\ and\ \citenamefont
  {Zurek}}]{Kearney:2015vba}%
  \BibitemOpen
  \bibfield  {author} {\bibinfo {author} {\bibfnamefont {J.}~\bibnamefont
  {Kearney}}, \bibinfo {author} {\bibfnamefont {H.}~\bibnamefont {Yoo}},\ and\
  \bibinfo {author} {\bibfnamefont {K.~M.}\ \bibnamefont {Zurek}},\ }\bibfield
  {title} {\bibinfo {title} {Is a {Higgs} vacuum instability fatal for
  high-scale inflation?},\ }\href {https://doi.org/10.1103/PhysRevD.91.123537}
  {\bibfield  {journal} {\bibinfo  {journal} {Phys. Rev. D}\ }\textbf {\bibinfo
  {volume} {91}},\ \bibinfo {pages} {123537} (\bibinfo {year} {2015})},\
  \Eprint {https://arxiv.org/abs/1503.05193} {arXiv:1503.05193} \BibitemShut
  {NoStop}%
\bibitem [{\citenamefont {Espinosa}\ \emph {et~al.}(2015)\citenamefont
  {Espinosa}, \citenamefont {Giudice}, \citenamefont {Morgante}, \citenamefont
  {Riotto}, \citenamefont {Senatore}, \citenamefont {Strumia},\ and\
  \citenamefont {Tetradis}}]{Espinosa:2015qea}%
  \BibitemOpen
  \bibfield  {author} {\bibinfo {author} {\bibfnamefont {J.~R.}\ \bibnamefont
  {Espinosa}}, \bibinfo {author} {\bibfnamefont {G.~F.}\ \bibnamefont
  {Giudice}}, \bibinfo {author} {\bibfnamefont {E.}~\bibnamefont {Morgante}},
  \bibinfo {author} {\bibfnamefont {A.}~\bibnamefont {Riotto}}, \bibinfo
  {author} {\bibfnamefont {L.}~\bibnamefont {Senatore}}, \bibinfo {author}
  {\bibfnamefont {A.}~\bibnamefont {Strumia}},\ and\ \bibinfo {author}
  {\bibfnamefont {N.}~\bibnamefont {Tetradis}},\ }\bibfield  {title} {\bibinfo
  {title} {The cosmological {Higgstory} of the vacuum instability},\ }\href
  {https://doi.org/10.1007/JHEP09(2015)174} {\bibfield  {journal} {\bibinfo
  {journal} {JHEP}\ }\textbf {\bibinfo {volume} {2015}}\bibfield  {number}
  {\bibinfo  {number} { (09)},\ \bibinfo {pages} {174}},\ }\Eprint
  {https://arxiv.org/abs/1505.04825} {arXiv:1505.04825} \BibitemShut {NoStop}%
\bibitem [{\citenamefont {Joti}\ \emph {et~al.}(2017)\citenamefont {Joti},
  \citenamefont {Katsis}, \citenamefont {Loupas}, \citenamefont {Salvio},
  \citenamefont {Strumia}, \citenamefont {Tetradis},\ and\ \citenamefont
  {Urbano}}]{Joti:2017fwe}%
  \BibitemOpen
  \bibfield  {author} {\bibinfo {author} {\bibfnamefont {A.}~\bibnamefont
  {Joti}}, \bibinfo {author} {\bibfnamefont {A.}~\bibnamefont {Katsis}},
  \bibinfo {author} {\bibfnamefont {D.}~\bibnamefont {Loupas}}, \bibinfo
  {author} {\bibfnamefont {A.}~\bibnamefont {Salvio}}, \bibinfo {author}
  {\bibfnamefont {A.}~\bibnamefont {Strumia}}, \bibinfo {author} {\bibfnamefont
  {N.}~\bibnamefont {Tetradis}},\ and\ \bibinfo {author} {\bibfnamefont
  {A.}~\bibnamefont {Urbano}},\ }\bibfield  {title} {\bibinfo {title}
  {({Higgs}) vacuum decay during inflation},\ }\href
  {https://doi.org/10.1007/JHEP07(2017)058} {\bibfield  {journal} {\bibinfo
  {journal} {JHEP}\ }\textbf {\bibinfo {volume} {2017}}\bibfield  {number}
  {\bibinfo  {number} { (07)},\ \bibinfo {pages} {058}},\ }\Eprint
  {https://arxiv.org/abs/1706.00792} {arXiv:1706.00792} \BibitemShut {NoStop}%
\bibitem [{\citenamefont {Markkanen}\ \emph
  {et~al.}(2018{\natexlab{a}})\citenamefont {Markkanen}, \citenamefont
  {Rajantie},\ and\ \citenamefont {Stopyra}}]{Markkanen:2018pdo}%
  \BibitemOpen
  \bibfield  {author} {\bibinfo {author} {\bibfnamefont {T.}~\bibnamefont
  {Markkanen}}, \bibinfo {author} {\bibfnamefont {A.}~\bibnamefont
  {Rajantie}},\ and\ \bibinfo {author} {\bibfnamefont {S.}~\bibnamefont
  {Stopyra}},\ }\bibfield  {title} {\bibinfo {title} {Cosmological aspects of
  {Higgs} vacuum metastability},\ }\href
  {https://doi.org/10.3389/fspas.2018.00040} {\bibfield  {journal} {\bibinfo
  {journal} {Front. Astron. Space Sci.}\ }\textbf {\bibinfo {volume} {5}},\
  \bibinfo {pages} {40} (\bibinfo {year} {2018}{\natexlab{a}})},\ \Eprint
  {https://arxiv.org/abs/1809.06923} {arXiv:1809.06923} \BibitemShut {NoStop}%
\bibitem [{\citenamefont {Mantziris}\ \emph {et~al.}(2021)\citenamefont
  {Mantziris}, \citenamefont {Markkanen},\ and\ \citenamefont
  {Rajantie}}]{Mantziris:2020rzh}%
  \BibitemOpen
  \bibfield  {author} {\bibinfo {author} {\bibfnamefont {A.}~\bibnamefont
  {Mantziris}}, \bibinfo {author} {\bibfnamefont {T.}~\bibnamefont
  {Markkanen}},\ and\ \bibinfo {author} {\bibfnamefont {A.}~\bibnamefont
  {Rajantie}},\ }\bibfield  {title} {\bibinfo {title} {Vacuum decay constraints
  on the {Higgs} curvature coupling from inflation},\ }\href
  {https://doi.org/10.1088/1475-7516/2021/03/077} {\bibfield  {journal}
  {\bibinfo  {journal} {JCAP}\ }\textbf {\bibinfo {volume} {2021}}\bibfield
  {number} {\bibinfo  {number} { (03)},\ \bibinfo {pages} {077}},\ }\Eprint
  {https://arxiv.org/abs/2011.03763} {arXiv:2011.03763} \BibitemShut {NoStop}%
\bibitem [{\citenamefont {Mantziris}\ \emph {et~al.}(2022)\citenamefont
  {Mantziris}, \citenamefont {Markkanen},\ and\ \citenamefont
  {Rajantie}}]{Mantziris:2022fuu}%
  \BibitemOpen
  \bibfield  {author} {\bibinfo {author} {\bibfnamefont {A.}~\bibnamefont
  {Mantziris}}, \bibinfo {author} {\bibfnamefont {T.}~\bibnamefont
  {Markkanen}},\ and\ \bibinfo {author} {\bibfnamefont {A.}~\bibnamefont
  {Rajantie}},\ }\bibfield  {title} {\bibinfo {title} {The effective {Higgs}
  potential and vacuum decay in {Starobinsky} inflation},\ }\href
  {https://doi.org/10.1088/1475-7516/2022/10/073} {\bibfield  {journal}
  {\bibinfo  {journal} {JCAP}\ }\textbf {\bibinfo {volume} {2022}}\bibfield
  {number} {\bibinfo  {number} { (10)},\ \bibinfo {pages} {073}},\ }\Eprint
  {https://arxiv.org/abs/2207.00696} {arXiv:2207.00696} \BibitemShut {NoStop}%
\bibitem [{\citenamefont {Herranen}\ \emph {et~al.}(2015)\citenamefont
  {Herranen}, \citenamefont {Markkanen}, \citenamefont {Nurmi},\ and\
  \citenamefont {Rajantie}}]{Herranen:2015ima}%
  \BibitemOpen
  \bibfield  {author} {\bibinfo {author} {\bibfnamefont {M.}~\bibnamefont
  {Herranen}}, \bibinfo {author} {\bibfnamefont {T.}~\bibnamefont {Markkanen}},
  \bibinfo {author} {\bibfnamefont {S.}~\bibnamefont {Nurmi}},\ and\ \bibinfo
  {author} {\bibfnamefont {A.}~\bibnamefont {Rajantie}},\ }\bibfield  {title}
  {\bibinfo {title} {Spacetime curvature and {Higgs} stability after
  inflation},\ }\href {https://doi.org/10.1103/PhysRevLett.115.241301}
  {\bibfield  {journal} {\bibinfo  {journal} {Phys. Rev. Lett.}\ }\textbf
  {\bibinfo {volume} {115}},\ \bibinfo {pages} {241301} (\bibinfo {year}
  {2015})},\ \Eprint {https://arxiv.org/abs/1506.04065} {arXiv:1506.04065}
  \BibitemShut {NoStop}%
\bibitem [{\citenamefont {Bars}\ \emph {et~al.}(2013)\citenamefont {Bars},
  \citenamefont {Steinhardt},\ and\ \citenamefont {Turok}}]{Bars:2013vba}%
  \BibitemOpen
  \bibfield  {author} {\bibinfo {author} {\bibfnamefont {I.}~\bibnamefont
  {Bars}}, \bibinfo {author} {\bibfnamefont {P.~J.}\ \bibnamefont
  {Steinhardt}},\ and\ \bibinfo {author} {\bibfnamefont {N.}~\bibnamefont
  {Turok}},\ }\bibfield  {title} {\bibinfo {title} {Cyclic cosmology, conformal
  symmetry and the metastability of the {Higgs}},\ }\href
  {https://doi.org/10.1016/j.physletb.2013.08.071} {\bibfield  {journal}
  {\bibinfo  {journal} {Phys. Lett. B}\ }\textbf {\bibinfo {volume} {726}},\
  \bibinfo {pages} {50} (\bibinfo {year} {2013})},\ \Eprint
  {https://arxiv.org/abs/1307.8106} {arXiv:1307.8106} \BibitemShut {NoStop}%
\bibitem [{\citenamefont {Hassan}\ and\ \citenamefont
  {March-Russell}(2026)}]{Hassan:2026cmp}%
  \BibitemOpen
  \bibfield  {author} {\bibinfo {author} {\bibfnamefont {S.}~\bibnamefont
  {Hassan}}\ and\ \bibinfo {author} {\bibfnamefont {J.}~\bibnamefont
  {March-Russell}},\ }\href@noop {} {\bibinfo {title} {Compact space catalysis
  of false vacuum decay and {Schwinger} effect}} (\bibinfo {year} {2026}),\
  \Eprint {https://arxiv.org/abs/2605.11081} {arXiv:2605.11081} \BibitemShut
  {NoStop}%
\bibitem [{\citenamefont {Brown}(2018)}]{Brown:2015kgj}%
  \BibitemOpen
  \bibfield  {author} {\bibinfo {author} {\bibfnamefont {A.~R.}\ \bibnamefont
  {Brown}},\ }\bibfield  {title} {\bibinfo {title} {{Schwinger pair production
  at nonzero temperatures or in compact directions}},\ }\href
  {https://doi.org/10.1103/PhysRevD.98.036008} {\bibfield  {journal} {\bibinfo
  {journal} {Phys. Rev. D}\ }\textbf {\bibinfo {volume} {98}},\ \bibinfo
  {pages} {036008} (\bibinfo {year} {2018})},\ \Eprint
  {https://arxiv.org/abs/1512.05716} {arXiv:1512.05716 [hep-th]} \BibitemShut
  {NoStop}%
\bibitem [{\citenamefont {Antoniadis}\ \emph {et~al.}(2024)\citenamefont
  {Antoniadis}, \citenamefont {Bielli}, \citenamefont {Chatrabhuti},\ and\
  \citenamefont {Isono}}]{Antoniadis:2024ent}%
  \BibitemOpen
  \bibfield  {author} {\bibinfo {author} {\bibfnamefont {I.}~\bibnamefont
  {Antoniadis}}, \bibinfo {author} {\bibfnamefont {D.}~\bibnamefont {Bielli}},
  \bibinfo {author} {\bibfnamefont {A.}~\bibnamefont {Chatrabhuti}},\ and\
  \bibinfo {author} {\bibfnamefont {H.}~\bibnamefont {Isono}},\ }\bibfield
  {title} {\bibinfo {title} {{Thin-wall vacuum decay in the presence of a
  compact dimension}},\ }\href {https://doi.org/10.1007/JHEP09(2024)011}
  {\bibfield  {journal} {\bibinfo  {journal} {JHEP}\ }\textbf {\bibinfo
  {volume} {2024}}\bibfield  {number} {\bibinfo  {number} { (09)},\ \bibinfo
  {pages} {011}},\ }\Eprint {https://arxiv.org/abs/2405.16920}
  {arXiv:2405.16920 [hep-th]} \BibitemShut {NoStop}%
\bibitem [{\citenamefont {Zhang}(2026{\natexlab{a}})}]{v12PRL}%
  \BibitemOpen
  \bibfield  {author} {\bibinfo {author} {\bibfnamefont {A.}~\bibnamefont
  {Zhang}},\ }\href@noop {} {\bibinfo {title} {{Gauge Field Fluctuations
  Protect the Higgs Vacuum When a Small Universe Begins}}} (\bibinfo {year}
  {2026}{\natexlab{a}}),\ \Eprint {https://arxiv.org/abs/2610.28704}
  {arXiv:2610.28704 [hep-ph]} \BibitemShut {NoStop}%
\bibitem [{\citenamefont {Duff}(1994)}]{Duff:1993wm}%
  \BibitemOpen
  \bibfield  {author} {\bibinfo {author} {\bibfnamefont {M.~J.}\ \bibnamefont
  {Duff}},\ }\bibfield  {title} {\bibinfo {title} {{Twenty years of the Weyl
  anomaly}},\ }\href {https://doi.org/10.1088/0264-9381/11/6/004} {\bibfield
  {journal} {\bibinfo  {journal} {Class. Quant. Grav.}\ }\textbf {\bibinfo
  {volume} {11}},\ \bibinfo {pages} {1387} (\bibinfo {year} {1994})},\ \Eprint
  {https://arxiv.org/abs/hep-th/9308075} {arXiv:hep-th/9308075} \BibitemShut
  {NoStop}%
\bibitem [{\citenamefont {Starobinsky}(1980)}]{Starobinsky:1980te}%
  \BibitemOpen
  \bibfield  {author} {\bibinfo {author} {\bibfnamefont {A.~A.}\ \bibnamefont
  {Starobinsky}},\ }\bibfield  {title} {\bibinfo {title} {{A New Type of
  Isotropic Cosmological Models Without Singularity}},\ }\href
  {https://doi.org/10.1016/0370-2693(80)90670-X} {\bibfield  {journal}
  {\bibinfo  {journal} {Phys. Lett. B}\ }\textbf {\bibinfo {volume} {91}},\
  \bibinfo {pages} {99} (\bibinfo {year} {1980})}\BibitemShut {NoStop}%
\bibitem [{\citenamefont {Fischetti}\ \emph {et~al.}(1979)\citenamefont
  {Fischetti}, \citenamefont {Hartle},\ and\ \citenamefont
  {Hu}}]{Fischetti:1979ue}%
  \BibitemOpen
  \bibfield  {author} {\bibinfo {author} {\bibfnamefont {M.~V.}\ \bibnamefont
  {Fischetti}}, \bibinfo {author} {\bibfnamefont {J.~B.}\ \bibnamefont
  {Hartle}},\ and\ \bibinfo {author} {\bibfnamefont {B.~L.}\ \bibnamefont
  {Hu}},\ }\bibfield  {title} {\bibinfo {title} {{Quantum Effects in the Early
  Universe. 1. Influence of Trace Anomalies on Homogeneous, Isotropic,
  Classical Geometries}},\ }\href {https://doi.org/10.1103/PhysRevD.20.1757}
  {\bibfield  {journal} {\bibinfo  {journal} {Phys. Rev. D}\ }\textbf {\bibinfo
  {volume} {20}},\ \bibinfo {pages} {1757} (\bibinfo {year}
  {1979})}\BibitemShut {NoStop}%
\bibitem [{\citenamefont {Hu}\ and\ \citenamefont {Parker}(1978)}]{Hu:1978zd}%
  \BibitemOpen
  \bibfield  {author} {\bibinfo {author} {\bibfnamefont {B.~L.}\ \bibnamefont
  {Hu}}\ and\ \bibinfo {author} {\bibfnamefont {L.}~\bibnamefont {Parker}},\
  }\bibfield  {title} {\bibinfo {title} {{Anisotropy Damping Through Quantum
  Effects in the Early Universe}},\ }\href
  {https://doi.org/10.1103/PhysRevD.17.933} {\bibfield  {journal} {\bibinfo
  {journal} {Phys. Rev. D}\ }\textbf {\bibinfo {volume} {17}},\ \bibinfo
  {pages} {933} (\bibinfo {year} {1978})},\ \bibinfo {note} {erratum: Phys.
  Rev. D 17, 3292 (1978)}\BibitemShut {NoStop}%
\bibitem [{\citenamefont {Hartle}\ and\ \citenamefont
  {Hu}(1979)}]{Hartle:1979uf}%
  \BibitemOpen
  \bibfield  {author} {\bibinfo {author} {\bibfnamefont {J.~B.}\ \bibnamefont
  {Hartle}}\ and\ \bibinfo {author} {\bibfnamefont {B.~L.}\ \bibnamefont
  {Hu}},\ }\bibfield  {title} {\bibinfo {title} {{Quantum Effects in the Early
  Universe. 2. Effective Action for Scalar Fields in Homogeneous Cosmologies
  with Small Anisotropy}},\ }\href {https://doi.org/10.1103/PhysRevD.20.1772}
  {\bibfield  {journal} {\bibinfo  {journal} {Phys. Rev. D}\ }\textbf {\bibinfo
  {volume} {20}},\ \bibinfo {pages} {1772} (\bibinfo {year}
  {1979})}\BibitemShut {NoStop}%
\bibitem [{\citenamefont {Hartle}\ and\ \citenamefont
  {Hu}(1980)}]{Hartle:1980nn}%
  \BibitemOpen
  \bibfield  {author} {\bibinfo {author} {\bibfnamefont {J.~B.}\ \bibnamefont
  {Hartle}}\ and\ \bibinfo {author} {\bibfnamefont {B.~L.}\ \bibnamefont
  {Hu}},\ }\bibfield  {title} {\bibinfo {title} {{Quantum effects in the early
  universe. III. Dissipation of anisotropy by scalar particle production}},\
  }\href {https://doi.org/10.1103/PhysRevD.21.2756} {\bibfield  {journal}
  {\bibinfo  {journal} {Phys. Rev. D}\ }\textbf {\bibinfo {volume} {21}},\
  \bibinfo {pages} {2756} (\bibinfo {year} {1980})}\BibitemShut {NoStop}%
\bibitem [{\citenamefont {Herdeiro}\ and\ \citenamefont
  {Sampaio}(2006)}]{Herdeiro:2005zj}%
  \BibitemOpen
  \bibfield  {author} {\bibinfo {author} {\bibfnamefont {C.~A.~R.}\
  \bibnamefont {Herdeiro}}\ and\ \bibinfo {author} {\bibfnamefont
  {M.}~\bibnamefont {Sampaio}},\ }\bibfield  {title} {\bibinfo {title}
  {{Casimir energy and a cosmological bounce}},\ }\href
  {https://doi.org/10.1088/0264-9381/23/2/012} {\bibfield  {journal} {\bibinfo
  {journal} {Class. Quant. Grav.}\ }\textbf {\bibinfo {volume} {23}},\ \bibinfo
  {pages} {473} (\bibinfo {year} {2006})},\ \Eprint
  {https://arxiv.org/abs/hep-th/0510052} {arXiv:hep-th/0510052} \BibitemShut
  {NoStop}%
\bibitem [{\citenamefont {Godlowski}\ \emph {et~al.}(2008)\citenamefont
  {Godlowski}, \citenamefont {Szydlowski},\ and\ \citenamefont
  {Zhu}}]{Godlowski:2007gx}%
  \BibitemOpen
  \bibfield  {author} {\bibinfo {author} {\bibfnamefont {W.}~\bibnamefont
  {Godlowski}}, \bibinfo {author} {\bibfnamefont {M.}~\bibnamefont
  {Szydlowski}},\ and\ \bibinfo {author} {\bibfnamefont {Z.-H.}\ \bibnamefont
  {Zhu}},\ }\bibfield  {title} {\bibinfo {title} {{Constraining bouncing
  cosmology caused by Casimir effect}},\ }\href
  {https://doi.org/10.1134/S0202289308010039} {\bibfield  {journal} {\bibinfo
  {journal} {Grav. Cosmol.}\ }\textbf {\bibinfo {volume} {14}},\ \bibinfo
  {pages} {17} (\bibinfo {year} {2008})},\ \Eprint
  {https://arxiv.org/abs/astro-ph/0702237} {arXiv:astro-ph/0702237}
  \BibitemShut {NoStop}%
\bibitem [{\citenamefont {Arkani-Hamed}\ \emph {et~al.}(2007)\citenamefont
  {Arkani-Hamed}, \citenamefont {Dubovsky}, \citenamefont {Nicolis},\ and\
  \citenamefont {Villadoro}}]{Arkani-Hamed:2007ryu}%
  \BibitemOpen
  \bibfield  {author} {\bibinfo {author} {\bibfnamefont {N.}~\bibnamefont
  {Arkani-Hamed}}, \bibinfo {author} {\bibfnamefont {S.}~\bibnamefont
  {Dubovsky}}, \bibinfo {author} {\bibfnamefont {A.}~\bibnamefont {Nicolis}},\
  and\ \bibinfo {author} {\bibfnamefont {G.}~\bibnamefont {Villadoro}},\
  }\bibfield  {title} {\bibinfo {title} {{Quantum Horizons of the Standard
  Model Landscape}},\ }\href {https://doi.org/10.1088/1126-6708/2007/06/078}
  {\bibfield  {journal} {\bibinfo  {journal} {JHEP}\ }\textbf {\bibinfo
  {volume} {2007}}\bibfield  {number} {\bibinfo  {number} { (06)},\ \bibinfo
  {pages} {078}},\ }\Eprint {https://arxiv.org/abs/hep-th/0703067}
  {arXiv:hep-th/0703067} \BibitemShut {NoStop}%
\bibitem [{\citenamefont {Fornal}\ and\ \citenamefont
  {Wise}(2011)}]{Fornal:2011tw}%
  \BibitemOpen
  \bibfield  {author} {\bibinfo {author} {\bibfnamefont {B.}~\bibnamefont
  {Fornal}}\ and\ \bibinfo {author} {\bibfnamefont {M.~B.}\ \bibnamefont
  {Wise}},\ }\bibfield  {title} {\bibinfo {title} {{Standard model with
  compactified spatial dimensions}},\ }\href
  {https://doi.org/10.1007/JHEP07(2011)086} {\bibfield  {journal} {\bibinfo
  {journal} {JHEP}\ }\textbf {\bibinfo {volume} {2011}}\bibfield  {number}
  {\bibinfo  {number} { (07)},\ \bibinfo {pages} {086}},\ }\Eprint
  {https://arxiv.org/abs/1106.0890} {arXiv:1106.0890 [hep-th]} \BibitemShut
  {NoStop}%
\bibitem [{\citenamefont {Hamada}\ and\ \citenamefont
  {Shiu}(2017)}]{Hamada:2017yji}%
  \BibitemOpen
  \bibfield  {author} {\bibinfo {author} {\bibfnamefont {Y.}~\bibnamefont
  {Hamada}}\ and\ \bibinfo {author} {\bibfnamefont {G.}~\bibnamefont {Shiu}},\
  }\bibfield  {title} {\bibinfo {title} {{Weak Gravity Conjecture, Multiple
  Point Principle and the Standard Model Landscape}},\ }\href
  {https://doi.org/10.1007/JHEP11(2017)043} {\bibfield  {journal} {\bibinfo
  {journal} {JHEP}\ }\textbf {\bibinfo {volume} {2017}}\bibfield  {number}
  {\bibinfo  {number} { (11)},\ \bibinfo {pages} {043}},\ }\Eprint
  {https://arxiv.org/abs/1707.06326} {arXiv:1707.06326 [hep-th]} \BibitemShut
  {NoStop}%
\bibitem [{\citenamefont {L{\"u}scher}(1982)}]{Luscher:1982uv}%
  \BibitemOpen
  \bibfield  {author} {\bibinfo {author} {\bibfnamefont {M.}~\bibnamefont
  {L{\"u}scher}},\ }\bibfield  {title} {\bibinfo {title} {{A New Method to
  Compute the Spectrum of Low Lying States in Massless Asymptotically Free
  Field Theories}},\ }\href {https://doi.org/10.1016/0370-2693(82)90210-6}
  {\bibfield  {journal} {\bibinfo  {journal} {Phys. Lett. B}\ }\textbf
  {\bibinfo {volume} {118}},\ \bibinfo {pages} {391} (\bibinfo {year}
  {1982})}\BibitemShut {NoStop}%
\bibitem [{\citenamefont {L{\"u}scher}(1983)}]{Luscher:1982ma}%
  \BibitemOpen
  \bibfield  {author} {\bibinfo {author} {\bibfnamefont {M.}~\bibnamefont
  {L{\"u}scher}},\ }\bibfield  {title} {\bibinfo {title} {{Some Analytic
  Results Concerning the Mass Spectrum of Yang-Mills Gauge Theories on a
  Torus}},\ }\href {https://doi.org/10.1016/0550-3213(83)90436-4} {\bibfield
  {journal} {\bibinfo  {journal} {Nucl. Phys. B}\ }\textbf {\bibinfo {volume}
  {219}},\ \bibinfo {pages} {233} (\bibinfo {year} {1983})}\BibitemShut
  {NoStop}%
\bibitem [{\citenamefont {L{\"u}scher}\ and\ \citenamefont
  {M{\"u}nster}(1984)}]{Luscher:1983gm}%
  \BibitemOpen
  \bibfield  {author} {\bibinfo {author} {\bibfnamefont {M.}~\bibnamefont
  {L{\"u}scher}}\ and\ \bibinfo {author} {\bibfnamefont {G.}~\bibnamefont
  {M{\"u}nster}},\ }\bibfield  {title} {\bibinfo {title} {{Weak Coupling
  Expansion of the Low Lying Energy Values in the SU(2) Gauge Theory on a
  Torus}},\ }\href {https://doi.org/10.1016/0550-3213(84)90038-5} {\bibfield
  {journal} {\bibinfo  {journal} {Nucl. Phys. B}\ }\textbf {\bibinfo {volume}
  {232}},\ \bibinfo {pages} {445} (\bibinfo {year} {1984})}\BibitemShut
  {NoStop}%
\bibitem [{\citenamefont {Weisz}\ and\ \citenamefont
  {Ziemann}(1987)}]{Weisz:1986vu}%
  \BibitemOpen
  \bibfield  {author} {\bibinfo {author} {\bibfnamefont {P.}~\bibnamefont
  {Weisz}}\ and\ \bibinfo {author} {\bibfnamefont {V.}~\bibnamefont
  {Ziemann}},\ }\bibfield  {title} {\bibinfo {title} {{Weak Coupling Expansions
  of the Low Lying Energy Values in SU(3) Gauge Theory on a Torus}},\ }\href
  {https://doi.org/10.1016/0550-3213(87)90031-9} {\bibfield  {journal}
  {\bibinfo  {journal} {Nucl. Phys. B}\ }\textbf {\bibinfo {volume} {284}},\
  \bibinfo {pages} {157} (\bibinfo {year} {1987})}\BibitemShut {NoStop}%
\bibitem [{\citenamefont {van Baal}(1988)}]{vanBaal:1988va}%
  \BibitemOpen
  \bibfield  {author} {\bibinfo {author} {\bibfnamefont {P.}~\bibnamefont {van
  Baal}},\ }\bibfield  {title} {\bibinfo {title} {{The Small Volume Expansion
  of Gauge Theories Coupled to Massless Fermions}},\ }\href
  {https://doi.org/10.1016/0550-3213(88)90323-9} {\bibfield  {journal}
  {\bibinfo  {journal} {Nucl. Phys. B}\ }\textbf {\bibinfo {volume} {307}},\
  \bibinfo {pages} {274} (\bibinfo {year} {1988})},\ \bibinfo {note} {[Erratum:
  Nucl. Phys. B 312, 752 (1989)]}\BibitemShut {NoStop}%
\bibitem [{\citenamefont {van Baal}(2001)}]{vanBaal:2000zc}%
  \BibitemOpen
  \bibfield  {author} {\bibinfo {author} {\bibfnamefont {P.}~\bibnamefont {van
  Baal}},\ }\bibfield  {title} {\bibinfo {title} {{QCD in a finite volume}},\
  }in\ \href {https://doi.org/10.1142/9789812810458_0021} {\emph {\bibinfo
  {booktitle} {{At the Frontier of Particle Physics}}}}\ (\bibinfo  {publisher}
  {World Scientific},\ \bibinfo {year} {2001})\ pp.\ \bibinfo {pages}
  {683--760},\ \Eprint {https://arxiv.org/abs/hep-ph/0008206}
  {arXiv:hep-ph/0008206} \BibitemShut {NoStop}%
\bibitem [{\citenamefont {Mithani}\ and\ \citenamefont
  {Vilenkin}(2015)}]{Mithani:2014toa}%
  \BibitemOpen
  \bibfield  {author} {\bibinfo {author} {\bibfnamefont {A.~T.}\ \bibnamefont
  {Mithani}}\ and\ \bibinfo {author} {\bibfnamefont {A.}~\bibnamefont
  {Vilenkin}},\ }\bibfield  {title} {\bibinfo {title} {{Stabilizing oscillating
  universes against quantum decay}},\ }\href
  {https://doi.org/10.1088/1475-7516/2015/07/010} {\bibfield  {journal}
  {\bibinfo  {journal} {JCAP}\ }\textbf {\bibinfo {volume} {2015}}\bibfield
  {number} {\bibinfo  {number} { (07)},\ \bibinfo {pages} {010}},\ }\Eprint
  {https://arxiv.org/abs/1407.5361} {arXiv:1407.5361 [hep-th]} \BibitemShut
  {NoStop}%
\bibitem [{\citenamefont {Lachi{\`e}ze-Rey}\ and\ \citenamefont
  {Luminet}(1995)}]{Lachieze-Rey:1995qrb}%
  \BibitemOpen
  \bibfield  {author} {\bibinfo {author} {\bibfnamefont {M.}~\bibnamefont
  {Lachi{\`e}ze-Rey}}\ and\ \bibinfo {author} {\bibfnamefont {J.-P.}\
  \bibnamefont {Luminet}},\ }\bibfield  {title} {\bibinfo {title} {{Cosmic
  topology}},\ }\href {https://doi.org/10.1016/0370-1573(94)00085-H} {\bibfield
   {journal} {\bibinfo  {journal} {Phys. Rept.}\ }\textbf {\bibinfo {volume}
  {254}},\ \bibinfo {pages} {135} (\bibinfo {year} {1995})},\ \Eprint
  {https://arxiv.org/abs/gr-qc/9605010} {arXiv:gr-qc/9605010} \BibitemShut
  {NoStop}%
\bibitem [{\citenamefont {Cornish}\ \emph {et~al.}(1998)\citenamefont
  {Cornish}, \citenamefont {Spergel},\ and\ \citenamefont
  {Starkman}}]{Cornish:1997ab}%
  \BibitemOpen
  \bibfield  {author} {\bibinfo {author} {\bibfnamefont {N.~J.}\ \bibnamefont
  {Cornish}}, \bibinfo {author} {\bibfnamefont {D.~N.}\ \bibnamefont
  {Spergel}},\ and\ \bibinfo {author} {\bibfnamefont {G.~D.}\ \bibnamefont
  {Starkman}},\ }\bibfield  {title} {\bibinfo {title} {{Circles in the sky:
  Finding topology with the microwave background radiation}},\ }\href
  {https://doi.org/10.1088/0264-9381/15/9/013} {\bibfield  {journal} {\bibinfo
  {journal} {Class. Quant. Grav.}\ }\textbf {\bibinfo {volume} {15}},\ \bibinfo
  {pages} {2657} (\bibinfo {year} {1998})},\ \Eprint
  {https://arxiv.org/abs/astro-ph/9801212} {arXiv:astro-ph/9801212}
  \BibitemShut {NoStop}%
\bibitem [{\citenamefont {Vaudrevange}\ \emph {et~al.}(2012)\citenamefont
  {Vaudrevange}, \citenamefont {Starkman}, \citenamefont {Cornish},\ and\
  \citenamefont {Spergel}}]{Vaudrevange:2012da}%
  \BibitemOpen
  \bibfield  {author} {\bibinfo {author} {\bibfnamefont {P.~M.}\ \bibnamefont
  {Vaudrevange}}, \bibinfo {author} {\bibfnamefont {G.~D.}\ \bibnamefont
  {Starkman}}, \bibinfo {author} {\bibfnamefont {N.~J.}\ \bibnamefont
  {Cornish}},\ and\ \bibinfo {author} {\bibfnamefont {D.~N.}\ \bibnamefont
  {Spergel}},\ }\bibfield  {title} {\bibinfo {title} {{Constraints on the
  Topology of the Universe: Extension to General Geometries}},\ }\href
  {https://doi.org/10.1103/PhysRevD.86.083526} {\bibfield  {journal} {\bibinfo
  {journal} {Phys. Rev. D}\ }\textbf {\bibinfo {volume} {86}},\ \bibinfo
  {pages} {083526} (\bibinfo {year} {2012})},\ \Eprint
  {https://arxiv.org/abs/1206.2939} {arXiv:1206.2939 [astro-ph.CO]}
  \BibitemShut {NoStop}%
\bibitem [{\citenamefont {Petersen}\ \emph {et~al.}(2023)\citenamefont
  {Petersen} \emph {et~al.}}]{COMPACT:2022nsu}%
  \BibitemOpen
  \bibfield  {author} {\bibinfo {author} {\bibfnamefont {P.}~\bibnamefont
  {Petersen}} \emph {et~al.} (\bibinfo {collaboration} {COMPACT}),\ }\bibfield
  {title} {\bibinfo {title} {{Cosmic topology. Part~I. Limits on orientable
  Euclidean manifolds from circle searches}},\ }\href
  {https://doi.org/10.1088/1475-7516/2023/01/030} {\bibfield  {journal}
  {\bibinfo  {journal} {JCAP}\ }\textbf {\bibinfo {volume} {2023}}\bibfield
  {number} {\bibinfo  {number} { (01)},\ \bibinfo {pages} {030}},\ }\bibinfo
  {note} {erratum: JCAP 04, E01 (2024)},\ \Eprint
  {https://arxiv.org/abs/2211.02603} {arXiv:2211.02603 [astro-ph.CO]}
  \BibitemShut {NoStop}%
\bibitem [{\citenamefont {Eskilt}\ \emph {et~al.}(2024)\citenamefont {Eskilt}
  \emph {et~al.}}]{COMPACT:2023rkp}%
  \BibitemOpen
  \bibfield  {author} {\bibinfo {author} {\bibfnamefont {J.~R.}\ \bibnamefont
  {Eskilt}} \emph {et~al.} (\bibinfo {collaboration} {COMPACT}),\ }\bibfield
  {title} {\bibinfo {title} {{Cosmic topology. Part~IIa. Eigenmodes,
  correlation matrices, and detectability of orientable~Euclidean manifolds}},\
  }\href {https://doi.org/10.1088/1475-7516/2024/03/036} {\bibfield  {journal}
  {\bibinfo  {journal} {JCAP}\ }\textbf {\bibinfo {volume} {2024}}\bibfield
  {number} {\bibinfo  {number} { (03)},\ \bibinfo {pages} {036}},\ }\Eprint
  {https://arxiv.org/abs/2306.17112} {arXiv:2306.17112 [astro-ph.CO]}
  \BibitemShut {NoStop}%
\bibitem [{\citenamefont {Himeno}\ \emph {et~al.}(2026)\citenamefont {Himeno},
  \citenamefont {Nishizawa},\ and\ \citenamefont {Ichiki}}]{Himeno:2026lgh}%
  \BibitemOpen
  \bibfield  {author} {\bibinfo {author} {\bibfnamefont {R.}~\bibnamefont
  {Himeno}}, \bibinfo {author} {\bibfnamefont {A.~J.}\ \bibnamefont
  {Nishizawa}},\ and\ \bibinfo {author} {\bibfnamefont {K.}~\bibnamefont
  {Ichiki}},\ }\href@noop {} {\bibinfo {title} {{Limited Impact of CMB B-mode
  Polarization on Constraints on Cosmic Topology}}} (\bibinfo {year} {2026}),\
  \Eprint {https://arxiv.org/abs/2610.05960} {arXiv:2610.05960 [astro-ph.CO]}
  \BibitemShut {NoStop}%
\bibitem [{\citenamefont {Zhang}(2026{\natexlab{b}})}]{v12Report}%
  \BibitemOpen
  \bibfield  {author} {\bibinfo {author} {\bibfnamefont {A.}~\bibnamefont
  {Zhang}},\ }\href@noop {} {\bibinfo {title} {{Casimir Cosmology: Derivations
  and Numerical Validation}}},\ \bibinfo {howpublished} {{Companion technical
  report, Zenodo},
  doi:\href{https://doi.org/10.5281/zenodo.23434139}{10.5281/zenodo.23434139}}
  (\bibinfo {year} {2026}{\natexlab{b}}),\ \bibinfo {note} {{N}ot independently
  peer reviewed}\BibitemShut {NoStop}%
\bibitem [{\citenamefont {Zhang}(2026{\natexlab{c}})}]{DNECv12code}%
  \BibitemOpen
  \bibfield  {author} {\bibinfo {author} {\bibfnamefont {A.}~\bibnamefont
  {Zhang}},\ }\href {https://doi.org/10.5281/zenodo.23434067} {\bibinfo {title}
  {{Code and data for: A Casimir turning point as the beginning of a small flat
  universe}}},\ \bibinfo {howpublished} {{Zenodo},
  doi:\href{https://doi.org/10.5281/zenodo.23434067}{10.5281/zenodo.23434067}}
  (\bibinfo {year} {2026}{\natexlab{c}}),\ \bibinfo {note}
  {[{S}oftware]}\BibitemShut {NoStop}%
\bibitem [{\citenamefont {Hartle}\ and\ \citenamefont
  {Horowitz}(1981)}]{Hartle:1981zt}%
  \BibitemOpen
  \bibfield  {author} {\bibinfo {author} {\bibfnamefont {J.~B.}\ \bibnamefont
  {Hartle}}\ and\ \bibinfo {author} {\bibfnamefont {G.~T.}\ \bibnamefont
  {Horowitz}},\ }\bibfield  {title} {\bibinfo {title} {{Ground State
  Expectation Value of the Metric in the 1/$N$ or Semiclassical Approximation
  to Quantum Gravity}},\ }\href {https://doi.org/10.1103/PhysRevD.24.257}
  {\bibfield  {journal} {\bibinfo  {journal} {Phys. Rev. D}\ }\textbf {\bibinfo
  {volume} {24}},\ \bibinfo {pages} {257} (\bibinfo {year} {1981})}\BibitemShut
  {NoStop}%
\bibitem [{\citenamefont {Takahashi}\ \emph {et~al.}(2026)\citenamefont
  {Takahashi} \emph {et~al.}}]{ParticleDataGroup:2026mpi}%
  \BibitemOpen
  \bibfield  {author} {\bibinfo {author} {\bibfnamefont {F.}~\bibnamefont
  {Takahashi}} \emph {et~al.} (\bibinfo {collaboration} {Particle Data
  Group}),\ }\bibfield  {title} {\bibinfo {title} {{Review of Particle
  Physics}},\ }\href {https://doi.org/10.1142/s0217751x26300115} {\bibfield
  {journal} {\bibinfo  {journal} {Int. J. Mod. Phys. A}\ }\textbf {\bibinfo
  {volume} {41}},\ \bibinfo {pages} {2630011} (\bibinfo {year}
  {2026})}\BibitemShut {NoStop}%
\bibitem [{\citenamefont {Hoang}(2020)}]{Hoang:2020iah}%
  \BibitemOpen
  \bibfield  {author} {\bibinfo {author} {\bibfnamefont {A.~H.}\ \bibnamefont
  {Hoang}},\ }\bibfield  {title} {\bibinfo {title} {{What is the Top Quark
  Mass?}},\ }\href {https://doi.org/10.1146/annurev-nucl-101918-023530}
  {\bibfield  {journal} {\bibinfo  {journal} {Ann. Rev. Nucl. Part. Sci.}\
  }\textbf {\bibinfo {volume} {70}},\ \bibinfo {pages} {225} (\bibinfo {year}
  {2020})},\ \Eprint {https://arxiv.org/abs/2004.12915} {arXiv:2004.12915
  [hep-ph]} \BibitemShut {NoStop}%
\bibitem [{\citenamefont {Solomons}\ \emph {et~al.}(2006)\citenamefont
  {Solomons}, \citenamefont {Dunsby},\ and\ \citenamefont
  {Ellis}}]{Solomons:2001ef}%
  \BibitemOpen
  \bibfield  {author} {\bibinfo {author} {\bibfnamefont {D.~M.}\ \bibnamefont
  {Solomons}}, \bibinfo {author} {\bibfnamefont {P.}~\bibnamefont {Dunsby}},\
  and\ \bibinfo {author} {\bibfnamefont {G.}~\bibnamefont {Ellis}},\ }\bibfield
   {title} {\bibinfo {title} {{Bounce behaviour in Kantowski-Sachs and Bianchi
  cosmologies}},\ }\href {https://doi.org/10.1088/0264-9381/23/23/001}
  {\bibfield  {journal} {\bibinfo  {journal} {Class. Quant. Grav.}\ }\textbf
  {\bibinfo {volume} {23}},\ \bibinfo {pages} {6585} (\bibinfo {year}
  {2006})},\ \Eprint {https://arxiv.org/abs/gr-qc/0103087}
  {arXiv:gr-qc/0103087} \BibitemShut {NoStop}%
\bibitem [{\citenamefont {Hofman}\ and\ \citenamefont
  {Maldacena}(2008)}]{Hofman:2008ar}%
  \BibitemOpen
  \bibfield  {author} {\bibinfo {author} {\bibfnamefont {D.~M.}\ \bibnamefont
  {Hofman}}\ and\ \bibinfo {author} {\bibfnamefont {J.}~\bibnamefont
  {Maldacena}},\ }\bibfield  {title} {\bibinfo {title} {{Conformal collider
  physics: Energy and charge correlations}},\ }\href
  {https://doi.org/10.1088/1126-6708/2008/05/012} {\bibfield  {journal}
  {\bibinfo  {journal} {JHEP}\ }\textbf {\bibinfo {volume} {2008}}\bibfield
  {number} {\bibinfo  {number} { (05)},\ \bibinfo {pages} {012}},\ }\Eprint
  {https://arxiv.org/abs/0803.1467} {arXiv:0803.1467 [hep-th]} \BibitemShut
  {NoStop}%
\bibitem [{\citenamefont {Hofman}\ \emph {et~al.}(2016)\citenamefont {Hofman},
  \citenamefont {Li}, \citenamefont {Meltzer}, \citenamefont {Poland},\ and\
  \citenamefont {Rejon-Barrera}}]{Hofman:2016awc}%
  \BibitemOpen
  \bibfield  {author} {\bibinfo {author} {\bibfnamefont {D.~M.}\ \bibnamefont
  {Hofman}}, \bibinfo {author} {\bibfnamefont {D.}~\bibnamefont {Li}}, \bibinfo
  {author} {\bibfnamefont {D.}~\bibnamefont {Meltzer}}, \bibinfo {author}
  {\bibfnamefont {D.}~\bibnamefont {Poland}},\ and\ \bibinfo {author}
  {\bibfnamefont {F.}~\bibnamefont {Rejon-Barrera}},\ }\bibfield  {title}
  {\bibinfo {title} {{A Proof of the Conformal Collider Bounds}},\ }\href
  {https://doi.org/10.1007/JHEP06(2016)111} {\bibfield  {journal} {\bibinfo
  {journal} {JHEP}\ }\textbf {\bibinfo {volume} {2016}}\bibfield  {number}
  {\bibinfo  {number} { (06)},\ \bibinfo {pages} {111}},\ }\Eprint
  {https://arxiv.org/abs/1603.03771} {arXiv:1603.03771 [hep-th]} \BibitemShut
  {NoStop}%
\bibitem [{\citenamefont {Cesar~e Silva}\ and\ \citenamefont
  {Shapiro}(2024)}]{CesareSilva:2024nym}%
  \BibitemOpen
  \bibfield  {author} {\bibinfo {author} {\bibfnamefont {W.}~\bibnamefont
  {Cesar~e Silva}}\ and\ \bibinfo {author} {\bibfnamefont {I.~L.}\ \bibnamefont
  {Shapiro}},\ }\bibfield  {title} {\bibinfo {title} {{Semiclassical bounce
  with strong minimal assumptions}},\ }\href
  {https://doi.org/10.1103/PhysRevD.110.043540} {\bibfield  {journal} {\bibinfo
   {journal} {Phys. Rev. D}\ }\textbf {\bibinfo {volume} {110}},\ \bibinfo
  {pages} {043540} (\bibinfo {year} {2024})},\ \Eprint
  {https://arxiv.org/abs/2402.18785} {arXiv:2402.18785 [gr-qc]} \BibitemShut
  {NoStop}%
\bibitem [{\citenamefont {Cesar~e Silva}\ \emph {et~al.}(2026)\citenamefont
  {Cesar~e Silva}, \citenamefont {Bertini},\ and\ \citenamefont
  {Shapiro}}]{CesareSilva:2026zne}%
  \BibitemOpen
  \bibfield  {author} {\bibinfo {author} {\bibfnamefont {W.}~\bibnamefont
  {Cesar~e Silva}}, \bibinfo {author} {\bibfnamefont {N.~R.}\ \bibnamefont
  {Bertini}},\ and\ \bibinfo {author} {\bibfnamefont {I.~L.}\ \bibnamefont
  {Shapiro}},\ }\href@noop {} {\bibinfo {title} {{Low-energy limit in the
  anomaly-induced action and the semiclassical cosmological bounce}}} (\bibinfo
  {year} {2026}),\ \Eprint {https://arxiv.org/abs/2608.09720} {arXiv:2608.09720
  [gr-qc]} \BibitemShut {NoStop}%
\bibitem [{\citenamefont {Anderson}(1983)}]{Anderson:1983nq}%
  \BibitemOpen
  \bibfield  {author} {\bibinfo {author} {\bibfnamefont {P.}~\bibnamefont
  {Anderson}},\ }\bibfield  {title} {\bibinfo {title} {{Effects of quantum
  fields on singularities and particle horizons in the early universe}},\
  }\href {https://doi.org/10.1103/PhysRevD.28.271} {\bibfield  {journal}
  {\bibinfo  {journal} {Phys. Rev. D}\ }\textbf {\bibinfo {volume} {28}},\
  \bibinfo {pages} {271} (\bibinfo {year} {1983})}\BibitemShut {NoStop}%
\bibitem [{\citenamefont {De~Felice}\ and\ \citenamefont
  {Tsujikawa}(2010)}]{DeFelice:2010aj}%
  \BibitemOpen
  \bibfield  {author} {\bibinfo {author} {\bibfnamefont {A.}~\bibnamefont
  {De~Felice}}\ and\ \bibinfo {author} {\bibfnamefont {S.}~\bibnamefont
  {Tsujikawa}},\ }\bibfield  {title} {\bibinfo {title} {{f(R) theories}},\
  }\href {https://doi.org/10.12942/lrr-2010-3} {\bibfield  {journal} {\bibinfo
  {journal} {Living Rev. Rel.}\ }\textbf {\bibinfo {volume} {13}},\ \bibinfo
  {pages} {3} (\bibinfo {year} {2010})},\ \Eprint
  {https://arxiv.org/abs/1002.4928} {arXiv:1002.4928 [gr-qc]} \BibitemShut
  {NoStop}%
\bibitem [{\citenamefont {Banach}\ and\ \citenamefont
  {Dowker}(1979)}]{Banach:1979iy}%
  \BibitemOpen
  \bibfield  {author} {\bibinfo {author} {\bibfnamefont {R.}~\bibnamefont
  {Banach}}\ and\ \bibinfo {author} {\bibfnamefont {J.~S.}\ \bibnamefont
  {Dowker}},\ }\bibfield  {title} {\bibinfo {title} {{The Vacuum Stress Tensor
  for Automorphic Fields on Some Flat Space-times}},\ }\href
  {https://doi.org/10.1088/0305-4470/12/12/032} {\bibfield  {journal} {\bibinfo
   {journal} {J. Phys. A}\ }\textbf {\bibinfo {volume} {12}},\ \bibinfo {pages}
  {2545} (\bibinfo {year} {1979})}\BibitemShut {NoStop}%
\bibitem [{\citenamefont {Bellucci}\ and\ \citenamefont
  {Saharian}(2009)}]{Bellucci:2009jr}%
  \BibitemOpen
  \bibfield  {author} {\bibinfo {author} {\bibfnamefont {S.}~\bibnamefont
  {Bellucci}}\ and\ \bibinfo {author} {\bibfnamefont {A.~A.}\ \bibnamefont
  {Saharian}},\ }\bibfield  {title} {\bibinfo {title} {{Fermionic Casimir
  densities in toroidally compactified spacetimes with applications to
  nanotubes}},\ }\href {https://doi.org/10.1103/PhysRevD.79.085019} {\bibfield
  {journal} {\bibinfo  {journal} {Phys. Rev. D}\ }\textbf {\bibinfo {volume}
  {79}},\ \bibinfo {pages} {085019} (\bibinfo {year} {2009})},\ \Eprint
  {https://arxiv.org/abs/0902.3726} {arXiv:0902.3726 [hep-th]} \BibitemShut
  {NoStop}%
\bibitem [{\citenamefont {Ibanez}\ \emph {et~al.}(2017)\citenamefont {Ibanez},
  \citenamefont {Martin-Lozano},\ and\ \citenamefont
  {Valenzuela}}]{Ibanez:2017kvh}%
  \BibitemOpen
  \bibfield  {author} {\bibinfo {author} {\bibfnamefont {L.~E.}\ \bibnamefont
  {Ibanez}}, \bibinfo {author} {\bibfnamefont {V.}~\bibnamefont
  {Martin-Lozano}},\ and\ \bibinfo {author} {\bibfnamefont {I.}~\bibnamefont
  {Valenzuela}},\ }\bibfield  {title} {\bibinfo {title} {{Constraining Neutrino
  Masses, the Cosmological Constant and BSM Physics from the Weak Gravity
  Conjecture}},\ }\href {https://doi.org/10.1007/JHEP11(2017)066} {\bibfield
  {journal} {\bibinfo  {journal} {JHEP}\ }\textbf {\bibinfo {volume}
  {2017}}\bibfield  {number} {\bibinfo  {number} { (11)},\ \bibinfo {pages}
  {066}},\ }\Eprint {https://arxiv.org/abs/1706.05392} {arXiv:1706.05392
  [hep-th]} \BibitemShut {NoStop}%
\bibitem [{\citenamefont {Hosotani}(1983)}]{Hosotani:1983xw}%
  \BibitemOpen
  \bibfield  {author} {\bibinfo {author} {\bibfnamefont {Y.}~\bibnamefont
  {Hosotani}},\ }\bibfield  {title} {\bibinfo {title} {{Dynamical Mass
  Generation by Compact Extra Dimensions}},\ }\href
  {https://doi.org/10.1016/0370-2693(83)90170-3} {\bibfield  {journal}
  {\bibinfo  {journal} {Phys. Lett. B}\ }\textbf {\bibinfo {volume} {126}},\
  \bibinfo {pages} {309} (\bibinfo {year} {1983})}\BibitemShut {NoStop}%
\bibitem [{\citenamefont {Penrose}(1979)}]{Penrose:1980ge}%
  \BibitemOpen
  \bibfield  {author} {\bibinfo {author} {\bibfnamefont {R.}~\bibnamefont
  {Penrose}},\ }\bibfield  {title} {\bibinfo {title} {{Singularities and
  time-asymmetry}},\ }in\ \href@noop {} {\emph {\bibinfo {booktitle} {General
  Relativity: An Einstein Centenary Survey}}},\ \bibinfo {editor} {edited by\
  \bibinfo {editor} {\bibfnamefont {S.~W.}\ \bibnamefont {Hawking}}\ and\
  \bibinfo {editor} {\bibfnamefont {W.}~\bibnamefont {Israel}}}\ (\bibinfo
  {publisher} {Cambridge University Press},\ \bibinfo {address} {Cambridge},\
  \bibinfo {year} {1979})\ pp.\ \bibinfo {pages} {581--638}\BibitemShut
  {NoStop}%
\bibitem [{\citenamefont {Garc{\'\i}a~Etxebarria}\ \emph
  {et~al.}(2020)\citenamefont {Garc{\'\i}a~Etxebarria}, \citenamefont
  {Montero}, \citenamefont {Sousa},\ and\ \citenamefont
  {Valenzuela}}]{GarciaEtxebarria:2020xsr}%
  \BibitemOpen
  \bibfield  {author} {\bibinfo {author} {\bibfnamefont {I.}~\bibnamefont
  {Garc{\'\i}a~Etxebarria}}, \bibinfo {author} {\bibfnamefont {M.}~\bibnamefont
  {Montero}}, \bibinfo {author} {\bibfnamefont {K.}~\bibnamefont {Sousa}},\
  and\ \bibinfo {author} {\bibfnamefont {I.}~\bibnamefont {Valenzuela}},\
  }\bibfield  {title} {\bibinfo {title} {{Nothing is certain in string
  compactifications}},\ }\href {https://doi.org/10.1007/JHEP12(2020)032}
  {\bibfield  {journal} {\bibinfo  {journal} {JHEP}\ }\textbf {\bibinfo
  {volume} {2020}}\bibfield  {number} {\bibinfo  {number} { (12)},\ \bibinfo
  {pages} {032}},\ }\Eprint {https://arxiv.org/abs/2005.06494}
  {arXiv:2005.06494 [hep-th]} \BibitemShut {NoStop}%
\bibitem [{\citenamefont {Godet}(2026)}]{Godet:2026gqw}%
  \BibitemOpen
  \bibfield  {author} {\bibinfo {author} {\bibfnamefont {V.}~\bibnamefont
  {Godet}},\ }\href@noop {} {\bibinfo {title} {{Inflation and topology from the
  no-boundary state}}} (\bibinfo {year} {2026}),\ \Eprint
  {https://arxiv.org/abs/2605.05317} {arXiv:2605.05317 [hep-th]} \BibitemShut
  {NoStop}%
\bibitem [{\citenamefont {Goncharov}\ and\ \citenamefont
  {Bytsenko}(1985)}]{Goncharov:1985gi}%
  \BibitemOpen
  \bibfield  {author} {\bibinfo {author} {\bibfnamefont {Y.~P.}\ \bibnamefont
  {Goncharov}}\ and\ \bibinfo {author} {\bibfnamefont {A.~A.}\ \bibnamefont
  {Bytsenko}},\ }\bibfield  {title} {\bibinfo {title} {{The Supersymmetric
  Casimir Effect and Quantum Creation of the Universe With Nontrivial
  Topology}},\ }\href {https://doi.org/10.1016/0370-2693(85)90006-1} {\bibfield
   {journal} {\bibinfo  {journal} {Phys. Lett. B}\ }\textbf {\bibinfo {volume}
  {160}},\ \bibinfo {pages} {385} (\bibinfo {year} {1985})}\BibitemShut
  {NoStop}%
\bibitem [{\citenamefont {Negro}\ \emph
  {et~al.}(2026{\natexlab{a}})\citenamefont {Negro}, \citenamefont
  {Hinterbichler}, \citenamefont {Starkman}, \citenamefont {Akrami},
  \citenamefont {Anselmi}, \citenamefont {Carr{\'o}n~Duque}, \citenamefont
  {Martin~Barandiaran},\ and\ \citenamefont {Tamosiunas}}]{Negro:2026ooh}%
  \BibitemOpen
  \bibfield  {author} {\bibinfo {author} {\bibfnamefont {A.}~\bibnamefont
  {Negro}}, \bibinfo {author} {\bibfnamefont {K.}~\bibnamefont
  {Hinterbichler}}, \bibinfo {author} {\bibfnamefont {G.~D.}\ \bibnamefont
  {Starkman}}, \bibinfo {author} {\bibfnamefont {Y.}~\bibnamefont {Akrami}},
  \bibinfo {author} {\bibfnamefont {S.}~\bibnamefont {Anselmi}}, \bibinfo
  {author} {\bibfnamefont {J.}~\bibnamefont {Carr{\'o}n~Duque}}, \bibinfo
  {author} {\bibfnamefont {M.}~\bibnamefont {Martin~Barandiaran}},\ and\
  \bibinfo {author} {\bibfnamefont {A.}~\bibnamefont {Tamosiunas}} (\bibinfo
  {collaboration} {COMPACT}),\ }\href@noop {} {\bibinfo {title} {{Breaking de
  Sitter Symmetry with Cosmic Topology: Cosmological Role of Infrared Vacuum
  Freedom}}} (\bibinfo {year} {2026}{\natexlab{a}}),\ \Eprint
  {https://arxiv.org/abs/2609.28637} {arXiv:2609.28637 [hep-th]} \BibitemShut
  {NoStop}%
\bibitem [{\citenamefont {Pavel}(2021)}]{Pavel:2021mxn}%
  \BibitemOpen
  \bibfield  {author} {\bibinfo {author} {\bibfnamefont {H.-P.}\ \bibnamefont
  {Pavel}},\ }\href@noop {} {\bibinfo {title} {{Low-energy spectrum of SU(3)
  Yang-Mills Quantum Mechanics}}} (\bibinfo {year} {2021}),\ \Eprint
  {https://arxiv.org/abs/2112.06248} {arXiv:2112.06248 [hep-th]} \BibitemShut
  {NoStop}%
\bibitem [{\citenamefont {McInnes}(2006)}]{McInnes:2005sa}%
  \BibitemOpen
  \bibfield  {author} {\bibinfo {author} {\bibfnamefont {B.}~\bibnamefont
  {McInnes}},\ }\bibfield  {title} {\bibinfo {title} {{Pre-inflationary
  spacetime in string cosmology}},\ }\href
  {https://doi.org/10.1016/j.nuclphysb.2006.04.035} {\bibfield  {journal}
  {\bibinfo  {journal} {Nucl. Phys. B}\ }\textbf {\bibinfo {volume} {748}},\
  \bibinfo {pages} {309} (\bibinfo {year} {2006})},\ \Eprint
  {https://arxiv.org/abs/hep-th/0511227} {arXiv:hep-th/0511227} \BibitemShut
  {NoStop}%
\bibitem [{\citenamefont {Parker}(1969)}]{Parker:1969au}%
  \BibitemOpen
  \bibfield  {author} {\bibinfo {author} {\bibfnamefont {L.}~\bibnamefont
  {Parker}},\ }\bibfield  {title} {\bibinfo {title} {{Quantized fields and
  particle creation in expanding universes. 1.}},\ }\href
  {https://doi.org/10.1103/PhysRev.183.1057} {\bibfield  {journal} {\bibinfo
  {journal} {Phys. Rev.}\ }\textbf {\bibinfo {volume} {183}},\ \bibinfo {pages}
  {1057} (\bibinfo {year} {1969})}\BibitemShut {NoStop}%
\bibitem [{\citenamefont {Egan}\ and\ \citenamefont
  {Lineweaver}(2010)}]{Egan:2009yy}%
  \BibitemOpen
  \bibfield  {author} {\bibinfo {author} {\bibfnamefont {C.~A.}\ \bibnamefont
  {Egan}}\ and\ \bibinfo {author} {\bibfnamefont {C.~H.}\ \bibnamefont
  {Lineweaver}},\ }\bibfield  {title} {\bibinfo {title} {{A Larger Estimate of
  the Entropy of the Universe}},\ }\href
  {https://doi.org/10.1088/0004-637X/710/2/1825} {\bibfield  {journal}
  {\bibinfo  {journal} {Astrophys. J.}\ }\textbf {\bibinfo {volume} {710}},\
  \bibinfo {pages} {1825} (\bibinfo {year} {2010})},\ \Eprint
  {https://arxiv.org/abs/0909.3983} {arXiv:0909.3983 [astro-ph.CO]}
  \BibitemShut {NoStop}%
\bibitem [{\citenamefont {Carroll}\ and\ \citenamefont
  {Chen}(2004)}]{Carroll:2004pn}%
  \BibitemOpen
  \bibfield  {author} {\bibinfo {author} {\bibfnamefont {S.~M.}\ \bibnamefont
  {Carroll}}\ and\ \bibinfo {author} {\bibfnamefont {J.}~\bibnamefont {Chen}},\
  }\href@noop {} {\bibinfo {title} {{Spontaneous inflation and the origin of
  the arrow of time}}} (\bibinfo {year} {2004}),\ \Eprint
  {https://arxiv.org/abs/hep-th/0410270} {arXiv:hep-th/0410270} \BibitemShut
  {NoStop}%
\bibitem [{\citenamefont {Barbour}\ \emph {et~al.}(2014)\citenamefont
  {Barbour}, \citenamefont {Koslowski},\ and\ \citenamefont
  {Mercati}}]{Barbour:2014bga}%
  \BibitemOpen
  \bibfield  {author} {\bibinfo {author} {\bibfnamefont {J.}~\bibnamefont
  {Barbour}}, \bibinfo {author} {\bibfnamefont {T.}~\bibnamefont {Koslowski}},\
  and\ \bibinfo {author} {\bibfnamefont {F.}~\bibnamefont {Mercati}},\
  }\bibfield  {title} {\bibinfo {title} {{Identification of a gravitational
  arrow of time}},\ }\href {https://doi.org/10.1103/PhysRevLett.113.181101}
  {\bibfield  {journal} {\bibinfo  {journal} {Phys. Rev. Lett.}\ }\textbf
  {\bibinfo {volume} {113}},\ \bibinfo {pages} {181101} (\bibinfo {year}
  {2014})},\ \Eprint {https://arxiv.org/abs/1409.0917} {arXiv:1409.0917
  [gr-qc]} \BibitemShut {NoStop}%
\bibitem [{\citenamefont {Hartle}\ and\ \citenamefont
  {Hertog}(2012)}]{Hartle:2011rb}%
  \BibitemOpen
  \bibfield  {author} {\bibinfo {author} {\bibfnamefont {J.}~\bibnamefont
  {Hartle}}\ and\ \bibinfo {author} {\bibfnamefont {T.}~\bibnamefont
  {Hertog}},\ }\bibfield  {title} {\bibinfo {title} {{Arrows of Time in the
  Bouncing Universes of the No-boundary Quantum State}},\ }\href
  {https://doi.org/10.1103/PhysRevD.85.103524} {\bibfield  {journal} {\bibinfo
  {journal} {Phys. Rev. D}\ }\textbf {\bibinfo {volume} {85}},\ \bibinfo
  {pages} {103524} (\bibinfo {year} {2012})},\ \Eprint
  {https://arxiv.org/abs/1104.1733} {arXiv:1104.1733 [hep-th]} \BibitemShut
  {NoStop}%
\bibitem [{\citenamefont {Aguirre}\ and\ \citenamefont
  {Gratton}(2002)}]{Aguirre:2001ks}%
  \BibitemOpen
  \bibfield  {author} {\bibinfo {author} {\bibfnamefont {A.}~\bibnamefont
  {Aguirre}}\ and\ \bibinfo {author} {\bibfnamefont {S.}~\bibnamefont
  {Gratton}},\ }\bibfield  {title} {\bibinfo {title} {{Steady state eternal
  inflation}},\ }\href {https://doi.org/10.1103/PhysRevD.65.083507} {\bibfield
  {journal} {\bibinfo  {journal} {Phys. Rev. D}\ }\textbf {\bibinfo {volume}
  {65}},\ \bibinfo {pages} {083507} (\bibinfo {year} {2002})},\ \Eprint
  {https://arxiv.org/abs/astro-ph/0111191} {arXiv:astro-ph/0111191}
  \BibitemShut {NoStop}%
\bibitem [{\citenamefont {Aguirre}\ and\ \citenamefont
  {Gratton}(2003)}]{Aguirre:2003ck}%
  \BibitemOpen
  \bibfield  {author} {\bibinfo {author} {\bibfnamefont {A.}~\bibnamefont
  {Aguirre}}\ and\ \bibinfo {author} {\bibfnamefont {S.}~\bibnamefont
  {Gratton}},\ }\bibfield  {title} {\bibinfo {title} {{Inflation without a
  beginning: A Null boundary proposal}},\ }\href
  {https://doi.org/10.1103/PhysRevD.67.083515} {\bibfield  {journal} {\bibinfo
  {journal} {Phys. Rev. D}\ }\textbf {\bibinfo {volume} {67}},\ \bibinfo
  {pages} {083515} (\bibinfo {year} {2003})},\ \Eprint
  {https://arxiv.org/abs/gr-qc/0301042} {arXiv:gr-qc/0301042} \BibitemShut
  {NoStop}%
\bibitem [{\citenamefont {Boyle}\ \emph {et~al.}(2018)\citenamefont {Boyle},
  \citenamefont {Finn},\ and\ \citenamefont {Turok}}]{Boyle:2018tzc}%
  \BibitemOpen
  \bibfield  {author} {\bibinfo {author} {\bibfnamefont {L.}~\bibnamefont
  {Boyle}}, \bibinfo {author} {\bibfnamefont {K.}~\bibnamefont {Finn}},\ and\
  \bibinfo {author} {\bibfnamefont {N.}~\bibnamefont {Turok}},\ }\bibfield
  {title} {\bibinfo {title} {{CPT}-symmetric universe},\ }\href
  {https://doi.org/10.1103/PhysRevLett.121.251301} {\bibfield  {journal}
  {\bibinfo  {journal} {Phys. Rev. Lett.}\ }\textbf {\bibinfo {volume} {121}},\
  \bibinfo {pages} {251301} (\bibinfo {year} {2018})},\ \Eprint
  {https://arxiv.org/abs/1803.08928} {arXiv:1803.08928} \BibitemShut {NoStop}%
\bibitem [{\citenamefont {Vilenkin}(2013)}]{Vilenkin:2013rza}%
  \BibitemOpen
  \bibfield  {author} {\bibinfo {author} {\bibfnamefont {A.}~\bibnamefont
  {Vilenkin}},\ }\bibfield  {title} {\bibinfo {title} {{Arrows of time and the
  beginning of the universe}},\ }\href
  {https://doi.org/10.1103/PhysRevD.88.043516} {\bibfield  {journal} {\bibinfo
  {journal} {Phys. Rev. D}\ }\textbf {\bibinfo {volume} {88}},\ \bibinfo
  {pages} {043516} (\bibinfo {year} {2013})},\ \Eprint
  {https://arxiv.org/abs/1305.3836} {arXiv:1305.3836 [hep-th]} \BibitemShut
  {NoStop}%
\bibitem [{\citenamefont {Garcia}\ \emph {et~al.}(2024)\citenamefont {Garcia},
  \citenamefont {Kaneta}, \citenamefont {Ke}, \citenamefont {Mambrini},
  \citenamefont {Olive},\ and\ \citenamefont {Verner}}]{Garcia:2023obw}%
  \BibitemOpen
  \bibfield  {author} {\bibinfo {author} {\bibfnamefont {M.~A.~G.}\
  \bibnamefont {Garcia}}, \bibinfo {author} {\bibfnamefont {K.}~\bibnamefont
  {Kaneta}}, \bibinfo {author} {\bibfnamefont {W.}~\bibnamefont {Ke}}, \bibinfo
  {author} {\bibfnamefont {Y.}~\bibnamefont {Mambrini}}, \bibinfo {author}
  {\bibfnamefont {K.~A.}\ \bibnamefont {Olive}},\ and\ \bibinfo {author}
  {\bibfnamefont {S.}~\bibnamefont {Verner}},\ }\bibfield  {title} {\bibinfo
  {title} {{The role of vectors in reheating}},\ }\href
  {https://doi.org/10.1088/1475-7516/2024/06/014} {\bibfield  {journal}
  {\bibinfo  {journal} {JCAP}\ }\textbf {\bibinfo {volume} {2024}}\bibfield
  {number} {\bibinfo  {number} { (06)},\ \bibinfo {pages} {014}},\ }\Eprint
  {https://arxiv.org/abs/2311.14794} {arXiv:2311.14794 [hep-ph]} \BibitemShut
  {NoStop}%
\bibitem [{\citenamefont {Battefeld}\ and\ \citenamefont
  {Peter}(2015)}]{Battefeld:2014uga}%
  \BibitemOpen
  \bibfield  {author} {\bibinfo {author} {\bibfnamefont {D.}~\bibnamefont
  {Battefeld}}\ and\ \bibinfo {author} {\bibfnamefont {P.}~\bibnamefont
  {Peter}},\ }\bibfield  {title} {\bibinfo {title} {{A Critical Review of
  Classical Bouncing Cosmologies}},\ }\href
  {https://doi.org/10.1016/j.physrep.2014.12.004} {\bibfield  {journal}
  {\bibinfo  {journal} {Phys. Rept.}\ }\textbf {\bibinfo {volume} {571}},\
  \bibinfo {pages} {1} (\bibinfo {year} {2015})},\ \Eprint
  {https://arxiv.org/abs/1406.2790} {arXiv:1406.2790 [astro-ph.CO]}
  \BibitemShut {NoStop}%
\bibitem [{\citenamefont {Barvinsky}\ and\ \citenamefont
  {Kamenshchik}(2006)}]{Barvinsky:2006uh}%
  \BibitemOpen
  \bibfield  {author} {\bibinfo {author} {\bibfnamefont {A.~O.}\ \bibnamefont
  {Barvinsky}}\ and\ \bibinfo {author} {\bibfnamefont {A.~Y.}\ \bibnamefont
  {Kamenshchik}},\ }\bibfield  {title} {\bibinfo {title} {{Cosmological
  landscape from nothing: Some like it hot}},\ }\href
  {https://doi.org/10.1088/1475-7516/2006/09/014} {\bibfield  {journal}
  {\bibinfo  {journal} {JCAP}\ }\textbf {\bibinfo {volume} {2006}}\bibfield
  {number} {\bibinfo  {number} { (09)},\ \bibinfo {pages} {014}},\ }\Eprint
  {https://arxiv.org/abs/hep-th/0605132} {arXiv:hep-th/0605132} \BibitemShut
  {NoStop}%
\bibitem [{\citenamefont {Obers}\ and\ \citenamefont
  {Pioline}(2000)}]{Obers:1999um}%
  \BibitemOpen
  \bibfield  {author} {\bibinfo {author} {\bibfnamefont {N.~A.}\ \bibnamefont
  {Obers}}\ and\ \bibinfo {author} {\bibfnamefont {B.}~\bibnamefont
  {Pioline}},\ }\bibfield  {title} {\bibinfo {title} {{Eisenstein series and
  string thresholds}},\ }\href {https://doi.org/10.1007/s002200050022}
  {\bibfield  {journal} {\bibinfo  {journal} {Commun. Math. Phys.}\ }\textbf
  {\bibinfo {volume} {209}},\ \bibinfo {pages} {275} (\bibinfo {year}
  {2000})},\ \Eprint {https://arxiv.org/abs/hep-th/9903113}
  {arXiv:hep-th/9903113} \BibitemShut {NoStop}%
\bibitem [{\citenamefont {Ponton}\ and\ \citenamefont
  {Poppitz}(2001)}]{Ponton:2001hq}%
  \BibitemOpen
  \bibfield  {author} {\bibinfo {author} {\bibfnamefont {E.}~\bibnamefont
  {Ponton}}\ and\ \bibinfo {author} {\bibfnamefont {E.}~\bibnamefont
  {Poppitz}},\ }\bibfield  {title} {\bibinfo {title} {{Casimir energy and
  radius stabilization in five and six dimensional orbifolds}},\ }\href
  {https://doi.org/10.1088/1126-6708/2001/06/019} {\bibfield  {journal}
  {\bibinfo  {journal} {JHEP}\ }\textbf {\bibinfo {volume} {2001}}\bibfield
  {number} {\bibinfo  {number} { (06)},\ \bibinfo {pages} {019}},\ }\Eprint
  {https://arxiv.org/abs/hep-ph/0105021} {arXiv:hep-ph/0105021} \BibitemShut
  {NoStop}%
\bibitem [{\citenamefont {Luo}\ and\ \citenamefont {Wang}(2023)}]{Luo:2022tqy}%
  \BibitemOpen
  \bibfield  {author} {\bibinfo {author} {\bibfnamefont {C.}~\bibnamefont
  {Luo}}\ and\ \bibinfo {author} {\bibfnamefont {Y.}~\bibnamefont {Wang}},\
  }\bibfield  {title} {\bibinfo {title} {{Casimir energy and modularity in
  higher-dimensional conformal field theories}},\ }\href
  {https://doi.org/10.1007/JHEP07(2023)028} {\bibfield  {journal} {\bibinfo
  {journal} {JHEP}\ }\textbf {\bibinfo {volume} {2023}}\bibfield  {number}
  {\bibinfo  {number} { (07)},\ \bibinfo {pages} {028}},\ }\Eprint
  {https://arxiv.org/abs/2212.14866} {arXiv:2212.14866 [hep-th]} \BibitemShut
  {NoStop}%
\bibitem [{\citenamefont {Aparici}\ \emph {et~al.}(2026)\citenamefont
  {Aparici}, \citenamefont {Basile},\ and\ \citenamefont
  {Risso}}]{Aparici:2025kjj}%
  \BibitemOpen
  \bibfield  {author} {\bibinfo {author} {\bibfnamefont {M.}~\bibnamefont
  {Aparici}}, \bibinfo {author} {\bibfnamefont {I.}~\bibnamefont {Basile}},\
  and\ \bibinfo {author} {\bibfnamefont {N.}~\bibnamefont {Risso}},\ }\bibfield
   {title} {\bibinfo {title} {{Instabilities in scale-separated Casimir
  vacua}},\ }\href {https://doi.org/10.1007/JHEP04(2026)018} {\bibfield
  {journal} {\bibinfo  {journal} {JHEP}\ }\textbf {\bibinfo {volume}
  {2026}}\bibfield  {number} {\bibinfo  {number} { (04)},\ \bibinfo {pages}
  {018}},\ }\Eprint {https://arxiv.org/abs/2507.17802} {arXiv:2507.17802
  [hep-th]} \BibitemShut {NoStop}%
\bibitem [{\citenamefont {Seriu}(1996)}]{Seriu:1995hc}%
  \BibitemOpen
  \bibfield  {author} {\bibinfo {author} {\bibfnamefont {M.}~\bibnamefont
  {Seriu}},\ }\bibfield  {title} {\bibinfo {title} {{Back reaction on the
  topological degrees of freedom in (2+1)-dimensional space-time}},\ }\href
  {https://doi.org/10.1103/PhysRevD.53.1889} {\bibfield  {journal} {\bibinfo
  {journal} {Phys. Rev. D}\ }\textbf {\bibinfo {volume} {53}},\ \bibinfo
  {pages} {1889} (\bibinfo {year} {1996})},\ \Eprint
  {https://arxiv.org/abs/gr-qc/9511046} {arXiv:gr-qc/9511046} \BibitemShut
  {NoStop}%
\bibitem [{\citenamefont {Negro}\ \emph
  {et~al.}(2026{\natexlab{b}})\citenamefont {Negro} \emph
  {et~al.}}]{COMPACT:2026vdj}%
  \BibitemOpen
  \bibfield  {author} {\bibinfo {author} {\bibfnamefont {A.}~\bibnamefont
  {Negro}} \emph {et~al.} (\bibinfo {collaboration} {COMPACT}),\ }\href@noop {}
  {\bibinfo {title} {{Quantum Signatures of Cosmic Topology: How Casimir
  Backreaction Transmits Isotropy Violation}}} (\bibinfo {year}
  {2026}{\natexlab{b}}),\ \Eprint {https://arxiv.org/abs/2603.12319}
  {arXiv:2603.12319 [hep-th]} \BibitemShut {NoStop}%
\bibitem [{\citenamefont {Ennola}(1964)}]{Ennola:1964epstein}%
  \BibitemOpen
  \bibfield  {author} {\bibinfo {author} {\bibfnamefont {V.}~\bibnamefont
  {Ennola}},\ }\bibfield  {title} {\bibinfo {title} {{On a problem about the
  Epstein zeta-function}},\ }\href {https://doi.org/10.1017/S0305004100038330}
  {\bibfield  {journal} {\bibinfo  {journal} {Math. Proc. Cambridge Philos.
  Soc.}\ }\textbf {\bibinfo {volume} {60}},\ \bibinfo {pages} {855} (\bibinfo
  {year} {1964})}\BibitemShut {NoStop}%
\bibitem [{\citenamefont {B{\'e}termin}(2019)}]{Betermin:2016cubic}%
  \BibitemOpen
  \bibfield  {author} {\bibinfo {author} {\bibfnamefont {L.}~\bibnamefont
  {B{\'e}termin}},\ }\bibfield  {title} {\bibinfo {title} {{Local optimality of
  cubic lattices for interaction energies}},\ }\href
  {https://doi.org/10.1007/s13324-017-0205-5} {\bibfield  {journal} {\bibinfo
  {journal} {Anal. Math. Phys.}\ }\textbf {\bibinfo {volume} {9}},\ \bibinfo
  {pages} {403} (\bibinfo {year} {2019})},\ \Eprint
  {https://arxiv.org/abs/1611.07798} {arXiv:1611.07798 [math-ph]} \BibitemShut
  {NoStop}%
\bibitem [{\citenamefont {Polarski}\ and\ \citenamefont
  {Starobinsky}(1996)}]{Polarski:1995jg}%
  \BibitemOpen
  \bibfield  {author} {\bibinfo {author} {\bibfnamefont {D.}~\bibnamefont
  {Polarski}}\ and\ \bibinfo {author} {\bibfnamefont {A.~A.}\ \bibnamefont
  {Starobinsky}},\ }\bibfield  {title} {\bibinfo {title} {{Semiclassicality and
  decoherence of cosmological perturbations}},\ }\href
  {https://doi.org/10.1088/0264-9381/13/3/006} {\bibfield  {journal} {\bibinfo
  {journal} {Class. Quant. Grav.}\ }\textbf {\bibinfo {volume} {13}},\ \bibinfo
  {pages} {377} (\bibinfo {year} {1996})},\ \Eprint
  {https://arxiv.org/abs/gr-qc/9504030} {arXiv:gr-qc/9504030} \BibitemShut
  {NoStop}%
\bibitem [{\citenamefont {Landau}\ and\ \citenamefont
  {Pollak}(1961)}]{LandauPollak1961}%
  \BibitemOpen
  \bibfield  {author} {\bibinfo {author} {\bibfnamefont {H.~J.}\ \bibnamefont
  {Landau}}\ and\ \bibinfo {author} {\bibfnamefont {H.~O.}\ \bibnamefont
  {Pollak}},\ }\bibfield  {title} {\bibinfo {title} {{Prolate spheroidal wave
  functions, Fourier analysis and uncertainty. II}},\ }\href
  {https://doi.org/10.1002/j.1538-7305.1961.tb03977.x} {\bibfield  {journal}
  {\bibinfo  {journal} {Bell Syst. Tech. J.}\ }\textbf {\bibinfo {volume}
  {40}},\ \bibinfo {pages} {65} (\bibinfo {year} {1961})}\BibitemShut {NoStop}%
\bibitem [{\citenamefont {Hawking}(1985)}]{Hawking:1985af}%
  \BibitemOpen
  \bibfield  {author} {\bibinfo {author} {\bibfnamefont {S.~W.}\ \bibnamefont
  {Hawking}},\ }\bibfield  {title} {\bibinfo {title} {The arrow of time in
  cosmology},\ }\href {https://doi.org/10.1103/PhysRevD.32.2489} {\bibfield
  {journal} {\bibinfo  {journal} {Phys. Rev. D}\ }\textbf {\bibinfo {volume}
  {32}},\ \bibinfo {pages} {2489} (\bibinfo {year} {1985})}\BibitemShut
  {NoStop}%
\bibitem [{\citenamefont {Page}(1985)}]{Page:1985ei}%
  \BibitemOpen
  \bibfield  {author} {\bibinfo {author} {\bibfnamefont {D.~N.}\ \bibnamefont
  {Page}},\ }\bibfield  {title} {\bibinfo {title} {{Will Entropy Decrease if
  the Universe Recollapses?}},\ }\href
  {https://doi.org/10.1103/PhysRevD.32.2496} {\bibfield  {journal} {\bibinfo
  {journal} {Phys. Rev. D}\ }\textbf {\bibinfo {volume} {32}},\ \bibinfo
  {pages} {2496} (\bibinfo {year} {1985})}\BibitemShut {NoStop}%
\bibitem [{\citenamefont {Kiefer}\ and\ \citenamefont
  {Zeh}(1995)}]{Kiefer:1994gp}%
  \BibitemOpen
  \bibfield  {author} {\bibinfo {author} {\bibfnamefont {C.}~\bibnamefont
  {Kiefer}}\ and\ \bibinfo {author} {\bibfnamefont {H.~D.}\ \bibnamefont
  {Zeh}},\ }\bibfield  {title} {\bibinfo {title} {Arrow of time in a
  recollapsing quantum universe},\ }\href
  {https://doi.org/10.1103/PhysRevD.51.4145} {\bibfield  {journal} {\bibinfo
  {journal} {Phys. Rev. D}\ }\textbf {\bibinfo {volume} {51}},\ \bibinfo
  {pages} {4145} (\bibinfo {year} {1995})},\ \Eprint
  {https://arxiv.org/abs/gr-qc/9402036} {arXiv:gr-qc/9402036} \BibitemShut
  {NoStop}%
\bibitem [{\citenamefont {Hartle}\ and\ \citenamefont
  {Hawking}(1983)}]{Hartle:1983ai}%
  \BibitemOpen
  \bibfield  {author} {\bibinfo {author} {\bibfnamefont {J.~B.}\ \bibnamefont
  {Hartle}}\ and\ \bibinfo {author} {\bibfnamefont {S.~W.}\ \bibnamefont
  {Hawking}},\ }\bibfield  {title} {\bibinfo {title} {{Wave Function of the
  Universe}},\ }\href {https://doi.org/10.1103/PhysRevD.28.2960} {\bibfield
  {journal} {\bibinfo  {journal} {Phys. Rev. D}\ }\textbf {\bibinfo {volume}
  {28}},\ \bibinfo {pages} {2960} (\bibinfo {year} {1983})}\BibitemShut
  {NoStop}%
\bibitem [{\citenamefont {Vilenkin}(1984)}]{Vilenkin:1984wp}%
  \BibitemOpen
  \bibfield  {author} {\bibinfo {author} {\bibfnamefont {A.}~\bibnamefont
  {Vilenkin}},\ }\bibfield  {title} {\bibinfo {title} {{Quantum Creation of
  Universes}},\ }\href {https://doi.org/10.1103/PhysRevD.30.509} {\bibfield
  {journal} {\bibinfo  {journal} {Phys. Rev. D}\ }\textbf {\bibinfo {volume}
  {30}},\ \bibinfo {pages} {509} (\bibinfo {year} {1984})}\BibitemShut
  {NoStop}%
\bibitem [{\citenamefont {Louko}\ and\ \citenamefont
  {Ruback}(1991)}]{Louko:1989up}%
  \BibitemOpen
  \bibfield  {author} {\bibinfo {author} {\bibfnamefont {J.}~\bibnamefont
  {Louko}}\ and\ \bibinfo {author} {\bibfnamefont {P.~J.}\ \bibnamefont
  {Ruback}},\ }\bibfield  {title} {\bibinfo {title} {{Spatially flat quantum
  cosmology}},\ }\href {https://doi.org/10.1088/0264-9381/8/1/013} {\bibfield
  {journal} {\bibinfo  {journal} {Class. Quant. Grav.}\ }\textbf {\bibinfo
  {volume} {8}},\ \bibinfo {pages} {91} (\bibinfo {year} {1991})}\BibitemShut
  {NoStop}%
\bibitem [{\citenamefont {Koller}\ and\ \citenamefont {van
  Baal}(1988)}]{Koller:1987fq}%
  \BibitemOpen
  \bibfield  {author} {\bibinfo {author} {\bibfnamefont {J.}~\bibnamefont
  {Koller}}\ and\ \bibinfo {author} {\bibfnamefont {P.}~\bibnamefont {van
  Baal}},\ }\bibfield  {title} {\bibinfo {title} {A nonperturbative analysis in
  finite volume gauge theory},\ }\href
  {https://doi.org/10.1016/0550-3213(88)90665-7} {\bibfield  {journal}
  {\bibinfo  {journal} {Nucl. Phys. B}\ }\textbf {\bibinfo {volume} {302}},\
  \bibinfo {pages} {1} (\bibinfo {year} {1988})}\BibitemShut {NoStop}%
\bibitem [{\citenamefont {Siegert}(1939)}]{Siegert:1939zz}%
  \BibitemOpen
  \bibfield  {author} {\bibinfo {author} {\bibfnamefont {A.~J.~F.}\
  \bibnamefont {Siegert}},\ }\bibfield  {title} {\bibinfo {title} {{On the
  Derivation of the Dispersion Formula for Nuclear Reactions}},\ }\href
  {https://doi.org/10.1103/PhysRev.56.750} {\bibfield  {journal} {\bibinfo
  {journal} {Phys. Rev.}\ }\textbf {\bibinfo {volume} {56}},\ \bibinfo {pages}
  {750} (\bibinfo {year} {1939})}\BibitemShut {NoStop}%
\bibitem [{\citenamefont {Gell-Mann}\ and\ \citenamefont
  {Hartle}(1994)}]{Gell-Mann:1991kdm}%
  \BibitemOpen
  \bibfield  {author} {\bibinfo {author} {\bibfnamefont {M.}~\bibnamefont
  {Gell-Mann}}\ and\ \bibinfo {author} {\bibfnamefont {J.~B.}\ \bibnamefont
  {Hartle}},\ }\bibfield  {title} {\bibinfo {title} {{Time symmetry and
  asymmetry in quantum mechanics and quantum cosmology}},\ }in\ \href@noop {}
  {\emph {\bibinfo {booktitle} {{Physical Origins of Time Asymmetry}}}},\
  \bibinfo {editor} {edited by\ \bibinfo {editor} {\bibfnamefont
  {J.}~\bibnamefont {Halliwell}}, \bibinfo {editor} {\bibfnamefont
  {J.}~\bibnamefont {Perez-Mercader}},\ and\ \bibinfo {editor} {\bibfnamefont
  {W.}~\bibnamefont {Zurek}}}\ (\bibinfo  {publisher} {Cambridge University
  Press},\ \bibinfo {address} {Cambridge},\ \bibinfo {year} {1994})\ \Eprint
  {https://arxiv.org/abs/gr-qc/9304023} {arXiv:gr-qc/9304023} \BibitemShut
  {NoStop}%
\bibitem [{\citenamefont {Goldstein}\ and\ \citenamefont
  {Page}(1995)}]{Goldstein:1994cp}%
  \BibitemOpen
  \bibfield  {author} {\bibinfo {author} {\bibfnamefont {S.}~\bibnamefont
  {Goldstein}}\ and\ \bibinfo {author} {\bibfnamefont {D.~N.}\ \bibnamefont
  {Page}},\ }\bibfield  {title} {\bibinfo {title} {{Linearly Positive
  Histories: Probabilities for a Robust Family of Sequences of Quantum
  Events}},\ }\href {https://doi.org/10.1103/PhysRevLett.74.3715} {\bibfield
  {journal} {\bibinfo  {journal} {Phys. Rev. Lett.}\ }\textbf {\bibinfo
  {volume} {74}},\ \bibinfo {pages} {3715} (\bibinfo {year} {1995})},\ \Eprint
  {https://arxiv.org/abs/gr-qc/9403055} {arXiv:gr-qc/9403055} \BibitemShut
  {NoStop}%
\bibitem [{\citenamefont {Butenschoen}\ \emph {et~al.}(2016)\citenamefont
  {Butenschoen}, \citenamefont {Dehnadi}, \citenamefont {Hoang}, \citenamefont
  {Mateu}, \citenamefont {Preisser},\ and\ \citenamefont
  {Stewart}}]{Butenschoen:2016lpz}%
  \BibitemOpen
  \bibfield  {author} {\bibinfo {author} {\bibfnamefont {M.}~\bibnamefont
  {Butenschoen}}, \bibinfo {author} {\bibfnamefont {B.}~\bibnamefont
  {Dehnadi}}, \bibinfo {author} {\bibfnamefont {A.~H.}\ \bibnamefont {Hoang}},
  \bibinfo {author} {\bibfnamefont {V.}~\bibnamefont {Mateu}}, \bibinfo
  {author} {\bibfnamefont {M.}~\bibnamefont {Preisser}},\ and\ \bibinfo
  {author} {\bibfnamefont {I.~W.}\ \bibnamefont {Stewart}},\ }\bibfield
  {title} {\bibinfo {title} {{Top Quark Mass Calibration for Monte Carlo Event
  Generators}},\ }\href {https://doi.org/10.1103/PhysRevLett.117.232001}
  {\bibfield  {journal} {\bibinfo  {journal} {Phys. Rev. Lett.}\ }\textbf
  {\bibinfo {volume} {117}},\ \bibinfo {pages} {232001} (\bibinfo {year}
  {2016})},\ \Eprint {https://arxiv.org/abs/1608.01318} {arXiv:1608.01318
  [hep-ph]} \BibitemShut {NoStop}%
\bibitem [{\citenamefont {Hoang}\ \emph {et~al.}(2018)\citenamefont {Hoang},
  \citenamefont {Jain}, \citenamefont {Lepenik}, \citenamefont {Mateu},
  \citenamefont {Preisser}, \citenamefont {Scimemi},\ and\ \citenamefont
  {Stewart}}]{Hoang:2017suc}%
  \BibitemOpen
  \bibfield  {author} {\bibinfo {author} {\bibfnamefont {A.~H.}\ \bibnamefont
  {Hoang}}, \bibinfo {author} {\bibfnamefont {A.}~\bibnamefont {Jain}},
  \bibinfo {author} {\bibfnamefont {C.}~\bibnamefont {Lepenik}}, \bibinfo
  {author} {\bibfnamefont {V.}~\bibnamefont {Mateu}}, \bibinfo {author}
  {\bibfnamefont {M.}~\bibnamefont {Preisser}}, \bibinfo {author}
  {\bibfnamefont {I.}~\bibnamefont {Scimemi}},\ and\ \bibinfo {author}
  {\bibfnamefont {I.~W.}\ \bibnamefont {Stewart}},\ }\bibfield  {title}
  {\bibinfo {title} {{The MSR mass and the $
  \mathcal{O}\left({\Lambda}_{\mathrm{QCD}}\right) $ renormalon sum rule}},\
  }\href {https://doi.org/10.1007/JHEP04(2018)003} {\bibfield  {journal}
  {\bibinfo  {journal} {JHEP}\ }\textbf {\bibinfo {volume} {2018}}\bibfield
  {number} {\bibinfo  {number} { (04)},\ \bibinfo {pages} {003}},\ }\Eprint
  {https://arxiv.org/abs/1704.01580} {arXiv:1704.01580 [hep-ph]} \BibitemShut
  {NoStop}%
\bibitem [{\citenamefont {Birrell}\ and\ \citenamefont
  {Davies}(1982)}]{Birrell:1982ix}%
  \BibitemOpen
  \bibfield  {author} {\bibinfo {author} {\bibfnamefont {N.~D.}\ \bibnamefont
  {Birrell}}\ and\ \bibinfo {author} {\bibfnamefont {P.~C.~W.}\ \bibnamefont
  {Davies}},\ }\href {https://doi.org/10.1017/CBO9780511622632} {\emph
  {\bibinfo {title} {{Quantum Fields in Curved Space}}}},\ Cambridge Monographs
  on Mathematical Physics\ (\bibinfo  {publisher} {Cambridge University
  Press},\ \bibinfo {address} {Cambridge, UK},\ \bibinfo {year}
  {1982})\BibitemShut {NoStop}%
\bibitem [{\citenamefont {Hook}\ \emph {et~al.}(2015)\citenamefont {Hook},
  \citenamefont {Kearney}, \citenamefont {Shakya},\ and\ \citenamefont
  {Zurek}}]{Hook:2014uia}%
  \BibitemOpen
  \bibfield  {author} {\bibinfo {author} {\bibfnamefont {A.}~\bibnamefont
  {Hook}}, \bibinfo {author} {\bibfnamefont {J.}~\bibnamefont {Kearney}},
  \bibinfo {author} {\bibfnamefont {B.}~\bibnamefont {Shakya}},\ and\ \bibinfo
  {author} {\bibfnamefont {K.~M.}\ \bibnamefont {Zurek}},\ }\bibfield  {title}
  {\bibinfo {title} {Probable or improbable universe? {C}orrelating electroweak
  vacuum instability with the scale of inflation},\ }\href
  {https://doi.org/10.1007/JHEP01(2015)061} {\bibfield  {journal} {\bibinfo
  {journal} {JHEP}\ }\textbf {\bibinfo {volume} {2015}}\bibfield  {number}
  {\bibinfo  {number} { (01)},\ \bibinfo {pages} {061}},\ }\Eprint
  {https://arxiv.org/abs/1404.5953} {arXiv:1404.5953} \BibitemShut {NoStop}%
\bibitem [{\citenamefont {Markkanen}\ \emph
  {et~al.}(2018{\natexlab{b}})\citenamefont {Markkanen}, \citenamefont {Nurmi},
  \citenamefont {Rajantie},\ and\ \citenamefont {Stopyra}}]{Markkanen:2018bfx}%
  \BibitemOpen
  \bibfield  {author} {\bibinfo {author} {\bibfnamefont {T.}~\bibnamefont
  {Markkanen}}, \bibinfo {author} {\bibfnamefont {S.}~\bibnamefont {Nurmi}},
  \bibinfo {author} {\bibfnamefont {A.}~\bibnamefont {Rajantie}},\ and\
  \bibinfo {author} {\bibfnamefont {S.}~\bibnamefont {Stopyra}},\ }\bibfield
  {title} {\bibinfo {title} {The 1-loop effective potential for the standard
  model in curved spacetime},\ }\href {https://doi.org/10.1007/JHEP06(2018)040}
  {\bibfield  {journal} {\bibinfo  {journal} {JHEP}\ }\textbf {\bibinfo
  {volume} {2018}}\bibfield  {number} {\bibinfo  {number} { (06)},\ \bibinfo
  {pages} {040}},\ }\Eprint {https://arxiv.org/abs/1804.02020}
  {arXiv:1804.02020} \BibitemShut {NoStop}%
\bibitem [{\citenamefont {Hawking}\ and\ \citenamefont
  {Moss}(1982)}]{Hawking:1981fz}%
  \BibitemOpen
  \bibfield  {author} {\bibinfo {author} {\bibfnamefont {S.~W.}\ \bibnamefont
  {Hawking}}\ and\ \bibinfo {author} {\bibfnamefont {I.~G.}\ \bibnamefont
  {Moss}},\ }\bibfield  {title} {\bibinfo {title} {{Supercooled Phase
  Transitions in the Very Early Universe}},\ }\href
  {https://doi.org/10.1016/0370-2693(82)90946-7} {\bibfield  {journal}
  {\bibinfo  {journal} {Phys. Lett. B}\ }\textbf {\bibinfo {volume} {110}},\
  \bibinfo {pages} {35} (\bibinfo {year} {1982})}\BibitemShut {NoStop}%
\bibitem [{\citenamefont {Martin}\ and\ \citenamefont
  {Robertson}(2019)}]{Martin:2019lqd}%
  \BibitemOpen
  \bibfield  {author} {\bibinfo {author} {\bibfnamefont {S.~P.}\ \bibnamefont
  {Martin}}\ and\ \bibinfo {author} {\bibfnamefont {D.~G.}\ \bibnamefont
  {Robertson}},\ }\bibfield  {title} {\bibinfo {title} {{Standard model
  parameters in the tadpole-free pure $\overline{\rm{MS}}$ scheme}},\ }\href
  {https://doi.org/10.1103/PhysRevD.100.073004} {\bibfield  {journal} {\bibinfo
   {journal} {Phys. Rev. D}\ }\textbf {\bibinfo {volume} {100}},\ \bibinfo
  {pages} {073004} (\bibinfo {year} {2019})},\ \Eprint
  {https://arxiv.org/abs/1907.02500} {arXiv:1907.02500 [hep-ph]} \BibitemShut
  {NoStop}%
\bibitem [{\citenamefont {Mihaila}\ \emph {et~al.}(2012)\citenamefont
  {Mihaila}, \citenamefont {Salomon},\ and\ \citenamefont
  {Steinhauser}}]{Mihaila:2012fm}%
  \BibitemOpen
  \bibfield  {author} {\bibinfo {author} {\bibfnamefont {L.~N.}\ \bibnamefont
  {Mihaila}}, \bibinfo {author} {\bibfnamefont {J.}~\bibnamefont {Salomon}},\
  and\ \bibinfo {author} {\bibfnamefont {M.}~\bibnamefont {Steinhauser}},\
  }\bibfield  {title} {\bibinfo {title} {{Gauge Coupling Beta Functions in the
  Standard Model to Three Loops}},\ }\href
  {https://doi.org/10.1103/PhysRevLett.108.151602} {\bibfield  {journal}
  {\bibinfo  {journal} {Phys. Rev. Lett.}\ }\textbf {\bibinfo {volume} {108}},\
  \bibinfo {pages} {151602} (\bibinfo {year} {2012})},\ \Eprint
  {https://arxiv.org/abs/1201.5868} {arXiv:1201.5868 [hep-ph]} \BibitemShut
  {NoStop}%
\bibitem [{\citenamefont {Bednyakov}\ \emph
  {et~al.}(2013{\natexlab{a}})\citenamefont {Bednyakov}, \citenamefont
  {Pikelner},\ and\ \citenamefont {Velizhanin}}]{Bednyakov:2012rb}%
  \BibitemOpen
  \bibfield  {author} {\bibinfo {author} {\bibfnamefont {A.~V.}\ \bibnamefont
  {Bednyakov}}, \bibinfo {author} {\bibfnamefont {A.~F.}\ \bibnamefont
  {Pikelner}},\ and\ \bibinfo {author} {\bibfnamefont {V.~N.}\ \bibnamefont
  {Velizhanin}},\ }\bibfield  {title} {\bibinfo {title} {{Anomalous dimensions
  of gauge fields and gauge coupling beta-functions in the Standard Model at
  three loops}},\ }\href {https://doi.org/10.1007/JHEP01(2013)017} {\bibfield
  {journal} {\bibinfo  {journal} {JHEP}\ }\textbf {\bibinfo {volume}
  {2013}}\bibfield  {number} {\bibinfo  {number} { (01)},\ \bibinfo {pages}
  {017}},\ }\Eprint {https://arxiv.org/abs/1210.6873} {arXiv:1210.6873
  [hep-ph]} \BibitemShut {NoStop}%
\bibitem [{\citenamefont {Bednyakov}\ \emph
  {et~al.}(2013{\natexlab{b}})\citenamefont {Bednyakov}, \citenamefont
  {Pikelner},\ and\ \citenamefont {Velizhanin}}]{Bednyakov:2012en}%
  \BibitemOpen
  \bibfield  {author} {\bibinfo {author} {\bibfnamefont {A.~V.}\ \bibnamefont
  {Bednyakov}}, \bibinfo {author} {\bibfnamefont {A.~F.}\ \bibnamefont
  {Pikelner}},\ and\ \bibinfo {author} {\bibfnamefont {V.~N.}\ \bibnamefont
  {Velizhanin}},\ }\bibfield  {title} {\bibinfo {title} {{Yukawa coupling
  beta-functions in the Standard Model at three loops}},\ }\href
  {https://doi.org/10.1016/j.physletb.2013.04.038} {\bibfield  {journal}
  {\bibinfo  {journal} {Phys. Lett. B}\ }\textbf {\bibinfo {volume} {722}},\
  \bibinfo {pages} {336} (\bibinfo {year} {2013}{\natexlab{b}})},\ \Eprint
  {https://arxiv.org/abs/1212.6829} {arXiv:1212.6829 [hep-ph]} \BibitemShut
  {NoStop}%
\bibitem [{\citenamefont {Bednyakov}\ \emph
  {et~al.}(2013{\natexlab{c}})\citenamefont {Bednyakov}, \citenamefont
  {Pikelner},\ and\ \citenamefont {Velizhanin}}]{Bednyakov:2013eba}%
  \BibitemOpen
  \bibfield  {author} {\bibinfo {author} {\bibfnamefont {A.~V.}\ \bibnamefont
  {Bednyakov}}, \bibinfo {author} {\bibfnamefont {A.~F.}\ \bibnamefont
  {Pikelner}},\ and\ \bibinfo {author} {\bibfnamefont {V.~N.}\ \bibnamefont
  {Velizhanin}},\ }\bibfield  {title} {\bibinfo {title} {{Higgs self-coupling
  beta-function in the Standard Model at three loops}},\ }\href
  {https://doi.org/10.1016/j.nuclphysb.2013.07.015} {\bibfield  {journal}
  {\bibinfo  {journal} {Nucl. Phys. B}\ }\textbf {\bibinfo {volume} {875}},\
  \bibinfo {pages} {552} (\bibinfo {year} {2013}{\natexlab{c}})},\ \Eprint
  {https://arxiv.org/abs/1303.4364} {arXiv:1303.4364 [hep-ph]} \BibitemShut
  {NoStop}%
\bibitem [{\citenamefont {Chetyrkin}\ and\ \citenamefont
  {Zoller}(2012)}]{Chetyrkin:2012rz}%
  \BibitemOpen
  \bibfield  {author} {\bibinfo {author} {\bibfnamefont {K.~G.}\ \bibnamefont
  {Chetyrkin}}\ and\ \bibinfo {author} {\bibfnamefont {M.~F.}\ \bibnamefont
  {Zoller}},\ }\bibfield  {title} {\bibinfo {title} {{Three-loop
  $\beta$-functions for top-Yukawa and the Higgs self-interaction in the
  Standard Model}},\ }\href {https://doi.org/10.1007/JHEP06(2012)033}
  {\bibfield  {journal} {\bibinfo  {journal} {JHEP}\ }\textbf {\bibinfo
  {volume} {2012}}\bibfield  {number} {\bibinfo  {number} { (06)},\ \bibinfo
  {pages} {033}},\ }\Eprint {https://arxiv.org/abs/1205.2892} {arXiv:1205.2892
  [hep-ph]} \BibitemShut {NoStop}%
\bibitem [{\citenamefont {Chetyrkin}\ and\ \citenamefont
  {Zoller}(2013)}]{Chetyrkin:2013wya}%
  \BibitemOpen
  \bibfield  {author} {\bibinfo {author} {\bibfnamefont {K.~G.}\ \bibnamefont
  {Chetyrkin}}\ and\ \bibinfo {author} {\bibfnamefont {M.~F.}\ \bibnamefont
  {Zoller}},\ }\bibfield  {title} {\bibinfo {title} {{$\beta$-function for the
  Higgs self-interaction in the Standard Model at three-loop level}},\ }\href
  {https://doi.org/10.1007/JHEP04(2013)091} {\bibfield  {journal} {\bibinfo
  {journal} {JHEP}\ }\textbf {\bibinfo {volume} {2013}}\bibfield  {number}
  {\bibinfo  {number} { (04)},\ \bibinfo {pages} {091}},\ }\bibinfo {note}
  {[Erratum: JHEP 09, 155 (2013)]},\ \Eprint {https://arxiv.org/abs/1303.2890}
  {arXiv:1303.2890 [hep-ph]} \BibitemShut {NoStop}%
\bibitem [{\citenamefont {Bednyakov}\ and\ \citenamefont
  {Pikelner}(2016)}]{Bednyakov:2015ooa}%
  \BibitemOpen
  \bibfield  {author} {\bibinfo {author} {\bibfnamefont {A.~V.}\ \bibnamefont
  {Bednyakov}}\ and\ \bibinfo {author} {\bibfnamefont {A.~F.}\ \bibnamefont
  {Pikelner}},\ }\bibfield  {title} {\bibinfo {title} {{Four-loop strong
  coupling beta-function in the Standard Model}},\ }\href
  {https://doi.org/10.1016/j.physletb.2016.09.007} {\bibfield  {journal}
  {\bibinfo  {journal} {Phys. Lett. B}\ }\textbf {\bibinfo {volume} {762}},\
  \bibinfo {pages} {151} (\bibinfo {year} {2016})},\ \Eprint
  {https://arxiv.org/abs/1508.02680} {arXiv:1508.02680 [hep-ph]} \BibitemShut
  {NoStop}%
\bibitem [{\citenamefont {Zoller}(2016)}]{Zoller:2015tha}%
  \BibitemOpen
  \bibfield  {author} {\bibinfo {author} {\bibfnamefont {M.~F.}\ \bibnamefont
  {Zoller}},\ }\bibfield  {title} {\bibinfo {title} {{Top-Yukawa effects on the
  $\beta$-function of the strong coupling in the SM at four-loop level}},\
  }\href {https://doi.org/10.1007/JHEP02(2016)095} {\bibfield  {journal}
  {\bibinfo  {journal} {JHEP}\ }\textbf {\bibinfo {volume} {2016}}\bibfield
  {number} {\bibinfo  {number} { (02)},\ \bibinfo {pages} {095}},\ }\Eprint
  {https://arxiv.org/abs/1508.03624} {arXiv:1508.03624 [hep-ph]} \BibitemShut
  {NoStop}%
\bibitem [{\citenamefont {Davies}\ \emph {et~al.}(2020)\citenamefont {Davies},
  \citenamefont {Herren}, \citenamefont {Poole}, \citenamefont {Steinhauser},\
  and\ \citenamefont {Thomsen}}]{Davies:2019onf}%
  \BibitemOpen
  \bibfield  {author} {\bibinfo {author} {\bibfnamefont {J.}~\bibnamefont
  {Davies}}, \bibinfo {author} {\bibfnamefont {F.}~\bibnamefont {Herren}},
  \bibinfo {author} {\bibfnamefont {C.}~\bibnamefont {Poole}}, \bibinfo
  {author} {\bibfnamefont {M.}~\bibnamefont {Steinhauser}},\ and\ \bibinfo
  {author} {\bibfnamefont {A.~E.}\ \bibnamefont {Thomsen}},\ }\bibfield
  {title} {\bibinfo {title} {{Gauge Coupling $\beta$ Functions to Four-Loop
  Order in the Standard Model}},\ }\href
  {https://doi.org/10.1103/PhysRevLett.124.071803} {\bibfield  {journal}
  {\bibinfo  {journal} {Phys. Rev. Lett.}\ }\textbf {\bibinfo {volume} {124}},\
  \bibinfo {pages} {071803} (\bibinfo {year} {2020})},\ \Eprint
  {https://arxiv.org/abs/1912.07624} {arXiv:1912.07624 [hep-ph]} \BibitemShut
  {NoStop}%
\bibitem [{\citenamefont {Martin}(2015)}]{Martin:2015eia}%
  \BibitemOpen
  \bibfield  {author} {\bibinfo {author} {\bibfnamefont {S.~P.}\ \bibnamefont
  {Martin}},\ }\bibfield  {title} {\bibinfo {title} {{Four-Loop Standard Model
  Effective Potential at Leading Order in QCD}},\ }\href
  {https://doi.org/10.1103/PhysRevD.92.054029} {\bibfield  {journal} {\bibinfo
  {journal} {Phys. Rev. D}\ }\textbf {\bibinfo {volume} {92}},\ \bibinfo
  {pages} {054029} (\bibinfo {year} {2015})},\ \Eprint
  {https://arxiv.org/abs/1508.00912} {arXiv:1508.00912 [hep-ph]} \BibitemShut
  {NoStop}%
\bibitem [{\citenamefont {Chetyrkin}\ and\ \citenamefont
  {Zoller}(2016)}]{Chetyrkin:2016ruf}%
  \BibitemOpen
  \bibfield  {author} {\bibinfo {author} {\bibfnamefont {K.~G.}\ \bibnamefont
  {Chetyrkin}}\ and\ \bibinfo {author} {\bibfnamefont {M.~F.}\ \bibnamefont
  {Zoller}},\ }\bibfield  {title} {\bibinfo {title} {{Leading QCD-induced
  four-loop contributions to the {\ensuremath{\beta}}-function of the Higgs
  self-coupling in the SM and vacuum stability}},\ }\href
  {https://doi.org/10.1007/JHEP06(2016)175} {\bibfield  {journal} {\bibinfo
  {journal} {JHEP}\ }\textbf {\bibinfo {volume} {2016}}\bibfield  {number}
  {\bibinfo  {number} { (06)},\ \bibinfo {pages} {175}},\ }\Eprint
  {https://arxiv.org/abs/1604.00853} {arXiv:1604.00853 [hep-ph]} \BibitemShut
  {NoStop}%
\bibitem [{\citenamefont {Baikov}\ \emph {et~al.}(2014)\citenamefont {Baikov},
  \citenamefont {Chetyrkin},\ and\ \citenamefont {K{\"u}hn}}]{Baikov:2014qja}%
  \BibitemOpen
  \bibfield  {author} {\bibinfo {author} {\bibfnamefont {P.~A.}\ \bibnamefont
  {Baikov}}, \bibinfo {author} {\bibfnamefont {K.~G.}\ \bibnamefont
  {Chetyrkin}},\ and\ \bibinfo {author} {\bibfnamefont {J.~H.}\ \bibnamefont
  {K{\"u}hn}},\ }\bibfield  {title} {\bibinfo {title} {{Quark Mass and Field
  Anomalous Dimensions to ${\cal O}(\alpha_s^5)$}},\ }\href
  {https://doi.org/10.1007/JHEP10(2014)076} {\bibfield  {journal} {\bibinfo
  {journal} {JHEP}\ }\textbf {\bibinfo {volume} {2014}}\bibfield  {number}
  {\bibinfo  {number} { (10)},\ \bibinfo {pages} {076}},\ }\Eprint
  {https://arxiv.org/abs/1402.6611} {arXiv:1402.6611 [hep-ph]} \BibitemShut
  {NoStop}%
\bibitem [{\citenamefont {Baikov}\ \emph {et~al.}(2017)\citenamefont {Baikov},
  \citenamefont {Chetyrkin},\ and\ \citenamefont {K{\"u}hn}}]{Baikov:2016tgj}%
  \BibitemOpen
  \bibfield  {author} {\bibinfo {author} {\bibfnamefont {P.~A.}\ \bibnamefont
  {Baikov}}, \bibinfo {author} {\bibfnamefont {K.~G.}\ \bibnamefont
  {Chetyrkin}},\ and\ \bibinfo {author} {\bibfnamefont {J.~H.}\ \bibnamefont
  {K{\"u}hn}},\ }\bibfield  {title} {\bibinfo {title} {{Five-Loop Running of
  the QCD Coupling Constant}},\ }\href
  {https://doi.org/10.1103/PhysRevLett.118.082002} {\bibfield  {journal}
  {\bibinfo  {journal} {Phys. Rev. Lett.}\ }\textbf {\bibinfo {volume} {118}},\
  \bibinfo {pages} {082002} (\bibinfo {year} {2017})},\ \Eprint
  {https://arxiv.org/abs/1606.08659} {arXiv:1606.08659 [hep-ph]} \BibitemShut
  {NoStop}%
\bibitem [{\citenamefont {Herzog}\ \emph {et~al.}(2017)\citenamefont {Herzog},
  \citenamefont {Ruijl}, \citenamefont {Ueda}, \citenamefont {Vermaseren},\
  and\ \citenamefont {Vogt}}]{Herzog:2017ohr}%
  \BibitemOpen
  \bibfield  {author} {\bibinfo {author} {\bibfnamefont {F.}~\bibnamefont
  {Herzog}}, \bibinfo {author} {\bibfnamefont {B.}~\bibnamefont {Ruijl}},
  \bibinfo {author} {\bibfnamefont {T.}~\bibnamefont {Ueda}}, \bibinfo {author}
  {\bibfnamefont {J.~A.~M.}\ \bibnamefont {Vermaseren}},\ and\ \bibinfo
  {author} {\bibfnamefont {A.}~\bibnamefont {Vogt}},\ }\bibfield  {title}
  {\bibinfo {title} {{The five-loop beta function of Yang-Mills theory with
  fermions}},\ }\href {https://doi.org/10.1007/JHEP02(2017)090} {\bibfield
  {journal} {\bibinfo  {journal} {JHEP}\ }\textbf {\bibinfo {volume}
  {2017}}\bibfield  {number} {\bibinfo  {number} { (02)},\ \bibinfo {pages}
  {090}},\ }\Eprint {https://arxiv.org/abs/1701.01404} {arXiv:1701.01404
  [hep-ph]} \BibitemShut {NoStop}%
\bibitem [{\citenamefont {Marquard}\ \emph {et~al.}(2015)\citenamefont
  {Marquard}, \citenamefont {Smirnov}, \citenamefont {Smirnov},\ and\
  \citenamefont {Steinhauser}}]{Marquard:2015qpa}%
  \BibitemOpen
  \bibfield  {author} {\bibinfo {author} {\bibfnamefont {P.}~\bibnamefont
  {Marquard}}, \bibinfo {author} {\bibfnamefont {A.~V.}\ \bibnamefont
  {Smirnov}}, \bibinfo {author} {\bibfnamefont {V.~A.}\ \bibnamefont
  {Smirnov}},\ and\ \bibinfo {author} {\bibfnamefont {M.}~\bibnamefont
  {Steinhauser}},\ }\bibfield  {title} {\bibinfo {title} {{Quark Mass Relations
  to Four-Loop Order in Perturbative QCD}},\ }\href
  {https://doi.org/10.1103/PhysRevLett.114.142002} {\bibfield  {journal}
  {\bibinfo  {journal} {Phys. Rev. Lett.}\ }\textbf {\bibinfo {volume} {114}},\
  \bibinfo {pages} {142002} (\bibinfo {year} {2015})},\ \Eprint
  {https://arxiv.org/abs/1502.01030} {arXiv:1502.01030 [hep-ph]} \BibitemShut
  {NoStop}%
\bibitem [{\citenamefont {Beneke}\ \emph {et~al.}(2017)\citenamefont {Beneke},
  \citenamefont {Marquard}, \citenamefont {Nason},\ and\ \citenamefont
  {Steinhauser}}]{Beneke:2016cbu}%
  \BibitemOpen
  \bibfield  {author} {\bibinfo {author} {\bibfnamefont {M.}~\bibnamefont
  {Beneke}}, \bibinfo {author} {\bibfnamefont {P.}~\bibnamefont {Marquard}},
  \bibinfo {author} {\bibfnamefont {P.}~\bibnamefont {Nason}},\ and\ \bibinfo
  {author} {\bibfnamefont {M.}~\bibnamefont {Steinhauser}},\ }\bibfield
  {title} {\bibinfo {title} {{On the ultimate uncertainty of the top quark pole
  mass}},\ }\href {https://doi.org/10.1016/j.physletb.2017.10.054} {\bibfield
  {journal} {\bibinfo  {journal} {Phys. Lett. B}\ }\textbf {\bibinfo {volume}
  {775}},\ \bibinfo {pages} {63} (\bibinfo {year} {2017})},\ \Eprint
  {https://arxiv.org/abs/1605.03609} {arXiv:1605.03609 [hep-ph]} \BibitemShut
  {NoStop}%
\bibitem [{\citenamefont {Camphuis}\ \emph {et~al.}(2026)\citenamefont
  {Camphuis} \emph {et~al.}}]{SPT-3G:2025bzu}%
  \BibitemOpen
  \bibfield  {author} {\bibinfo {author} {\bibfnamefont {E.}~\bibnamefont
  {Camphuis}} \emph {et~al.} (\bibinfo {collaboration} {SPT-3G}),\ }\bibfield
  {title} {\bibinfo {title} {{SPT-3G D1: CMB temperature and polarization power
  spectra and cosmology from 2019 and 2020 observations of the SPT-3G main
  field}},\ }\href {https://doi.org/10.1103/7wt3-9v2y} {\bibfield  {journal}
  {\bibinfo  {journal} {Phys. Rev. D}\ }\textbf {\bibinfo {volume} {113}},\
  \bibinfo {pages} {083504} (\bibinfo {year} {2026})},\ \Eprint
  {https://arxiv.org/abs/2506.20707} {arXiv:2506.20707 [astro-ph.CO]}
  \BibitemShut {NoStop}%
\bibitem [{\citenamefont {Abdul~Karim}\ \emph {et~al.}(2025)\citenamefont
  {Abdul~Karim} \emph {et~al.}}]{DESI:2025zgx}%
  \BibitemOpen
  \bibfield  {author} {\bibinfo {author} {\bibfnamefont {M.}~\bibnamefont
  {Abdul~Karim}} \emph {et~al.} (\bibinfo {collaboration} {DESI}),\ }\bibfield
  {title} {\bibinfo {title} {{DESI DR2 results. II. Measurements of baryon
  acoustic oscillations and cosmological constraints}},\ }\href
  {https://doi.org/10.1103/tr6y-kpc6} {\bibfield  {journal} {\bibinfo
  {journal} {Phys. Rev. D}\ }\textbf {\bibinfo {volume} {112}},\ \bibinfo
  {pages} {083515} (\bibinfo {year} {2025})},\ \Eprint
  {https://arxiv.org/abs/2503.14738} {arXiv:2503.14738 [astro-ph.CO]}
  \BibitemShut {NoStop}%
\bibitem [{\citenamefont {Louis}\ \emph {et~al.}(2025)\citenamefont {Louis}
  \emph {et~al.}}]{AtacamaCosmologyTelescope:2025blo}%
  \BibitemOpen
  \bibfield  {author} {\bibinfo {author} {\bibfnamefont {T.}~\bibnamefont
  {Louis}} \emph {et~al.} (\bibinfo {collaboration} {Atacama Cosmology
  Telescope}),\ }\bibfield  {title} {\bibinfo {title} {{The Atacama Cosmology
  Telescope: DR6 power spectra, likelihoods and {\ensuremath{\Lambda}}CDM
  parameters}},\ }\href {https://doi.org/10.1088/1475-7516/2025/11/062}
  {\bibfield  {journal} {\bibinfo  {journal} {JCAP}\ }\textbf {\bibinfo
  {volume} {2025}}\bibfield  {number} {\bibinfo  {number} { (11)},\ \bibinfo
  {pages} {062}},\ }\Eprint {https://arxiv.org/abs/2503.14452}
  {arXiv:2503.14452 [astro-ph.CO]} \BibitemShut {NoStop}%
\bibitem [{\citenamefont {Ade}\ \emph {et~al.}(2021)\citenamefont {Ade} \emph
  {et~al.}}]{BICEP:2021xfz}%
  \BibitemOpen
  \bibfield  {author} {\bibinfo {author} {\bibfnamefont {P.~A.~R.}\
  \bibnamefont {Ade}} \emph {et~al.} (\bibinfo {collaboration} {BICEP, Keck}),\
  }\bibfield  {title} {\bibinfo {title} {{Improved Constraints on Primordial
  Gravitational Waves using Planck, WMAP, and BICEP/Keck Observations through
  the 2018 Observing Season}},\ }\href
  {https://doi.org/10.1103/PhysRevLett.127.151301} {\bibfield  {journal}
  {\bibinfo  {journal} {Phys. Rev. Lett.}\ }\textbf {\bibinfo {volume} {127}},\
  \bibinfo {pages} {151301} (\bibinfo {year} {2021})},\ \Eprint
  {https://arxiv.org/abs/2110.00483} {arXiv:2110.00483 [astro-ph.CO]}
  \BibitemShut {NoStop}%
\bibitem [{\citenamefont {Allys}\ \emph {et~al.}(2023)\citenamefont {Allys}
  \emph {et~al.}}]{LiteBIRD:2022cnt}%
  \BibitemOpen
  \bibfield  {author} {\bibinfo {author} {\bibfnamefont {E.}~\bibnamefont
  {Allys}} \emph {et~al.} (\bibinfo {collaboration} {LiteBIRD}),\ }\bibfield
  {title} {\bibinfo {title} {{Probing Cosmic Inflation with the LiteBIRD Cosmic
  Microwave Background Polarization Survey}},\ }\href
  {https://doi.org/10.1093/ptep/ptac150} {\bibfield  {journal} {\bibinfo
  {journal} {PTEP}\ }\textbf {\bibinfo {volume} {2023}},\ \bibinfo {pages}
  {042F01} (\bibinfo {year} {2023})},\ \Eprint
  {https://arxiv.org/abs/2202.02773} {arXiv:2202.02773 [astro-ph.IM]}
  \BibitemShut {NoStop}%
\bibitem [{\citenamefont {Lewis}\ \emph {et~al.}(2000)\citenamefont {Lewis},
  \citenamefont {Challinor},\ and\ \citenamefont {Lasenby}}]{Lewis:1999bs}%
  \BibitemOpen
  \bibfield  {author} {\bibinfo {author} {\bibfnamefont {A.}~\bibnamefont
  {Lewis}}, \bibinfo {author} {\bibfnamefont {A.}~\bibnamefont {Challinor}},\
  and\ \bibinfo {author} {\bibfnamefont {A.}~\bibnamefont {Lasenby}},\
  }\bibfield  {title} {\bibinfo {title} {{Efficient computation of CMB
  anisotropies in closed FRW models}},\ }\href {https://doi.org/10.1086/309179}
  {\bibfield  {journal} {\bibinfo  {journal} {Astrophys. J.}\ }\textbf
  {\bibinfo {volume} {538}},\ \bibinfo {pages} {473} (\bibinfo {year}
  {2000})},\ \Eprint {https://arxiv.org/abs/astro-ph/9911177}
  {arXiv:astro-ph/9911177} \BibitemShut {NoStop}%
\bibitem [{\citenamefont {Aghanim}\ \emph {et~al.}(2020)\citenamefont {Aghanim}
  \emph {et~al.}}]{Planck:2018vyg}%
  \BibitemOpen
  \bibfield  {author} {\bibinfo {author} {\bibfnamefont {N.}~\bibnamefont
  {Aghanim}} \emph {et~al.} (\bibinfo {collaboration} {Planck}),\ }\bibfield
  {title} {\bibinfo {title} {{Planck 2018 results. VI. Cosmological
  parameters}},\ }\href {https://doi.org/10.1051/0004-6361/201833910}
  {\bibfield  {journal} {\bibinfo  {journal} {Astron. Astrophys.}\ }\textbf
  {\bibinfo {volume} {641}},\ \bibinfo {pages} {A6} (\bibinfo {year} {2020})},\
  \bibinfo {note} {erratum: Astron. Astrophys. 652, C4 (2021)},\ \Eprint
  {https://arxiv.org/abs/1807.06209} {arXiv:1807.06209 [astro-ph.CO]}
  \BibitemShut {NoStop}%
\bibitem [{\citenamefont {Ade}\ \emph {et~al.}(2014)\citenamefont {Ade} \emph
  {et~al.}}]{Planck:2013okc}%
  \BibitemOpen
  \bibfield  {author} {\bibinfo {author} {\bibfnamefont {P.~A.~R.}\
  \bibnamefont {Ade}} \emph {et~al.} (\bibinfo {collaboration} {Planck}),\
  }\bibfield  {title} {\bibinfo {title} {{Planck 2013 results. XXVI. Background
  geometry and topology of the Universe}},\ }\href
  {https://doi.org/10.1051/0004-6361/201321546} {\bibfield  {journal} {\bibinfo
   {journal} {Astron. Astrophys.}\ }\textbf {\bibinfo {volume} {571}},\
  \bibinfo {pages} {A26} (\bibinfo {year} {2014})},\ \Eprint
  {https://arxiv.org/abs/1303.5086} {arXiv:1303.5086 [astro-ph.CO]}
  \BibitemShut {NoStop}%
\bibitem [{\citenamefont {Ade}\ \emph {et~al.}(2016)\citenamefont {Ade} \emph
  {et~al.}}]{Planck:2015gmu}%
  \BibitemOpen
  \bibfield  {author} {\bibinfo {author} {\bibfnamefont {P.~A.~R.}\
  \bibnamefont {Ade}} \emph {et~al.} (\bibinfo {collaboration} {Planck}),\
  }\bibfield  {title} {\bibinfo {title} {{Planck 2015 results - XVIII.
  Background geometry and topology of the Universe}},\ }\href
  {https://doi.org/10.1051/0004-6361/201525829} {\bibfield  {journal} {\bibinfo
   {journal} {Astron. Astrophys.}\ }\textbf {\bibinfo {volume} {594}},\
  \bibinfo {pages} {A18} (\bibinfo {year} {2016})},\ \Eprint
  {https://arxiv.org/abs/1502.01593} {arXiv:1502.01593 [astro-ph.CO]}
  \BibitemShut {NoStop}%
\bibitem [{\citenamefont {Freitas}\ \emph {et~al.}(2019)\citenamefont {Freitas}
  \emph {et~al.}}]{Freitas:2019bre}%
  \BibitemOpen
  \bibfield  {author} {\bibinfo {author} {\bibfnamefont {A.}~\bibnamefont
  {Freitas}} \emph {et~al.},\ }\href@noop {} {\bibinfo {title} {{Theoretical
  uncertainties for electroweak and Higgs-boson precision measurements at
  FCC-ee}}} (\bibinfo {year} {2019}),\ \Eprint
  {https://arxiv.org/abs/1906.05379} {arXiv:1906.05379 [hep-ph]} \BibitemShut
  {NoStop}%
\bibitem [{\citenamefont {Benedikt}\ \emph {et~al.}(2025)\citenamefont
  {Benedikt} \emph {et~al.}}]{FCC:2025lpp}%
  \BibitemOpen
  \bibfield  {author} {\bibinfo {author} {\bibfnamefont {M.}~\bibnamefont
  {Benedikt}} \emph {et~al.} (\bibinfo {collaboration} {FCC}),\ }\bibfield
  {title} {\bibinfo {title} {{Future Circular Collider Feasibility Study
  Report: Volume 1, Physics, Experiments, Detectors}},\ }\href
  {https://doi.org/10.1140/epjc/s10052-025-15077-x} {\bibfield  {journal}
  {\bibinfo  {journal} {Eur. Phys. J. C}\ }\textbf {\bibinfo {volume} {85}},\
  \bibinfo {pages} {1468} (\bibinfo {year} {2025})},\ \bibinfo {note}
  {[Erratum: Eur. Phys. J. C 86, 844 (2026)]},\ \Eprint
  {https://arxiv.org/abs/2505.00272} {arXiv:2505.00272 [hep-ex]} \BibitemShut
  {NoStop}%
\bibitem [{\citenamefont {Watanabe}\ \emph {et~al.}(2009)\citenamefont
  {Watanabe}, \citenamefont {Kanno},\ and\ \citenamefont
  {Soda}}]{Watanabe:2009ct}%
  \BibitemOpen
  \bibfield  {author} {\bibinfo {author} {\bibfnamefont {M.-a.}\ \bibnamefont
  {Watanabe}}, \bibinfo {author} {\bibfnamefont {S.}~\bibnamefont {Kanno}},\
  and\ \bibinfo {author} {\bibfnamefont {J.}~\bibnamefont {Soda}},\ }\bibfield
  {title} {\bibinfo {title} {{Inflationary Universe with Anisotropic Hair}},\
  }\href {https://doi.org/10.1103/PhysRevLett.102.191302} {\bibfield  {journal}
  {\bibinfo  {journal} {Phys. Rev. Lett.}\ }\textbf {\bibinfo {volume} {102}},\
  \bibinfo {pages} {191302} (\bibinfo {year} {2009})},\ \Eprint
  {https://arxiv.org/abs/0902.2833} {arXiv:0902.2833 [hep-th]} \BibitemShut
  {NoStop}%
\bibitem [{\citenamefont {Maleknejad}\ and\ \citenamefont
  {Sheikh-Jabbari}(2013)}]{Maleknejad:2011jw}%
  \BibitemOpen
  \bibfield  {author} {\bibinfo {author} {\bibfnamefont {A.}~\bibnamefont
  {Maleknejad}}\ and\ \bibinfo {author} {\bibfnamefont {M.~M.}\ \bibnamefont
  {Sheikh-Jabbari}},\ }\bibfield  {title} {\bibinfo {title} {Gauge-flation:
  Inflation from non-{Abelian} gauge fields},\ }\href
  {https://doi.org/10.1016/j.physletb.2013.05.001} {\bibfield  {journal}
  {\bibinfo  {journal} {Phys. Lett. B}\ }\textbf {\bibinfo {volume} {723}},\
  \bibinfo {pages} {224} (\bibinfo {year} {2013})},\ \Eprint
  {https://arxiv.org/abs/1102.1513} {arXiv:1102.1513} \BibitemShut {NoStop}%
\bibitem [{\citenamefont {Adshead}\ and\ \citenamefont
  {Wyman}(2012)}]{Adshead:2012kp}%
  \BibitemOpen
  \bibfield  {author} {\bibinfo {author} {\bibfnamefont {P.}~\bibnamefont
  {Adshead}}\ and\ \bibinfo {author} {\bibfnamefont {M.}~\bibnamefont
  {Wyman}},\ }\bibfield  {title} {\bibinfo {title} {Chromo-natural inflation:
  Natural inflation on a steep potential with classical non-{Abelian} gauge
  fields},\ }\href {https://doi.org/10.1103/PhysRevLett.108.261302} {\bibfield
  {journal} {\bibinfo  {journal} {Phys. Rev. Lett.}\ }\textbf {\bibinfo
  {volume} {108}},\ \bibinfo {pages} {261302} (\bibinfo {year} {2012})},\
  \Eprint {https://arxiv.org/abs/1202.2366} {arXiv:1202.2366} \BibitemShut
  {NoStop}%
\bibitem [{\citenamefont {Endlich}\ \emph {et~al.}(2013)\citenamefont
  {Endlich}, \citenamefont {Nicolis},\ and\ \citenamefont
  {Wang}}]{Endlich:2012pz}%
  \BibitemOpen
  \bibfield  {author} {\bibinfo {author} {\bibfnamefont {S.}~\bibnamefont
  {Endlich}}, \bibinfo {author} {\bibfnamefont {A.}~\bibnamefont {Nicolis}},\
  and\ \bibinfo {author} {\bibfnamefont {J.}~\bibnamefont {Wang}},\ }\bibfield
  {title} {\bibinfo {title} {{Solid Inflation}},\ }\href
  {https://doi.org/10.1088/1475-7516/2013/10/011} {\bibfield  {journal}
  {\bibinfo  {journal} {JCAP}\ }\textbf {\bibinfo {volume} {2013}}\bibfield
  {number} {\bibinfo  {number} { (10)},\ \bibinfo {pages} {011}},\ }\Eprint
  {https://arxiv.org/abs/1210.0569} {arXiv:1210.0569 [hep-th]} \BibitemShut
  {NoStop}%
\bibitem [{\citenamefont {Slepian}(1964)}]{Slepian1964}%
  \BibitemOpen
  \bibfield  {author} {\bibinfo {author} {\bibfnamefont {D.}~\bibnamefont
  {Slepian}},\ }\bibfield  {title} {\bibinfo {title} {{Prolate spheroidal wave
  functions, Fourier analysis and uncertainty. IV: Extensions to many
  dimensions; generalized prolate spheroidal functions}},\ }\href
  {https://doi.org/10.1002/j.1538-7305.1964.tb01037.x} {\bibfield  {journal}
  {\bibinfo  {journal} {Bell Syst. Tech. J.}\ }\textbf {\bibinfo {volume}
  {43}},\ \bibinfo {pages} {3009} (\bibinfo {year} {1964})}\BibitemShut
  {NoStop}%
\bibitem [{\citenamefont {Coleman}\ and\ \citenamefont
  {Weinberg}(1973)}]{Coleman:1973jx}%
  \BibitemOpen
  \bibfield  {author} {\bibinfo {author} {\bibfnamefont {S.~R.}\ \bibnamefont
  {Coleman}}\ and\ \bibinfo {author} {\bibfnamefont {E.~J.}\ \bibnamefont
  {Weinberg}},\ }\bibfield  {title} {\bibinfo {title} {{Radiative Corrections
  as the Origin of Spontaneous Symmetry Breaking}},\ }\href
  {https://doi.org/10.1103/PhysRevD.7.1888} {\bibfield  {journal} {\bibinfo
  {journal} {Phys. Rev. D}\ }\textbf {\bibinfo {volume} {7}},\ \bibinfo {pages}
  {1888} (\bibinfo {year} {1973})}\BibitemShut {NoStop}%
\end{thebibliography}%
